\documentclass[12pt]{article}

\usepackage{graphicx}
\usepackage{float}
\usepackage{cite}
\usepackage{amsfonts}
\usepackage{amssymb}
\usepackage{relsize}
\usepackage[compatibility=false]{caption}
\usepackage{subcaption}

\def\gtwid{\mathrel{\raise.3ex\hbox{$>$\kern-.75em\lower1ex\hbox{$\sim$}}}}
\def\ltwid{\mathrel{\raise.3ex\hbox{$<$\kern-.75em\lower1ex\hbox{$\sim$}}}}
\def\square{\kern1pt\vbox{\hrule height 1.2pt\hbox{\vrule width 1.2pt\hskip 3pt
   \vbox{\vskip 6pt}\hskip 3pt\vrule width 0.6pt}\hrule height 0.6pt}\kern1pt}

\begin{document}

\begin{titlepage}

\begin{flushright}
UFIFT-QG-20-06
\end{flushright}

\vskip 0.2cm

\begin{center}
{\bf Inflaton Effective Potential from Photons for General $\epsilon$}
\end{center}

\vskip 0.2cm

\begin{center}
S. Katuwal$^{1*}$, S. P. Miao$^{2\star}$ and R. P. Woodard$^{1\dagger}$
\end{center}

\begin{center}
\it{$^{1}$ Department of Physics, University of Florida,\\
Gainesville, FL 32611, UNITED STATES}
\end{center}

\begin{center}
\it{$^{2}$ Department of Physics, National Cheng Kung University, \\
No. 1 University Road, Tainan City 70101, TAIWAN}
\end{center}

\vspace{0.2cm}

\begin{center}
ABSTRACT
\end{center}
We accurately approximate the contribution that photons make to the 
effective potential of a charged inflaton for inflationary geometries 
with an arbitrary first slow roll parameter $\epsilon$. We find a 
small, nonlocal contribution and a numerically larger, local part. 
The local part involves first and second derivatives of $\epsilon$, 
coming exclusively from the constrained part of the electromagnetic 
field which carries the long range interaction. This causes the 
effective potential induced by electromagnetism to respond more 
strongly to geometrical evolution than for either scalars, which have
no derivatives, or spin one half particles, which have only one
derivative. For $\epsilon = 0$ our final result agrees with that 
of Allen \cite{Allen:1983dg} on de Sitter background, while the flat 
space limit agrees with the classic result of Coleman and Weinberg 
\cite{Coleman:1973jx}.

\begin{flushleft}
PACS numbers: 04.50.Kd, 95.35.+d, 98.62.-g
\end{flushleft}

\vspace{0.2cm}

\begin{flushleft}
$^{*}$ e-mail: sanjib.katuwal@ufl.edu \\
$^{\star}$ email: spmiao5@mail.ncku.edu.tw \\
$^{\dagger}$ e-mail: woodard@phys.ufl.edu
\end{flushleft}

\end{titlepage}

\section{Introduction}

No one knows what caused primordial inflation but the data \cite{Akrami:2018odb}
are consistent with a minimally coupled, complex scalar inflaton $\varphi$,
\begin{equation}
\mathcal{L} = -\partial_{\mu} \varphi \partial_{\nu} \varphi^* g^{\mu\nu} 
\sqrt{-g} - V(\varphi \varphi^*) \sqrt{-g} \; .
\end{equation}
If the inflaton couples only to gravity the loop corrections to its effective
potential come only from quantum gravity and are suppressed by powers of the
loop-counting parameter $G H^2 \ltwid 10^{-10}$, where $G$ is Newton's constant
and $H$ is the Hubble parameter during inflation. In that case the classical
evolution suffers little disturbance but reheating is very slow. 

Efficient reheating requires coupling the inflaton to normal matter such as 
electromagnetism with a non-infinitesimal charge $q$,
\begin{eqnarray}
\lefteqn{\mathcal{L} = - \Bigl( \partial_{\mu} - i q A_{\mu}\Bigr) \varphi 
\Bigl( \partial_{\nu} + i q A_{\nu}\Bigr) \varphi^* g^{\mu\nu} \sqrt{-g} }
\nonumber \\
& & \hspace{5cm} - V(\varphi \varphi^*) \sqrt{-g} - \frac14 F_{\mu\nu} 
F_{\rho\sigma} g^{\mu\rho} g^{\nu\sigma} \sqrt{-g} \; . \qquad 
\label{Lagrangian}
\end{eqnarray}
But the price of efficient reheating is significant one loop corrections to 
the inflaton effective potential \cite{Green:2007gs}. For large fields these
corrections approach the Coleman-Weinberg form of flat space $\Delta V
\longrightarrow \frac{3}{16 \pi^2} (q^2 \varphi \varphi^*)^2 \ln(q^2 \varphi 
\varphi^*/s^2)$, where $s$ is the renormalization scale \cite{Coleman:1973jx}.
However, cosmological Coleman-Weinberg potentials generally depend in a 
complicated way on the geometry of inflation \cite{Miao:2015oba},
\begin{equation}
ds^2 = a^2 \Bigl[-d\eta^2 + d\vec{x} \!\cdot\! d\vec{x} \Bigr] \qquad 
\Longrightarrow \qquad H \equiv \frac{\partial_0 a}{a^2} \;\; , \;\; 
\epsilon(t) \equiv -\frac{\partial_0 H}{a H^2} \; . \label{geometry}
\end{equation}
For the special case of de Sitter (with constant $H$ and $\epsilon = 0$)
the result takes the form \cite{Allen:1983dg,Prokopec:2007ak,Miao:2019bnq},
\begin{equation}
\Delta V\Bigl\vert_{\epsilon=0} = \frac{3 H^4}{16 \pi^2} \Biggl\{ b\Bigl(
\frac{q^2 \varphi \varphi^*}{H^2}\Bigr) + \Bigl(\frac{q^2 \varphi \varphi^*}{H^2}
\Bigr) \ln\Bigl( \frac{H^2}{s^2}\Bigr) + \frac12 \Bigl( 
\frac{q^2 \varphi \varphi^*}{H^2}\Bigr)^2 \ln\Bigl( \frac{H^2}{s^2}\Bigr) 
\Biggr\} , \label{deSitterA}
\end{equation}
where the function $b(z)$ (whose $z$ and $z^2$ terms depend on renormalization 
conventions) is,
\begin{eqnarray}
\lefteqn{ b(z) = \Bigl(-1 + 2 \gamma\Bigr) z + \Bigl(-\frac32 + \gamma\Bigr) z^2}
\nonumber \\
& & \hspace{2cm} + \int_{0}^{z} \!\!\!\! dx \, (1 \!+\! x) \Biggl[ \psi\Bigl(
\frac32 \!+\! \frac12 \sqrt{1 \!-\! 8 x} \, \Bigr) + \psi\Bigl( \frac32 \!-\!
\frac12 \sqrt{1 \!-\! 8 x} \, \Bigr) \Biggr] . \qquad \label{deSitterB} 
\end{eqnarray}

Cosmological Coleman-Weinberg potentials are problematic because they make 
large corrections which cannot be completely subtracted using allowed local 
counterterms \cite{Miao:2015oba}. The classical evolution of inflation is 
subject to unacceptable modifications when partial subtractions are restricted 
to just functions of the inflaton \cite{Liao:2018sci}, or functions of the 
inflaton and the Ricci scalar \cite{Miao:2019bnq}. No other local subtractions
are permitted \cite{Woodard:2006nt} but it has been suggested that an acceptably
small distortion of classical inflation might result from cancellations between
the effective potentials induced by fermions and by bosons \cite{Miao:2020zeh}. 
The purpose of this paper is facilitate study of this scheme by developing an 
accurate approximation for extending the de Sitter results 
(\ref{deSitterA}-\ref{deSitterB}) to a general cosmological geometry 
(\ref{geometry}).

As before on flat space \cite{Coleman:1973jx}, and on de Sitter background
\cite{Prokopec:2007ak}, we define the derivative of the one loop effective 
potential through the equation,
\begin{equation}
\Delta V'(\varphi \varphi^*) = \delta \xi R + \frac12 \delta \lambda \varphi
\varphi^* + q^2 g^{\mu\nu} i\Bigl[ \mbox{}_{\mu} \Delta_{\nu}\Bigr](x;x) \; .
\label{DeltaVdef}
\end{equation}
Here $i [\mbox{}_{\mu} \Delta_{\nu}](x;x')$ is the propagator of a vector gauge 
field, in Lorentz gauge, which acquires its mass through the Higgs mechanism
rather than being a fundamental Proca field \cite{Tsamis:2006gj},
\begin{equation}
\Bigl[ \square_{\mu}^{~\nu} - R_{\mu}^{~\nu} - M^2 \delta_{\mu}^{~\nu}\Bigr] 
i\Bigl[\mbox{}_{\nu} \Delta_{\rho}\Bigr](x;x') = 
\frac{g_{\mu \rho} \, i \delta^D(x \!-\! x')}{\sqrt{-g}} + \partial_{\mu} 
\partial'_{\rho} i\Delta_t(x;x') \; . \label{propeqn}
\end{equation}
Here $\square_{\mu}^{~\nu}$ is the covariant vector d'Alembertian, $M^2 
\equiv 2 q^2 \varphi \varphi^*$ is the photon mass-squared, which is assumed 
to be constant (in spite of the background evolution) as per the definition of
``effective potential'', and $i\Delta_t(x;x')$ is the propagator of a massless, 
minimally coupled (MMC) scalar. We regulate the ultraviolet by working in $D$ 
spacetime dimensions.

In section 2 we express the photon propagator as an exact spatial Fourier 
mode sum involving massive temporal and spatially transverse vectors, along
with gradients of the MMC scalar. Section 3 begins by converting the various
mode equations to a dimensionless form, then these are approximated. Each
approximation is checked against explicit numerical evolution, both for the
simple quadratic potential, which is excluded by the lower bound on the 
tensor-to-scalar ratio \cite{Aghanim:2018eyx}, and for a plateau potential 
\cite{Starobinsky:1980te} that is in good agreement with all data. In section 
4 our approximations are applied to relation (\ref{DeltaVdef}) to compute the 
one loop effective potential. This consists of a local part which depends on 
the instantaneous geometry and a numerically smaller nonlocal part which
depends on the past geometry. Exact expressions are obtained, as well as 
expansions in the large field and small field regimes. Our conclusions are
given in section 5.

\section{Photon Mode Sum}

The purpose of this section is to express the Lorentz gauge propagator for 
a massive photon as a spatial Fourier mode sum. We begin by expressing the
right hand side of the propagator equation (\ref{propeqn}) as mode sum. Then
the various transverse vector modes are introduced. Next these modes are
combined so as to enforce the propagator equation. The section closes by
checking the de Sitter and flat space correspondence limits.

\subsection{Lessons from the Propagator Equation}

If we exploit Lorentz gauge, the $\mu = 0$ component of (\ref{propeqn}) reads,
\begin{eqnarray}
\lefteqn{ -\frac1{a^2} \Bigl[ -\partial^2 + (D\!-\!2) \partial_0 a H + a^2 M^2
\Bigr] i \Bigl[\mbox{}_0 \Delta_{\rho}\Bigr](x;x') } \nonumber \\
& & \hspace{5.5cm} = -\frac{\delta^0_{~\rho} i\delta^D(x \!-\! x')}{a^{D-2}} +
\partial_0 \partial'_{\rho} i\Delta_t(x;x') \; , \qquad \label{mutime}
\end{eqnarray}
where $\partial^2 \equiv \eta^{\mu\nu} \partial_{\mu} \partial_{\nu}$ is the 
flat space d'Alembertian. The $\mu = m$ component of equation (\ref{propeqn}) 
reads,
\begin{eqnarray}
\lefteqn{ -\frac1{a^2} \Biggl\{ \Bigl[ -\partial^2 + (D\!-\!4) a H \partial_0 
+ a^2 M^2 \Bigr] i \Bigl[\mbox{}_m \Delta_{\rho}\Bigr](x;x') + 2 a H \partial_m
i \Bigl[\mbox{}_0 \Delta_{\rho}\Bigr](x;x') \Biggr\} } \nonumber \\
& & \hspace{5.5cm} = \frac{\delta_{m \rho} i\delta^D(x \!-\! x')}{a^{D-2}} +
\partial_m \partial'_{\rho} i\Delta_t(x;x') \; . \qquad \label{muspace} 
\end{eqnarray}
We begin by writing the right hand sides of expressions (\ref{mutime}) and
(\ref{muspace}) as Fourier mode sums.

The MMC scalar propagator $i\Delta_t(x;x')$ can be expressed as a Fourier mode 
sum over functions $t(\eta,k)$ whose wave equation and Wronskian are,
\begin{equation}
\Bigl[\partial_0^2 + (D\!-\! 2) a H \partial_0 + k^2\Bigr] t(\eta,k) = 0
\qquad , \qquad t \!\cdot\! \partial_0 t^* - \partial_0 t \!\cdot\! t^* = 
\frac{i}{a^{D-2}} \; . \label{MMCeqn}
\end{equation}
Although no closed form solution exists to the $t(\eta,k)$ wave equation for
a general scale factor, relations (\ref{MMCeqn}) do define a unique solution 
when combined with the early time asymptotic form,
\begin{equation}
k \gg a H \qquad \Longrightarrow \qquad t(\eta,k) \longrightarrow 
\frac{e^{-i k \eta}}{\sqrt{2 k a^{D-2}}} \; . \label{asformt}
\end{equation}
Up to infrared corrections \cite{Janssen:2008px}, which are irrelevant owing to
the derivatives in expressions (\ref{propeqn}) and (\ref{mutime}), the Fourier
mode sum for $i\Delta_t(x;x')$ is,
\begin{eqnarray}
\lefteqn{ i\Delta_t(x;x') = \int \!\! \frac{d^{D-1}k}{(2\pi)^{D-1}} \Biggl\{ 
\theta(\Delta \eta) \, t(\eta,k) t^*(\eta',k) e^{i \vec{k} \cdot \Delta \vec{x}}
} \nonumber \\
& & \hspace{6cm} + \theta(-\Delta \eta) \, t^*(\eta,k) t(\eta',k) 
e^{-i\vec{k} \cdot \Delta \vec{x}} \Biggr\} , \qquad \label{MMCprop}
\end{eqnarray}
where $\Delta \eta \equiv \eta - \eta'$ and $\Delta \vec{x} \equiv \vec{x} - 
\vec{x}'$. Acting $\partial_0 \partial_{\rho}'$ on (\ref{MMCprop}) produces 
a term proportional to $\delta^0_{~\rho} \delta(\Delta \eta)$, which the 
Wronskian (\ref{MMCeqn}) and the change of variable $\vec{k} \rightarrow -\vec{k}$ 
allows us to recognize as a $D$-dimensional delta function,
\begin{eqnarray}
\lefteqn{ \partial_0 \partial_{\rho}' i\Delta_t(x;x') = \!\int \!\! 
\frac{d^{D-1}k}{(2\pi)^{D-1}} \Biggl\{ \delta^0_{~\rho} \delta(\Delta \eta)
\Bigl[t \!\cdot\! \partial_0 t^* \!-\! \partial_0 t \!\cdot\! t^*\Bigr] 
e^{i \vec{k} \cdot \Delta \vec{x}} + \theta(\Delta \eta) \partial_0 
\partial_{\rho}' } \nonumber \\
& & \hspace{0.7cm} \times \Bigl[ t(\eta,k)
 t^*(\eta',k) e^{i \vec{k} \cdot \Delta \vec{x}} \Bigr] + 
\theta(-\Delta \eta) \partial_0 \partial_{\rho}' \Bigl[t^*(\eta,k) t(\eta',k) 
e^{-i \vec{k} \cdot \Delta \vec{x}} \Bigr] \Biggr\} , \qquad \\
& & \hspace{-0.5cm} = \frac{\delta^0_{~\rho} i \delta^D(x \!-\! x')}{a^{D-2}}
+ \int \!\! \frac{d^{D-1}k}{(2\pi)^{D-1}} \Biggl\{ \theta(\Delta \eta) \,
T_{0}(x,\vec{k}) T^*_{\rho}(x',\vec{k}) \nonumber \\
& & \hspace{7cm} + \theta(-\Delta \eta) \, 
T^*_0(x,\vec{k}) T_{\rho}(x',\vec{k}) \Biggr\} . \qquad \label{d0drho}
\end{eqnarray}
Here we define $T_{\mu}(x,\vec{k}) \equiv \partial_{\mu} [ t(\eta,k) 
e^{i \vec{k} \cdot \vec{x}} \,]$.

Substituting (\ref{d0drho}) in the right hand side of (\ref{mutime}) gives,
\begin{eqnarray}
\lefteqn{ -\frac1{a^2} \Bigl[ -\partial^2 + (D\!-\!2) \partial_0 a H + a^2 M^2
\Bigr] i \Bigl[\mbox{}_0 \Delta_{\rho}\Bigr](x;x') } \nonumber \\
& & \hspace{0.2cm} = \! \int \!\! \frac{d^{D-1}k}{(2\pi)^{D-1}} \Biggl\{ 
\theta(\Delta \eta) \, T_{0}(x,\vec{k}) T^*_{\rho}(x',\vec{k}) + 
\theta(-\Delta \eta) \, T^*_{0}(x,\vec{k}) T_{\rho}(x',\vec{k}) \Biggr\} . 
\qquad \label{proptime}
\end{eqnarray}
The corresponding expression for (\ref{muspace}) is,
\begin{eqnarray}
\lefteqn{ -\frac1{a^2} \Biggl\{ \Bigl[ -\partial^2 + (D\!-\!4) a H \partial_0 
+ a^2 M^2 \Bigr] i \Bigl[\mbox{}_m \Delta_{\rho}\Bigr](x;x') + 2 a H \partial_m
i \Bigl[\mbox{}_0 \Delta_{\rho}\Bigr](x;x') \Biggr\} } \nonumber \\
& & \hspace{0.5cm} = \! \int \!\! \frac{d^{D-1}k}{(2\pi)^{D-1}} \Biggl\{
\frac{\delta_{m \rho} i\delta(\Delta \eta) e^{i \vec{k} \cdot \Delta \vec{x}}}{
a^{D-2}} + \theta(\Delta \eta) \, T_{m}(x,\vec{k}) T^*_{\rho}(x',\vec{k}) 
\nonumber \\
& & \hspace{6.9cm} + \theta(-\Delta \eta) \, T^*_{m}(x,\vec{k}) 
T_{\rho}(x',\vec{k}) \Biggr\} . \qquad \label{propspace}
\end{eqnarray}
The right hand sides of (\ref{proptime}) and (\ref{propspace}) are the Fourier 
mode sums that will guide us in constructing the photon propagator.

\subsection{Transverse Vector Mode Functions}

In the cosmological geometry (\ref{geometry}) a transverse (Lorentz gauge) 
vector field $F_{\mu}(x)$ obeys,
\begin{equation}
0 = D^{\mu} F_{\mu}(x) = \frac1{a^2} \Bigl[ -\Bigl( \partial_0 \!+\! (D\!-\!2)
a H \Bigr) F_0 + \partial_i F_i \Bigr] \equiv \frac1{a^2} \Bigl[ -\mathcal{D}
F_0 + \partial_i F_i \Bigr] \; . \label{transversality}
\end{equation}
We seek to express the photon propagator as a Fourier mode sum over a linear 
combination of transverse vector mode functions. Expressions 
(\ref{proptime}-\ref{propspace}) imply that one of these must be the gradient 
of a MMC scalar plane wave,
\begin{equation}
T_{\mu}(x,\vec{k}) \equiv \partial_{\mu} \Bigl[ t(\eta,k) e^{i \vec{k} \cdot 
\vec{x}} \Bigr] \; . \label{Tdef}
\end{equation}
Its transversality follows from the MMC mode equation (\ref{MMCeqn}),
\begin{equation}
-\mathcal{D} T_0 + \partial_i T_i = -\Bigl[\partial_0^2 \!+\! (D\!-\!2) a H 
\partial_0 \!+\! k^2 \Bigr] t(\eta,k) e^{i \vec{k} \cdot \vec{x}} = 0 \; .
\end{equation}

In $D$ spacetime dimensions there are $D-2$ purely spatial and transverse 
massive vector modes of the form,
\begin{equation}
V_{\mu}(x,\vec{k},\lambda,M) \equiv \epsilon_{\mu}(\vec{k},\lambda) 
\!\times\! v(\eta,k) e^{i \vec{k} \cdot \vec{x}} \qquad , \qquad \epsilon_0 
= 0 = k_i \epsilon_i \; .
\end{equation}
The polarization vectors $\epsilon_{\mu}(\vec{k},\lambda)$ are the same as those
of flat space, and their polarization sum is,
\begin{equation}
\sum_{\lambda} \epsilon_{\mu}(\vec{k},\lambda) \epsilon^*_{\rho}(\vec{k},\lambda)
= \left( \matrix{0 & 0 \cr 0 & \delta_{mr} - \widehat{k}_m \widehat{k}_r}\right)
\equiv \overline{\Pi}_{\mu\rho}(\vec{k}) \; . \label{polsum}
\end{equation}
The wave equation and Wronskian of $v(\eta,k)$ are,
\begin{equation}
\Bigl[ \partial_0^2 + (D \!-\! 4) a H \partial_0 + k^2 + a^2 M^2\Bigr] 
v(\eta,k) = 0 \quad , \quad v \cdot \partial_0 v^* - \partial_0 v \cdot
v^* = \frac{i}{a^{D-4}} \; . \label{spaceeqn}
\end{equation}
Relations (\ref{spaceeqn}) define a unique solution when coupled with the 
form for asymptotically early times,
\begin{equation}
k \gg \Bigl\{ a H, a M\Bigr\} \qquad \Longrightarrow \qquad v(\eta,k)
\longrightarrow \frac{a e^{-ik \eta}}{\sqrt{2 k a^{D-2}}} \; . \label{asformv}
\end{equation}

The spatially transverse vector modes $V_{\mu}(x,\vec{k},\lambda,M)$ 
represent dynamical photons. There is also a single temporal-longitudinal 
mode which represents the constrained part of the electromagnetic field.
It is a combination of $T_{\mu}(x,\vec{k})$ with a transverse vector 
formed from the $\mu = 0$ component $u(\eta,k,M)$ of a massive vector,
\begin{equation}
\Bigl[ \partial_0^2 + (D\!-\!2) \partial_0 a H + k^2 + a^2 M^2\Bigr] 
u(\eta,k,M) = 0 \;\; , \;\; u \cdot \partial_0 u^* - \partial_0 u
\cdot u^* = \frac{i}{a^{D-2}} \; . \label{tempeqn}
\end{equation} 
Relations (\ref{tempeqn}) define a unique solution when combined with the
early time asymptotic form,
\begin{equation}
k \gg \Bigl\{ a H, a M\Bigr\} \qquad \Longrightarrow \qquad u(\eta,k) 
\longrightarrow \frac{e^{-ik \eta}}{\sqrt{2 k a^{D-2}}} \; . \label{asformu}
\end{equation}
One converts $u(\eta,k,M)$ to a transverse vector $U_{\mu}(x,\vec{k},M)$,
\begin{equation}
U_{\mu}(x,\vec{k},M) \equiv \overline{\partial}_{\mu} \Bigl[ u(\eta,k)
e^{i \vec{k} \cdot \vec{x}} \Bigr] \; ,
\end{equation}
where the differential operator $\overline{\partial}_{\mu}$ has the $3+1$
decomposition,
\begin{equation}
\overline{\partial}_0 \equiv \sqrt{-\nabla^2} \longrightarrow k \qquad , 
\qquad \overline{\partial}_i \equiv -\frac{\partial_i \mathcal{D}}{
\sqrt{-\nabla^2}} \longrightarrow -i\widehat{k}_i \mathcal{D} \; .
\end{equation}

\subsection{Enforcing the Propagator Equation}

We have seen that the photon propagator $i[\mbox{}_{\rho} \Delta_{\rho}](x;x')$ 
is the spatial Fourier integral of contributions from the three transverse 
vector modes, each having the general form of constants times,
\begin{equation}
\mathcal{F}_{\mu\rho}(x;x') = \theta(\Delta \eta) F_{\mu}(x) F^*_{\rho}(x')
+ \theta(-\Delta \eta) F^*_{\mu}(x) F_{\rho}(x') \;\; , \;\;
F_{\mu} \in \Bigl\{ T_{\mu} , U_{\mu} , V_{\mu}\Bigr\} \; . \label{genform}
\end{equation}
We might anticipate that the spatially transverse modes contribute with unit
amplitude but the MMC scalar and temporal photon modes must be multiplied by 
the square of an inverse mass to even have the correct dimensions. The
multiplicative factors are chosen to enforce the propagator equation
(\ref{propeqn}).

To check the temporal components (\ref{proptime}) of the propagator equation we
must compute,
\begin{equation}
-\frac1{a^2} \Bigl[-\partial^2 + (D \!-\! 2) \partial_0 a H + a^2 M^2\Bigr]
\mathcal{F}_{0\rho}(x;x') \; . \label{proptimeF}
\end{equation}
To check the spatial components (\ref{propspace}) we need,
\begin{equation}
-\frac1{a^2} \Bigl[-\partial^2 + (D \!-\! 4) a H \partial_0 + a^2 M^2\Bigr]
\mathcal{F}_{m\rho}(x;x') -\frac1{a^2} \!\times\! 2 a H \partial_m 
\mathcal{F}_{0\rho}(x;x') \; . \label{propspaceF}
\end{equation}
The factors of $\partial_0$ in the differential operators of 
(\ref{proptimeF}-\ref{propspaceF}) can act on the theta functions or on the mode 
functions. When all derivatives act on the MMC contribution, the result is $-M^2$ 
times the original mode function,
\begin{eqnarray}
\lefteqn{-\frac1{a^2} \Bigl[-\partial^2 + (D \!-\! 2) \partial_0 a H + a^2 M^2\Bigr]
T_{0}(x) = -M^2 T_{0}(x) \; , } \label{MMCtime} \\
\lefteqn{-\frac1{a^2} \Bigl[-\partial^2 + (D \!-\! 4) a H \partial_0 + a^2 M^2\Bigr]
T_{m}(x) } \nonumber \\
& & \hspace{5.9cm} -\frac1{a^2} \!\times\! 2 a H \partial_m T_{0}(x) = -M^2 T_{m}(x) 
\; . \qquad \label{MMCspace}
\end{eqnarray}
This suggests that the MMC contribution enters the mode sum with a multiplicative
factor of $-M^{-2}$. No further information comes from acting the full differential 
operators on the other modes,
\begin{eqnarray}
-\frac1{a^2} \Bigl[-\partial^2 + (D \!-\! 2) \partial_0 a H + a^2 M^2\Bigr]
U_{0}(x) = 0 \; , \label{temptime} \\
-\frac1{a^2} \Bigl[-\partial^2 + (D \!-\! 4) a H \partial_0 + a^2 M^2\Bigr]
U_{m}(x) -\frac1{a^2} \!\times\! 2 a H \partial_m U_{0}(x) = 0 \; , 
\label{tempspace} \\
-\frac1{a^2} \Bigl[-\partial^2 + (D \!-\! 2) \partial_0 a H + a^2 M^2\Bigr]
V_{0}(x) = 0 \; , \label{transtime} \\
-\frac1{a^2} \Bigl[-\partial^2 + (D \!-\! 4) a H \partial_0 + a^2 M^2\Bigr]
V_{m}(x) -\frac1{a^2} \!\times\! 2 a H \partial_m V_{0}(x) = 0 \; .
\label{transspace}
\end{eqnarray}

It remains to check what happens when one or two factors of $\partial_0$ from
the differential operators in (\ref{proptimeF}-\ref{propspaceF}) act on the 
factors of $\theta(\pm \Delta \eta)$. A single conformal time derivative gives,
\begin{equation}
\partial_0 \mathcal{F}_{\mu\rho}(x;x') = \theta(\Delta \eta) \partial_0 F_{\mu}
F^*_{\rho} + \theta(-\Delta \eta) \partial_0 F^*_{\mu} F_{\rho} +
\delta(\Delta \eta) \Bigl[ F_{\mu} F^*_{\rho} - F^*_{\mu} F_{\rho}\Bigr] \; .
\label{firstetader}
\end{equation}
If we change the Fourier integration variable $\vec{k}$ to $-\vec{k}$ in the 
second of the delta function terms, the result for the MMC modes is,
\begin{eqnarray}
T_{\mu} T^*_{\rho} - T^*_{\mu} T_{\rho} \Bigl\vert_{\vec{k} \rightarrow -\vec{k}} 
&\!\!\!\!\! = \!\!\!\!\!& \left( \matrix{ [\partial_0 t \partial_0 t^* - 
\partial_0 t^* \, \partial_0 t] & -i k_r [\partial_0 t \, t^* - \partial_0 t^* 
\, t] \cr i k_m [t \, \partial t^* - t^* \partial_0 t] & k_m k_r [t \, t^* - t^* t]
} \right) e^{i \vec{k} \cdot \Delta \vec{x}} . \qquad \\
&\!\!\!\!\! = \!\!\!\!\!& \left( \matrix{ 0 & -k_r \cr -k_m & 0} \right) 
\frac{e^{i \vec{k} \cdot \Delta \vec{x}}}{a^{D-2}} . \qquad \label{firstMMC}
\end{eqnarray}
The temporal photon modes make exactly the same contribution,
\begin{eqnarray}
\lefteqn{U_{\mu} U^*_{\rho} - U^*_{\mu} U_{\rho} \Bigl\vert_{\vec{k} \rightarrow 
-\vec{k}} } \nonumber \\
& & \hspace{1cm} = \left( \matrix{ k^2 [u \, u^* - u^* u] & i k_r [u \, \mathcal{D}
u^* - u^* \mathcal{D} u] \cr -i k_m [\mathcal{D} u \, u^* - \mathcal{D} u^* u] & 
\widehat{k}_m \widehat{k}_r [\mathcal{D} u \, \mathcal{D} u^* - \mathcal{D} u^* \,
\mathcal{D} u] } \right) e^{i \vec{k} \cdot \Delta \vec{x}} . \qquad \\
& & \hspace{1cm} = \left( \matrix{ 0 & -k_r \cr -k_m & 0} \right) 
\frac{e^{i \vec{k} \cdot \Delta \vec{x}}}{a^{D-2}} . \qquad \label{firsttemp}
\end{eqnarray}
Canceling (\ref{firsttemp}) against (\ref{firstMMC}) --- whose multiplicative 
coefficient is $-M^{-2}$ --- fixes the multiplicative coefficient for the temporal 
photons as $+ M^{-2}$. The delta function term in (\ref{firstetader}) vanishes for 
the spatially transverse modes.

We turn now to second derivative which come from $-\partial^2 = \partial_0^2 -
\nabla^2$,
\begin{eqnarray}
\lefteqn{ \partial_0^2 \mathcal{F}_{\mu\rho}(x;x') = \theta(\Delta \eta) \, 
\partial_0^2 F_{\mu}(x) \, F^*_{\rho}(x') + \theta(-\Delta \eta) \, \partial_0^2 
F^*_{\mu}(x) \, F_{\rho}(x') } \nonumber \\
& & \hspace{1.5cm} + \delta(\Delta \eta) \Bigl[ \partial_0 F_{\mu} F^*_{\rho} -
\partial_0 F^*_{\mu} F_{\rho} \Bigr] + \partial_0 \Biggl\{ \delta(\Delta \eta) 
\Bigl[ F_{\mu} F^*_{\rho} - F^*_{\mu} F_{\rho}\Bigr] \Biggr\} . \qquad 
\label{secondetader}
\end{eqnarray}
We have already arranged for the cancellation of the final term in 
(\ref{secondetader}). For the new delta function term the MMC modes give,
\begin{eqnarray}
\lefteqn{\partial_0 T_{\mu} T^*_{\rho} - \partial_0 T^*_{\mu} T_{\rho} 
\Bigr\vert_{\vec{k} \rightarrow -\vec{k}} } \nonumber \\
& & \hspace{1.8cm} = \left( \matrix{ [\partial_0^2 t \, \partial_0 t^* - 
\partial_0^2 t^* \partial_0 t] & -i k_r [\partial_0^2 t \, t^* - \partial_0^2 
t^* t] \cr i k_m [ \partial_0 t \, \partial_0 t^* - \partial_0 t^* \partial_0 t] 
& k_m k_r [ \partial_0 t \, t^* - \partial_0 t^* t]}\right) 
e^{i \vec{k} \cdot \Delta \vec{x}} \; , \qquad \\
& & \hspace{1.8cm} = -i \left( \matrix{ k^2 & i k_r (D\!-\! 2) a H \cr 0 & k_m k_r}
\right) \frac{e^{i \vec{k} \cdot \Delta \vec{x}}}{a^{D-2}} \; , 
\label{secondMMCS} \qquad
\end{eqnarray}
where we have used $\partial_0^2 t = -[(D\!-\!2) a H \partial_0 + k^2] t$. The
corresponding contribution for the temporal modes is,
\begin{eqnarray}
\lefteqn{\partial_0 U_{\mu} U^*_{\rho} - \partial_0 U^*_{\mu} U_{\rho} 
\Bigr\vert_{\vec{k} \rightarrow -\vec{k}} } \nonumber \\
& & \hspace{-0.1cm} = \left( \matrix{ k^2 [\partial_0 u \, u^* - \partial_0 u^* u] 
\!&\! i k_r [\partial_0 u \, \mathcal{D} u^* - \partial_0 u^* \mathcal{D} u] \cr 
-i k_m [\partial_0 \mathcal{D} u \, u^* - \partial_0 \mathcal{D} u^* u] \!&\! 
\widehat{k}_r \widehat{k}_m [ \partial_0 \mathcal{D} u \, \mathcal{D} u^* - 
\partial_0 \mathcal{D} u^* \mathcal{D} u]}\right) e^{i\vec{k} \cdot \Delta \vec{x}} 
, \qquad \\
& & \hspace{-0.1cm} = -i \left( \matrix{ k^2 & i k_r (D\!-\!2) a H \cr 0 & 
\widehat{k}_m \widehat{k}_r (k^2 + a^2 M^2) }\right) 
\frac{e^{i\vec{k} \cdot \Delta \vec{x}}}{a^{D-2}} \; , \label{secondtemp} \qquad
\end{eqnarray}
where we have used $\partial_0 \mathcal{D} u_0 = -(k^2 + a^2 M^2) u_0$. And each 
of the spatially transverse modes gives,
\begin{eqnarray}
\partial_0 V_{\mu} V^*_{\rho} - \partial_0 V^*_{\mu} V_{\rho} 
\Bigr\vert_{\vec{k} \rightarrow -\vec{k}} & = & \left( \matrix{ 0 & 0 \cr 0 & 
\epsilon_m \epsilon_r^* [ \partial_0 v \, v* - \partial_0 v^* v]}\right) 
e^{i\vec{k} \cdot \Delta \vec{x}} \; , \qquad \\
& = & -i \left( \matrix{ 0 & 0 \cr 0 & \epsilon_m \epsilon_r^*}\right)
\frac{e^{i \vec{k} \cdot \Delta \vec{x}}}{a^{D-4}} \; . \qquad \label{secondtrans} 
\end{eqnarray}

The second conformal time derivatives in both expression (\ref{proptimeF}) 
and the corresponding spatial relation (\ref{propspaceF}) come in the form 
$-\frac1{a^2} \times \partial_0^2$. Including the multiplicative factors, we see 
that the temporal delta functions which are induced consist of $\frac1{a^2 M^2}$ 
times (\ref{secondMMCS}) minus the same factor times (\ref{secondtemp}), plus the 
polarization sum (\ref{polsum}) over (\ref{secondtrans}),
\begin{eqnarray}
\lefteqn{ \frac{i}{M^2} \left( \matrix{ k^2 & i k_r (D\!-\!2) a H \cr 0 & k_m k_r}
\right) \frac{e^{i \vec{k} \cdot \Delta \vec{x}}}{a^{D}} - \frac{i}{M^2} \left( 
\matrix{ k^2 & i k_r (D\!-\!2) a H \cr 0 & \widehat{k}_m \widehat{k}_r (k^2 + a^2 M^2)}
\right) \frac{e^{i \vec{k} \cdot \Delta \vec{x}}}{a^{D}} } \nonumber \\
& & \hspace{2.5cm} -i \left( \matrix{0 & 0 \cr 0 & \delta_{mr} - \widehat{k}_m
\widehat{k}_r } \right) \frac{e^{i \vec{k} \cdot \Delta \vec{x}}}{a^{D-2}} = -i
\left( \matrix{0 & 0 \cr 0 & \delta_{mr}} \right) 
\frac{e^{i \vec{k} \cdot \Delta \vec{x}}}{a^{D-2}} \; . \qquad \label{deltaterms} 
\end{eqnarray}
With $-\frac1{M^2}$ times expressions (\ref{MMCtime}-\ref{MMCspace}) we see that 
the propagator equations (\ref{proptime}-\ref{propspace}) are obeyed by the
Fourier mode sum,
\begin{eqnarray}
\lefteqn{ i \Bigl[\mbox{}_{\mu} \Delta_{\rho} \Bigr](x;x') = \!\!\int\!\!\! 
\frac{d^{D-1}k}{(2\pi)^{D-1}} \Biggl\{ \!\theta(\Delta \eta) \!\Biggl[ 
\frac{U_{\mu}(x,\vec{k},M) U^*_{\rho}(x',\vec{k},M) \!-\! T_{\mu}(x,\vec{k}) 
T^*_{\rho}(x',\vec{k})}{M^2} } \nonumber \\
& & \hspace{1cm} + \overline{\Pi}_{\mu\rho}(\vec{k}) v(\eta,k) v^*(\eta',k) 
e^{i \vec{k} \cdot \Delta \vec{x}} \Biggr] + \theta(-\Delta \eta) \Biggl[ 
\frac{U^*_{\mu}(x,\vec{k},M) U_{\rho}(x',\vec{k},M)}{M^2} \nonumber \\
& & \hspace{2.5cm} - \frac{T^*_{\mu}(x,\vec{k}) T_{\rho}(x',\vec{k})}{M^2} + 
\overline{\Pi}_{\mu\rho}(\vec{k}) v^*(\eta,k) v(\eta',k) 
e^{-i \vec{k} \cdot \Delta \vec{x}} \Biggr] \Biggr\} . \qquad \label{propagator}
\end{eqnarray}
Note that the $U_{\mu}(x,\vec{k},M)$ and $T_{\mu}(x,\vec{k})$ modes combine
to form a vector integrated propagator analogous to the scalar ones
introduced in \cite{Miao:2011fc}.

The photon propagator can also be expressed as the sum of three bi-vector 
differential operators acting on a scalar propagator,
\begin{eqnarray}
\lefteqn{i\Bigl[\mbox{}_{\mu} \Delta_{\rho}\Bigr](x;x') = \frac1{M^2} \Bigl[
-\eta_{\mu\rho} + \overline{\Pi}_{\mu\rho}\Bigr] \frac{i \delta^D(x \!-\! x')}{
a^{D-2}} } \nonumber \\
& & \hspace{2cm} + \frac1{M^2} \Bigl[ \overline{\partial}_{\mu} 
\overline{\partial}_{\rho}' i\Delta_u(x;x') - \partial_{\mu} \partial_{\rho}' 
i\Delta_t(x;x')\Bigr] + \overline{\Pi}_{\mu\rho} i\Delta_v(x;x') \; . \qquad
\label{scalarsum}
\end{eqnarray}
The Fourier mode sum for the MMC scalar propagator $i\Delta_t(x;x')$ was given
in expression (\ref{MMCprop}). The mode sum for the temporal propagator
$i\Delta_u(x;x')$ comes from replacing $t(\eta,k)$ with $u(\eta,k)$ in 
(\ref{MMCprop}), and the mode sum for the transverse spatial propagator 
$i\Delta_v(x;x')$ is obtained by replacing $t(\eta,k)$ with $v(\eta,k)$. The
resulting lowest order (free) field strength correlators are,
\begin{eqnarray}
\lefteqn{\Bigl\langle \Omega \Bigl\vert T^*\Bigl[ F_{0j}(x) F_{0\ell}(x')\Bigr]
\Bigr\vert \Omega \Bigr\rangle = \frac{\partial_j \partial_{\ell}}{\nabla^2}
\frac{i \delta^D(x \!-\! x')}{a^{D-4}} } \nonumber \\
& & \hspace{3.9cm} + a^2 {a'}^2 M^2 \frac{\partial_j \partial_{\ell}}{\nabla^2}
i\Delta_u(x;x') + \overline{\Pi}_{j\ell} \partial_0 \partial_0' i \Delta_v(x;x') 
\; , \qquad \\
\lefteqn{\Bigl\langle \Omega \Bigl\vert T^*\Bigl[ F_{0j}(x) F_{k\ell}(x')\Bigr]
\Bigr\vert \Omega \Bigr\rangle = \Bigl[\delta_{jk} \partial_{\ell} \!-\! 
\delta_{j\ell} \partial_k\Bigr] \partial_0 i \Delta_v(x;x') \; , } \\
\lefteqn{\Bigl\langle \Omega \Bigl\vert T^*\Bigl[ F_{ij}(x) F_{k\ell}(x')\Bigr]
\Bigr\vert \Omega \Bigr\rangle } \nonumber \\
& & \hspace{3cm} = - \Bigl[\delta_{ik} \partial_j \partial_{\ell} \!-\! 
\delta_{kj} \partial_{\ell} \partial_i \!+\! \delta_{j\ell} \partial_i 
\partial_k \!-\! \delta_{\ell i} \partial_k \partial_j \Bigr] i\Delta_v(x;x') 
\; . \qquad
\end{eqnarray}
The $T^*$-ordering symbol in these correlators indicates that the derivatives in 
forming the field strength tensor, $F_{\mu\nu}(x) \equiv \partial_{\mu} A_{\nu}(x) 
- \partial_{\nu} A_{\mu}(x)$, are taken outside the time-ordering symbol.

An important simplification is,
\begin{equation}
T_{\mu}(x,\vec{k}) = -i \lim_{M \rightarrow 0} U_{\mu}(x,\vec{k},M) \; . 
\label{MMCtotemp}
\end{equation}
Comparing equations (\ref{MMCtime}) with (\ref{temptime}), and (\ref{MMCspace}) 
with (\ref{tempspace}), shows that both sides of relation (\ref{MMCtotemp}) obey
the same wave equation for $M = 0$. That they are identical follows from 
$t(\eta,k)$ and $u(\eta,k)$ having the same asymptotic forms (\ref{asformt}) and 
(\ref{asformu}). Relation (\ref{MMCtotemp}) is of great importance because it
guarantees that the propagator has no $\frac1{M^2}$ pole.

\subsection{The de Sitter Limit}

In the limit of $\epsilon = 0$ the mode functions have closed form 
solutions,\footnote{In the phase factors for $u(\eta,k,M)$ and $v(\eta,k,M)$ one 
must regard $\nu_b$ as a real number, even if $M^2 > \frac14 (D-3)^2 H^2$.}
\begin{eqnarray}
t(\eta,k) &\!\!\! \longrightarrow \!\!\!& e^{\frac{i \pi}{2} (\nu_A + \frac12)} 
\sqrt{\frac{\pi}{4 H a^{D-1}}} \!\times\! H^{(1)}_{\nu_A}\Bigl( \frac{k}{Ha}\Bigr) 
\; , \qquad \\
u(\eta,k,M) &\!\!\! \longrightarrow \!\!\!& e^{\frac{i \pi}{2} (\nu_b + \frac12)} 
\sqrt{\frac{\pi}{4 H a^{D-1}}} \!\times\! H^{(1)}_{\nu_b}\Bigl( \frac{k}{Ha}\Bigr) 
\; , \qquad \\
v(\eta,k,M) & \longrightarrow & e^{\frac{i \pi}{2} (\nu_b + \frac12)} 
\sqrt{\frac{\pi}{4 H a^{D-3}}} \!\times\! H^{(1)}_{\nu_b}\Bigl( \frac{k}{Ha}\Bigr) 
\; , \qquad 
\end{eqnarray}
where the indices are,
\begin{equation}
\nu_A \equiv \Bigl( \frac{D\!-\!1}{2}\Bigr) \qquad , \qquad \nu_b \equiv 
\sqrt{ \Bigl(\frac{D\!-\!3}{2}\Bigr)^2 \!-\! \frac{M^2}{H^2} } \; .
\label{indices}
\end{equation}
The Fourier mode sums for the three propagators can be mostly expressed in terms
of the de Sitter length function $y(x;x')$,
\begin{equation}
y(x;x') \equiv \Bigl\Vert \vec{x} \!-\! \vec{x}' \Bigr\Vert^2 - \Bigl(
\vert \eta \!-\! \eta'\vert \!-\! i \varepsilon \Bigr)^2 \; . \label{ydef}
\end{equation}
The de Sitter limit of the temporal photon propagator is a Hypergeometric 
function,
\begin{equation}
i\Delta_u(x;x') \longrightarrow \frac{H^{D-2}}{(4\pi)^{\frac{D}2}} 
\frac{\Gamma(\nu_a \!+\! \nu_b) \Gamma(\nu_A \!-\! \nu_b)}{\Gamma(\frac{D}2)}
\mbox{}_{2}F_{1}\Bigl(\nu_A \!+\! \nu_b, \nu_A \!-\! \nu_b , \frac{D}2 ;1 \!-\!
\frac{y}{4}\Bigr) \equiv b(y) \; . \label{dSu}
\end{equation}
The de Sitter limit of the spatially transverse photon propagator is closely
related,
\begin{equation}
i\Delta_v(x;x') \longrightarrow a a' b(y) \; . \label{dSv}
\end{equation}
However, infrared divergences break de Sitter invariance in the MMC scalar 
propagator \cite{Vilenkin:1982wt,Linde:1982uu,Starobinsky:1982ee}. The result
for the noncoincident propagator takes the form \cite{Onemli:2002hr,Onemli:2004mb},
\begin{equation}
i\Delta_t(x;x') \longrightarrow A(y) + \frac{H^{D-2}}{(4\pi)^{\frac{D}2}}
\frac{\Gamma(D \!-\! 1)}{\Gamma(\frac{D}2)} \ln(a a') \; , \label{dSt}
\end{equation}
where we only need derivatives of the function $A(y)$ \cite{Miao:2009hb},
\begin{eqnarray}
A'(y) & = & \frac12 (2 \!-\! y) B'(y) - \frac12 (D \!-\! 2) B(y) \; , 
\label{Adef} \\
B(y) & \equiv & \frac{\Gamma(D \!-\! 2) \Gamma(1)}{\Gamma(\frac{D}2)} 
\, \mbox{}_{2}F_1\Bigl(D \!-\! 2, 1,\frac{D}2 ; 1 \!-\! \frac{y}{4}\Bigr) \; .
\label{Bdef}
\end{eqnarray}
It is useful to note that the functions $B(y)$ and $b(y)$ obey,
\begin{eqnarray}
0 & = & (4 y \!-\! y^2) B''(y) + D (2 \!-\! y) B'(y) - (D\!-\!2) B(y) \; , \\
0 & = & (4 y \!-\! y^2) b''(y) + D (2 \!-\! y) b'(y) - (D\!-\!2) b(y) 
- \frac{M^2}{H} b(y) \; . 
\end{eqnarray}

A direct computation of the photon propagator on de Sitter background gives
\cite{Tsamis:2006gj},
\begin{eqnarray}
\lefteqn{ i\Bigl[\mbox{}_{\mu} \Delta_{\rho}\Bigr](x;x') \longrightarrow -
\frac{\partial^2 y}{\partial x^{\mu} \partial {x'}^{\rho}} \Bigl[ (4 y \!-\! y^2)
\frac{\partial}{\partial y} + (D \!-\! 1) (2 \!-\! y) \Bigr] \Bigl[ 
\frac{b'(y) \!-\! B'(y)}{2 M^2} \Bigr] } \nonumber \\
& & \hspace{3.5cm} + \frac{\partial y}{\partial x^{\mu}} 
\frac{\partial y}{\partial {x'}^{\rho}} \Bigl[ (2 \!-\! y) 
\frac{\partial}{\partial y} - (D \!-\! 1) \Bigr] \Bigl[ 
\frac{b'(y) \!-\! B'(y)}{2 M^2} \Bigr] . \qquad \label{directdS}
\end{eqnarray}
To see that the de Sitter limit of our mode sum (\ref{scalarsum}) agrees 
with (\ref{directdS}) we substitute the de Sitter limits (\ref{dSt}), (\ref{dSu}) 
and (\ref{dSv}) and make some tedious reorganizations. This is simplest for the 
MMC scalar contribution,
\begin{eqnarray}
\lefteqn{ \frac{ \delta^0_{~\mu} \delta^0_{~\rho} i \delta^D(x \!-\! x')}{M^2 
a^{D-2}} - \frac{ \partial_{\mu} \partial_{\rho}' i\Delta_t(x;x')}{M^2} 
\longrightarrow - \frac{\partial^2 y}{\partial x^{\mu} \partial {x'}^{\rho}} 
\frac{A'}{M^2} - \frac{\partial y}{\partial x^{\mu}} \frac{\partial y}{\partial 
{x'}^{\rho}} \frac{A''}{M^2} , } \\
& & \hspace{-0.6cm} = - \frac{\partial^2 y}{\partial x^{\mu} \partial {x'}^{\rho}}
\Bigl[ \frac{(2 \!-\! y) B' \!-\! (D \!-\! 2) B}{2 M^2} \Bigr] - 
\frac{\partial y}{\partial x^{\mu}} \frac{\partial y}{\partial {x'}^{\rho}}
\Bigl[ \frac{(2 \!-\! y) B'' \!-\! (D \!-\! 1) B'}{2 M^2} \Bigr] , \qquad \\
& & \hspace{-0.6cm} = \frac{\partial^2 y}{\partial x^{\mu} \partial {x'}^{\rho}}
\Bigl[ \frac{(4 y \!-\! y^2) B'' \!+\! (D \!-\! 1) (2 \!-\! y) B'}{2 M^2} \Bigr]
\nonumber \\
& & \hspace{5.9cm} - \frac{\partial y}{\partial x^{\mu}} 
\frac{\partial y}{\partial {x'}^{\rho}} \Bigl[ \frac{(2 \!-\! y) B'' \!-\! 
(D \!-\! 1) B'}{2 M^2} \Bigr] . \qquad
\end{eqnarray}

Each tensor component of the temporal photon contribution requires a separate 
treatment. The case of $\mu = 0 = \rho$ gives,
\begin{eqnarray}
\lefteqn{ \overline{\partial}_{0} \overline{\partial}_{0}' \,
\frac{i \Delta_u(x;x')}{M^2} \longrightarrow -\nabla^2 \frac{b(y)}{M^2} =
-\nabla^2 y \, \frac{b'}{M^2} - \partial_i y \, \partial_i y \, \frac{b''}{M^2} } 
\\
& & \hspace{-0.5cm} = \frac{a a' H^2}{M^2} \Biggl\{ -2 (D \!-\! 1)b' + 4 \Bigl[
2 \!-\! y \!-\! \frac{a}{a'} \!-\! \frac{a'}{a}\Bigr] b'' \Biggr\} , \\
& & \hspace{-0.5cm} = \frac{a a' H^2}{2 M^2} \Biggl\{ \Bigl[-(2 \!-\! y) + 2 \Bigl( 
\frac{a}{a'} \!+\! \frac{a'}{a}\Bigr) \Bigr] \Bigl[- (4 y \!-\! y^2) b'' - (D \!-\! 1)
(2 \!-\! y) b' \Bigr] \nonumber \\
& & \hspace{1.5cm} + \Bigl[8 \!-\! 4 y \!+\! y^2 - 2 (2 \!-\! y) \Bigl( \frac{a}{a'} 
\!+\! \frac{a'}{a} \Bigr) \Bigr] \Bigl[ (2 \!-\! y) b'' - (D \!-\! 1) b'\Bigr] 
\Biggr\} , \qquad \\
& & \hspace{-0.5cm} = -\frac{\partial^2 y}{\partial x^0 \partial {x'}^0} 
\Bigl[ (4y \!-\! y^2) \frac{\partial}{\partial y} + (D\!-\!1) (2 \!-\! y)\Bigr] 
\frac{b'}{2 M^2} \nonumber \\
& & \hspace{5cm} + \frac{\partial y}{\partial x^0} \frac{\partial y}{\partial {x'}^0} 
\Bigl[ (2 \!-\! y) \frac{\partial}{\partial y} - (D \!-\!1)\Bigr] \frac{b'}{2 M^2} . 
\qquad 
\end{eqnarray}
For $\mu = 0$ and $\rho = r$ we have,
\begin{eqnarray}
\lefteqn{ \overline{\partial}_{0} \overline{\partial}_{r}' \,
\frac{i \Delta_u(x;x')}{M^2} \longrightarrow \partial_r \mathcal{D}' \frac{b(y)}{M^2} 
= \partial_r \mathcal{D}' y \, \frac{b'}{M^2} + \partial_r y \, \partial_0' y \, 
\frac{b''}{M^2} } \\
& & \hspace{-0.5cm} = \frac{a {a'}^2 H^3 \Delta x^r}{M^2} \Biggl\{ 2 (D \!-\! 1) b'
- 2 (2 \!-\! y) b'' + 4 \frac{a}{a'} b''\Biggr\} , \\
& & \hspace{-0.5cm} = \frac{a^2 a' H^3 \Delta x^r}{M^2} \Biggl\{ \Bigl[ (4 y \!-\! y^2)
b'' + (D \!-\! 1) (2 \!-\! y) b'\Bigr] \nonumber \\
& & \hspace{5cm} + \Bigl[2 \!-\! y - 2 \frac{a'}{a}\Bigr] \Bigl[ (2 \!-\! y) b'' -
(D \!-\! 1) b'\Bigr] \Biggr\} , \qquad \\
& & \hspace{-0.5cm} = -\frac{\partial^2 y}{\partial x^0 \partial {x'}^r} 
\Bigl[ (4y \!-\! y^2) \frac{\partial}{\partial y} + (D\!-\!1) (2 \!-\! y)\Bigr] 
\frac{b'}{2 M^2} \nonumber \\
& & \hspace{5cm} + \frac{\partial y}{\partial x^0} \frac{\partial y}{\partial {x'}^r} 
\Bigl[ (2 \!-\! y) \frac{\partial}{\partial y} - (D \!-\!1)\Bigr] \frac{b'}{2 M^2} . 
\qquad 
\end{eqnarray}
And the result for $\mu = m$ and $\rho = 0$ is,
\begin{eqnarray}
\lefteqn{ \overline{\partial}_{m} \overline{\partial}_{0}' \,
\frac{i \Delta_u(x;x')}{M^2} \longrightarrow -\partial_m \mathcal{D} \frac{b(y)}{M^2} 
= -\partial_m \mathcal{D} y \, \frac{b'}{M^2} - \partial_m y \, \partial_0 y \, 
\frac{b''}{M^2} } \\
& & \hspace{-0.5cm} = -\frac{a^2 a' H^3 \Delta x^m}{M^2} \Biggl\{ 2 (D \!-\! 1) b'
- 2 (2 \!-\! y) b'' + 4 \frac{a'}{a} b''\Biggr\} , \\
& & \hspace{-0.5cm} = -\frac{a {a'}^2 H^3 \Delta x^m}{M^2} \Biggl\{ \Bigl[ (4 y \!-\! y^2)
b'' + (D \!-\! 1) (2 \!-\! y) b'\Bigr] \nonumber \\
& & \hspace{5cm} + \Bigl[2 \!-\! y - 2 \frac{a}{a'}\Bigr] \Bigl[ (2 \!-\! y) b'' -
(D \!-\! 1) b'\Bigr] \Biggr\} , \qquad \\
& & \hspace{-0.5cm} = -\frac{\partial^2 y}{\partial x^m \partial {x'}^0} 
\Bigl[ (4y \!-\! y^2) \frac{\partial}{\partial y} + (D\!-\!1) (2 \!-\! y)\Bigr] 
\frac{b'}{2 M^2} \nonumber \\
& & \hspace{5cm} + \frac{\partial y}{\partial x^m} \frac{\partial y}{\partial {x'}^0} 
\Bigl[ (2 \!-\! y) \frac{\partial}{\partial y} - (D \!-\!1)\Bigr] \frac{b'}{2 M^2} . 
\qquad 
\end{eqnarray}

The case of $\mu = m$ and $\rho = r$ requires the most intricate analysis. It begins
with the observation,
\begin{equation}
\frac{\partial_m \partial_r}{\nabla^2} \frac{i \delta^D(x \!-\! x')}{a^{D-2}} + 
\overline{\partial}_m \overline{\partial}_r' \frac{i \Delta_u(x;x')}{M^2} 
\longrightarrow \frac{\partial_m \partial_r}{\nabla^2} \, \frac{\mathcal{D} 
\mathcal{D}' b(y)}{M^2} \; . \label{spaceu}
\end{equation}
This component combines with the contribution from spatially transverse photons,
\begin{equation}
\overline{\Pi}_{mr} i\Delta_v(x;x') \longrightarrow \Bigl( \delta_{mr} - 
\frac{\partial_m \partial_r}{\nabla^2} \Bigr) a a' b(y) \; . \label{spacev}
\end{equation} 
The $\partial_m \partial_r/\nabla^2$ terms from expressions (\ref{spaceu}) and
(\ref{spacev}) give,
\begin{eqnarray}
\lefteqn{ \mathcal{D} \mathcal{D}' b(y) - a a' M^2 b(y) = a a' H^2 \Biggl\{
\Bigl[8 \!-\! 4 y \!+\! y^2 - 2 (2 \!-\! y) \Bigl( \frac{a}{a'} \!+\! \frac{a'}{a}
\Bigr) \Bigr] b'' } \nonumber \\
& & \hspace{0.3cm} + \Bigl[-(2 D \!-\! 3) (2 \!-\! y) + 2 (D \!-\! 1) \Bigl(
\frac{a}{a'} \!+\! \frac{a'}{a}\Bigr) \Bigr] b' + \Bigl[ (D \!-\! 2)^2 -
\frac{M^2}{H^2}\Bigr] b \Biggr\} , \qquad \\ 
& & \hspace{-0.5cm} = a a' H^2 \Biggl\{ 2 (2 \!-\! y)^2 b'' - 3 (D\!-\!1) 
(2 \!-\! y) b' + (D\!-\!2) (D\!-\! 1) b \nonumber \\
& & \hspace{5cm} + 2 \Bigl( \frac{a}{a'} \!+\! \frac{a'}{a}\Bigr) \Bigl[-
(2 \!-\! y) b'' + (D \!-\!1) b'\Bigr] \Biggr\} , \qquad \\
& & \hspace{-0.5cm} = \frac12 \nabla^2 I\Bigl[-(2 \!-\! y) b' + (D \!-\!2) b\Bigr]
\; , \qquad \label{longitudinal}
\end{eqnarray}
where $I[f(y)]$ represents the indefinite integral of $f(y)$ with respect to $y$.

Substituting relation (\ref{longitudinal}) in (\ref{spaceu}) and (\ref{spacev})
gives,
\begin{eqnarray}
\lefteqn{ \frac{\partial_m \partial_r}{\nabla^2} \frac{i \delta^D(x \!-\! x')}{a^{D-2}}
+ \overline{\partial}_m \overline{\partial}_r' \, \frac{i \Delta_u(x;x')}{M^2}
+ \overline{\Pi}_{mr} i\Delta_v(x;x') } \nonumber \\
& & \hspace{3.5cm} \longrightarrow a a' \delta_{mr} b(y) + \frac{\partial_m 
\partial_r}{2 M^2} I\Bigl[-(2 \!-\! y) b' + (D\!-\!2) b\Bigr] ,  \qquad \\
& & \hspace{-0.5cm} = \frac{a a' H^2}{M^2} \Biggl\{ \delta_{mr} \Bigl[(4 y \!-\!y^2) 
b'' + (D\!-\!1) (2 \!-\! y) b' \Bigr] \nonumber \\
& & \hspace{4cm} + 2 a a' H^2 \Delta x^m \Delta x^r 
\Bigl[-(2 \!-\! y) b'' + (D \!-\! 1) b'\Bigr] \Biggr\} , \qquad \\
& & \hspace{-0.5cm} = -\frac{\partial^2 y}{\partial x^m \partial {x'}^r} 
\Bigl[ (4y \!-\! y^2) \frac{\partial}{\partial y} + (D\!-\!1) (2 \!-\! y)\Bigr] 
\frac{b'}{2 M^2} \nonumber \\
& & \hspace{5cm} + \frac{\partial y}{\partial x^m} \frac{\partial y}{\partial {x'}^r} 
\Bigl[ (2 \!-\! y) \frac{\partial}{\partial y} - (D \!-\!1)\Bigr] \frac{b'}{2 M^2} . 
\qquad 
\end{eqnarray}
This completes our demonstration that the de Sitter limit of our propagator agrees
with the direct calculation (\ref{directdS}). It should also be noted that taking 
$H \rightarrow 0$ in the de Sitter limit gives the well known flat space result 
\cite{Tsamis:2006gj}, so we have really checked two correspondence limits.

\section{Approximating the Amplitudes}

The results of the previous section are exact but they rely upon mode functions
$t(\eta,k)$, $u(\eta,k,M)$ and $v(\eta,k,M)$ for which no explicit solution
is known in a general cosmological geometry (\ref{geometry}). The purpose of 
this section is to develop approximations for the amplitudes (norm-squares) of 
these mode functions. We begin converting all the dependent and independent 
variables to dimensionless form. Then approximations are developed for each of 
the three amplitudes and checked against numerical evolution for the inflationary 
geometry of a simple quadratic potential which reproduces the scalar amplitude 
and spectral index but gives too large a value for the tensor-to-scalar ratio. 
The section closes by demonstrating that our approximations remain valid for 
the plateau potentials which agree with current data. 

\subsection{Dimensionless Formulation}

Time scales vary so much during cosmology that it is desirable to change the
independent variable from conformal time $\eta$ to the number of e-foldings since
the start of inflation $n$,
\begin{equation}
n \equiv \ln\Bigl[ \frac{a(\eta)}{a_i}\Bigr] \qquad \Longrightarrow \qquad 
\partial_0 = a H \partial_n \quad , \quad \partial_0^2 = a^2 H^2 \Bigl[ 
\partial_n^2 + (1 - \epsilon) \partial_n \Bigr] \; . \label{ndef}
\end{equation}
We convert the wave number $k$ and the mass $M$ to dimensionless parameters
using factors of $8 \pi G$,
\begin{equation}
\kappa \equiv \sqrt{8\pi G} \, k \qquad , \qquad \mu \equiv \sqrt{8\pi G} \, M
\; . \label{kmudef}
\end{equation}
And the dimensionless Hubble parameter, inflaton and classical potential are,
\begin{equation}
\chi(n) \equiv \sqrt{8\pi G} \, H(\eta) \;\; , \;\; \psi(n) \equiv 
\sqrt{8\pi G} \, \varphi(\eta) \;\; , \;\; U(\psi \psi^*) \equiv (8\pi G)^2
V(\varphi \varphi^*) \; . \label{chipsiUdef}
\end{equation}
The first slow roll parameter is already dimensionless and we consider it to be
a function of $n$,
\begin{equation}
\epsilon(n) \equiv -\frac{\chi'}{\chi} \; . \label{epsilondef}
\end{equation}

In terms of these dimensionless variables the nontrivial Einstein equations
are,
\begin{eqnarray}
\frac12 (D\!-\!2) (D\!-\!1) \chi^2 & = & \chi^2 \psi' {\psi'}^* + U(\psi \psi^*) 
\; , \label{Einstein1} \\
-\frac12 (D\!-\!2) \Bigl(D \!-\!1 \!-\! 2\epsilon\Bigr) \chi^2 & = & \chi^2
\psi' {\psi'}^* - U(\psi \psi^*) \; . \label{Einstein2}
\end{eqnarray}
The dimensionless inflaton evolution equation is,
\begin{equation}
\chi^2 \Bigl[ \psi'' + (D \!-\! 1 \!-\! \epsilon) \psi'\Bigr] + \psi \,
U'(\psi\psi^*) = 0 \; . \label{inflaton1}
\end{equation}
This can be expressed entirely in terms of $\psi$ and its derivatives,
\begin{equation}
\psi'' + \Bigl(D \!-\! 1 \!-\! \frac{2 \psi' {\psi'}^*}{D \!-\! 2} \Bigr) 
\Biggl[ \psi' + \frac{(D \!-\! 2) U'(\psi \psi^*) \psi}{2 U(\psi \psi^*)}
\Biggr] = 0 \; . \label{imflaton2}
\end{equation}

Although our analytic approximations apply for any model of inflation,
comparing them with exact numerical results of course requires an explicit
model of inflation. It is simplest to carry out most of the analysis using 
a quadratic model with $U(\psi) = c^2 \psi \psi^*$. Applying the slow 
roll approximation gives analytic expressions for the scalar, the 
dimensionless Hubble parameter and the first slow roll parameter,
\begin{equation}
\psi(n) \simeq \sqrt{\psi_0^2 \!-\! 2 n} \quad , \quad \chi(n) \simeq 
\frac{c}{\sqrt{3}} \sqrt{\psi_0^2 \!-\! 2 n} \quad , \quad \epsilon(n) 
\simeq \frac{1}{\psi_0^2 \!-\! 2 n} \; , \label{slowroll}
\end{equation}
Note also that $\chi(n) \simeq \chi_0 \sqrt{1 - 2 n/\psi_0^2}$. By starting 
from $\psi_0 = 10.6$ one gets somewhat over 50 e-foldings of inflation. 
Setting $c = 7.126 \times 10^{-6}$ makes this model consistent with the 
observed values of the scalar spectral index and the scalar amplitude 
\cite{Aghanim:2018eyx}, but the model's tensor-to-scalar ratio is about 
three times larger than the 95\% confidence upper limit. Although we exploit
the simple slow roll results (\ref{slowroll}) of this phenomenologically
excluded model to develop approximations, the section closes with a 
demonstration that our analytic approximations continue to apply for 
viable models.

We define the dimensionless MMC scalar amplitude,
\begin{equation}
\mathcal{T}(n,\kappa) \equiv \ln\Bigl[ \frac{\vert t(\eta,k)\vert^2}{\sqrt{8\pi G}}
\Bigr] \; . \label{scTdef}
\end{equation}
Following the procedure of \cite{Romania:2011ez,Romania:2012tb,Brooker:2015iya} we
convert the mode equation and Wronskian (\ref{MMCeqn}) into the nonlinear relation,
\begin{equation}
\mathcal{T}'' + \frac12 {\mathcal{T}'}^2 + (D \!-\! 1 \!-\! \epsilon) \mathcal{T}'
+ \frac{2 \kappa^2 e^{-2n}}{\chi^2} - \frac{e^{-2(D-1) n - 2 \mathcal{T}}}{2 \chi^2}
= 0 \; . \label{scTeqn}
\end{equation}
The asymptotic relation (\ref{asformt}) implies the initial conditions needed for
equation (\ref{scTeqn}) to produce a unique solution,
\begin{equation}
\mathcal{T}(0,\kappa) = -\ln(2 \kappa) \qquad , \qquad \mathcal{T}'(0,\kappa) =
-(D\!-\!2) \; . \label{scTinitial}
\end{equation}

The temporal photon and spatially transverse photon amplitudes are defined
analogously,
\begin{equation}
\mathcal{U}(n,\kappa,\mu) \equiv \ln\Bigl[ \frac{\vert u(\eta,k,M)\vert^2}{
\sqrt{8\pi G}}\Bigr] \qquad , \qquad 
\mathcal{V}(n,\kappa,\mu) \equiv \ln\Bigl[ \frac{\vert v(\eta,k,M)\vert^2}{
\sqrt{8\pi G}}\Bigr] \; . \label{scUVdef}
\end{equation}
Applying the same procedure \cite{Romania:2011ez,Romania:2012tb,Brooker:2015iya}
to the temporal photon mode equation and Wronskian (\ref{tempeqn}) gives,
\begin{eqnarray}
\lefteqn{\mathcal{U}'' + \frac12 {\mathcal{U}'}^2 + (D \!-\! 1 \!-\! \epsilon) 
\mathcal{U}' } \nonumber \\
& & \hspace{2cm} + \frac{2 \kappa^2 e^{-2n}}{\chi^2} + 2 (D\!-\!2) (1 \!-\!
\epsilon) + \frac{2 \mu^2}{\chi^2} - \frac{e^{-2 (D-1) n - 2 \mathcal{U}}}{
2 \chi^2} = 0 \; . \qquad \label{scUeqn}
\end{eqnarray}
And the initial conditions follow from (\ref{asformu}),
\begin{equation}
\mathcal{U}(0,\kappa,\mu) = -\ln(2 \kappa) \qquad , \qquad 
\mathcal{U}'(0,\kappa,\mu) = -(D \!-\! 2) \; . \label{scUinitial}
\end{equation}
The analogous transformation of the spatially transverse photon mode equation
and Wronskian (\ref{spaceeqn}) produces,
\begin{equation}
\mathcal{V}'' + \frac12 {\mathcal{V}'}^2 + (D \!-\! 3 \!-\! \epsilon) 
\mathcal{V}' + \frac{2 \kappa^2 e^{-2n}}{\chi^2} + \frac{2 \mu^2}{\chi^2} 
- \frac{e^{-2 (D-3) n - 2 \mathcal{V}}}{2 \chi^2} = 0 \; . \label{scVeqn}
\end{equation}
The initial conditions associated with (\ref{asformv}) are,
\begin{equation}
\mathcal{V}(0,\kappa,\mu) = -\ln(2 \kappa) \qquad , \qquad 
\mathcal{V}'(0,\kappa,\mu) = -(D \!-\! 4) \; . \label{scVinitial}
\end{equation}

\subsection{Massless, Minimally Coupled Scalar}

\begin{figure}[H]
\centering
\begin{subfigure}[b]{0.33\textwidth}
\centering
\includegraphics[width=\textwidth]{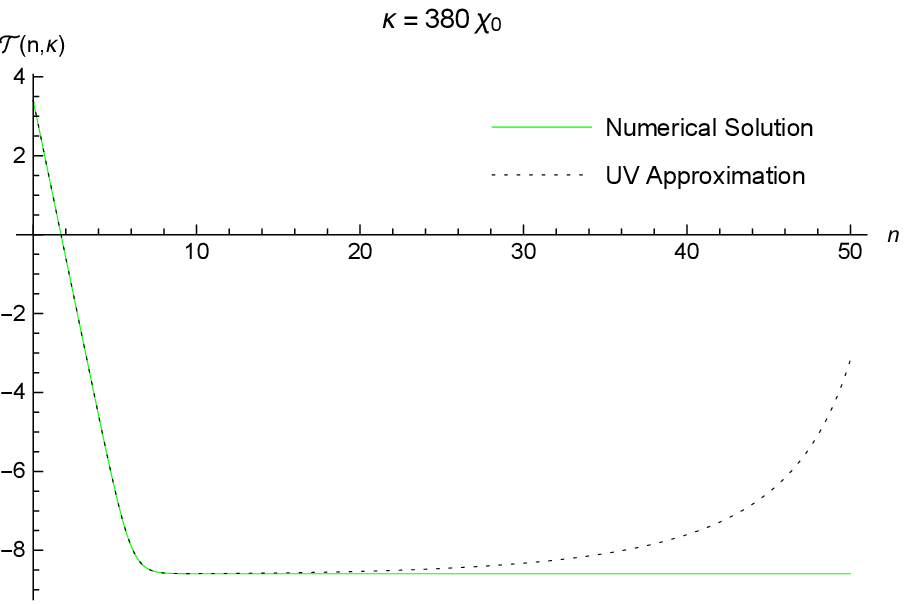}
\caption{$n_{\kappa} \simeq 6.0$}
\end{subfigure}
\begin{subfigure}[b]{0.33\textwidth}
\centering
\includegraphics[width=\textwidth]{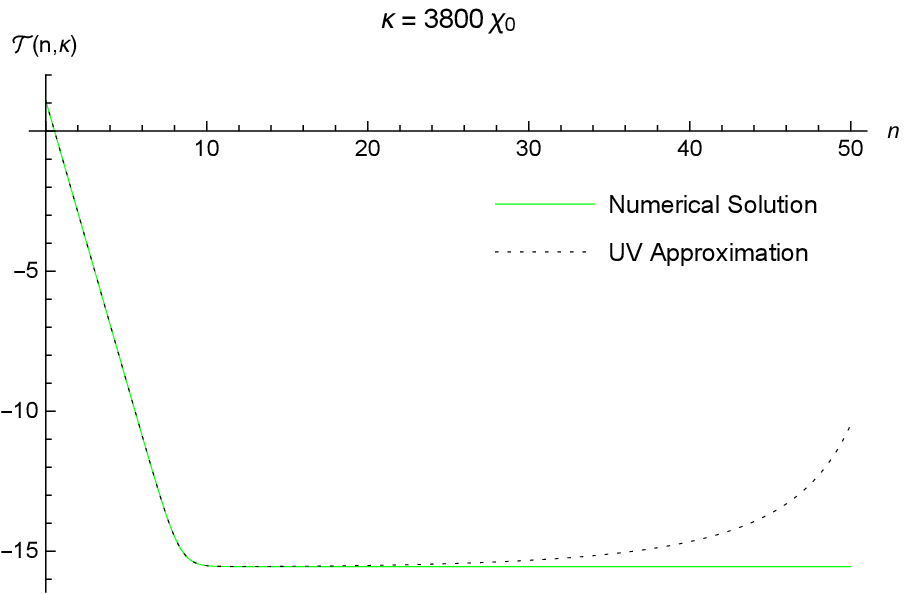}
\caption{$n_{\kappa} \simeq 8.3$}
\end{subfigure}
\begin{subfigure}[b]{0.33\textwidth}
\centering
\includegraphics[width=\textwidth]{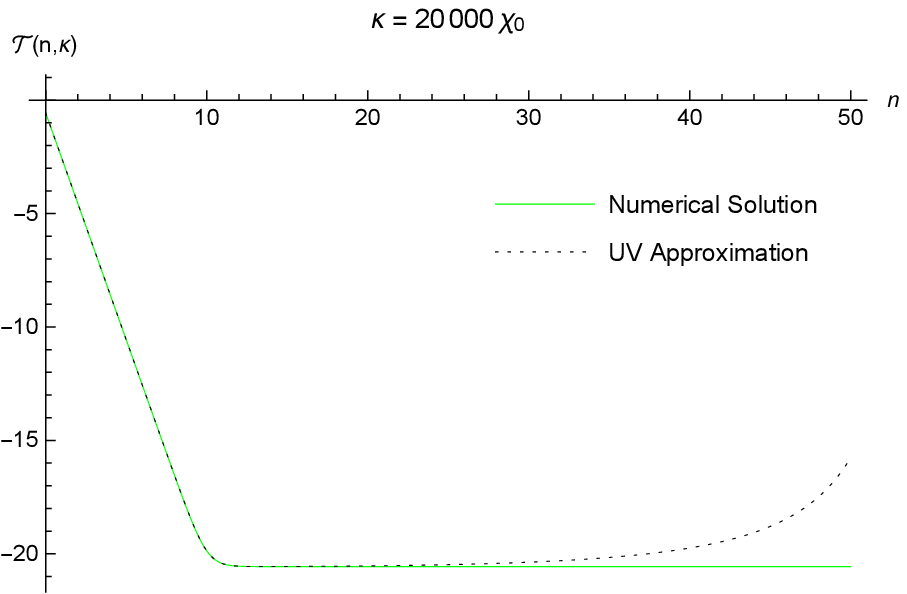}
\caption{$n_{\kappa} \simeq 10.0$}
\end{subfigure}%
\caption{\footnotesize Plots the massless, minimally coupled scalar amplitude 
$\mathcal{T}(n,\kappa)$ (in solid green) and the (black dashed) ultraviolet 
approximation (\ref{T1def}) versus the e-folding $n$ for three different 
values of $\kappa$.}
\label{MMCSUV}
\end{figure}

The MMC scalar amplitude is controlled by the relation between the physical wave
number $\kappa e^{-n}$ and the Hubble parameter $\chi(n)$. In the sub-horizon
regime of $\kappa > \chi(n) e^{n}$ the amplitude falls off roughly like
$\mathcal{T}(n,\kappa) \simeq -\ln(2 \kappa) - (D-2) n$, whereas it approaches
a constant in the super-horizon regime of $\kappa < \chi(n) e^{n}$. (The 
e-folding of first horizon crossing is $n_{\kappa}$ such that $\kappa =
\chi(n_{\kappa}) e^{n_{\kappa}}$.) Figure~\ref{MMCSUV} shows that both the 
sub-horizon regime, and also the initial phases of the super-horizon regime, 
are well described by the constant $\epsilon$ solution \cite{Brooker:2015iya},
\begin{equation}
\mathcal{T}_{1}(n,\kappa) \equiv \ln\Biggl[ \frac{\frac{\pi}{2} z(n,\kappa)}{2
\kappa e^{(D-2) n}} \Bigl\vert H^{(1)}_{\nu_t(n)}\Bigl( z(n,\kappa)\Bigr) 
\Bigr\vert^2 \Biggr] . \label{T1def}
\end{equation}
Here the ratio $z(n,\kappa)$ and the MMC scalar index $\nu_t(n)$ are,
\begin{equation}
z(n,\kappa) \equiv \frac{\kappa e^{-n}}{[1 \!-\! \epsilon(n)] \chi(n)} \qquad , 
\qquad \nu_t(n) \equiv \frac12 \Bigl( \frac{D \!-\! 1 \!-\! \epsilon(n)}{1 \!-\! 
\epsilon(n)} \Bigr) \; . \label{znudef}
\end{equation}

Of course expression (\ref{T1def}) is an approximation to the exact result.
Because we propose to use this to compute the divergent coincidence limit of
the propagator it is important to see how well $\mathcal{T}_1(n,\kappa)$ 
captures the ultraviolet behavior of $\mathcal{T}(n,\kappa)$. Because 
(\ref{T1def}) is exact for constant first slow roll parameter, the deviation 
must involve derivatives of $\epsilon(n)$. It turns out to fall off like 
$\kappa^{-4}$ \cite{Brooker:2015iya},
\begin{equation}
\mathcal{T}(n,\kappa) - \mathcal{T}_1(n,\kappa) = \Bigl( \frac{D\!-\!2}{16}
\Bigr) \Bigl[ (D + 5 - 7 \epsilon) \epsilon' + \epsilon''\Bigr] \Bigl( 
\frac{\chi e^n}{\kappa}\Bigr)^4 + O\Biggl( \Bigl(\frac{\chi e^n}{\kappa}
\Bigr)^6 \Biggr) . \label{Tasymp}
\end{equation}
We will see in section 4 that this suffices for an exact description of the
ultraviolet. 

\begin{figure}[H]
\centering
\begin{subfigure}[b]{0.33\textwidth}
\centering
\includegraphics[width=\textwidth]{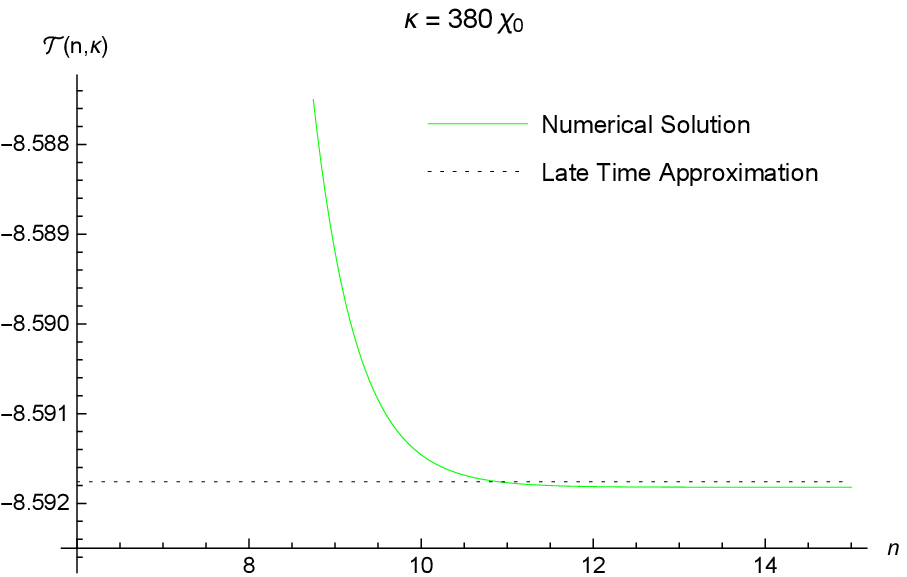}
\caption{$n_{\kappa} \simeq 6.0$}
\end{subfigure}
\begin{subfigure}[b]{0.33\textwidth}
\centering
\includegraphics[width=\textwidth]{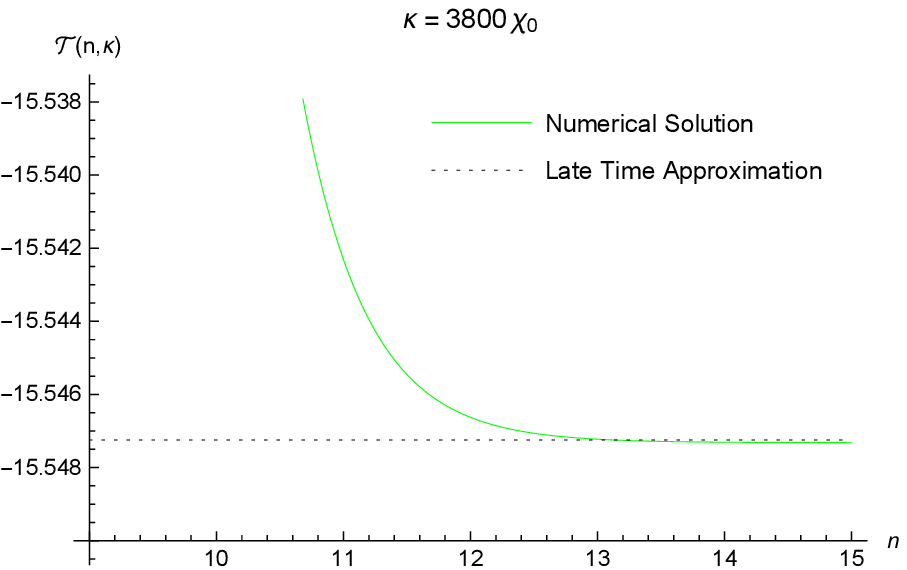}
\caption{$n_{\kappa} \simeq 8.3$}
\end{subfigure}
\begin{subfigure}[b]{0.33\textwidth}
\centering
\includegraphics[width=\textwidth]{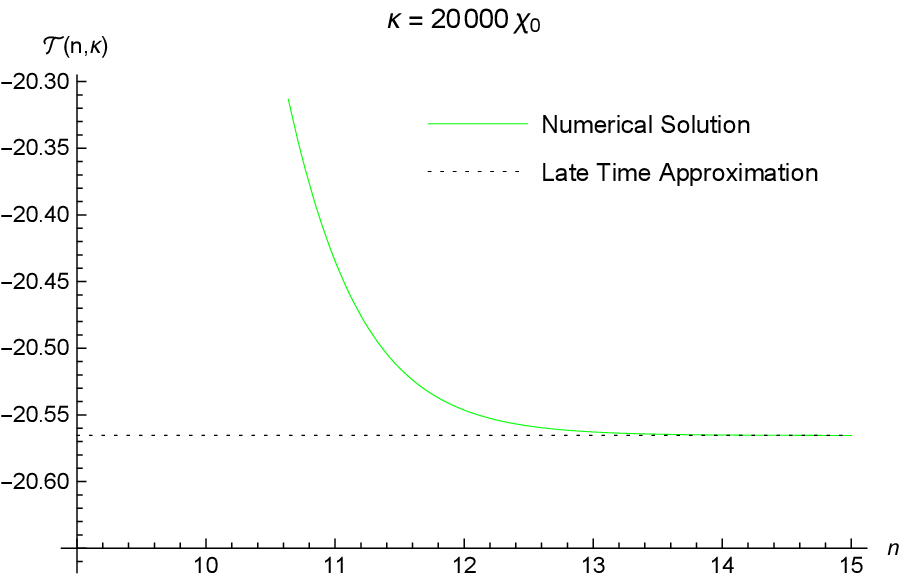}
\caption{$n_{\kappa} \simeq 10.0$}
\end{subfigure}%
\caption{\footnotesize Plots the massless, minimally coupled scalar amplitude 
$\mathcal{T}(n,\kappa)$ (in solid green) and the (black dashed) late time 
approximation (\ref{T2def}) versus the e-folding $n$ for three different 
values of $\kappa$.}
\label{MMCSlate}
\end{figure}

The discrepancy between $\mathcal{T}(n,\kappa)$ and $\mathcal{T}_1(n,\kappa)$
that is evident at late times in Figure~\ref{MMCSUV} is due to evolution of
the first slow roll parameter $\epsilon(n)$. Figure~\ref{MMCSlate} shows that
the asymptotic late time phase is captured with great accuracy by the form, 
\begin{equation}
\mathcal{T}_2(n,\kappa) = \ln\Biggl[ \frac{\chi^2(n_{\kappa})}{2 \kappa^3} 
\times C\Bigl( \epsilon(n_{\kappa})\Bigr) \Biggr] \; , \label{T2def}
\end{equation}
where the nearly unit correction factor $C(\epsilon)$ is,
\begin{equation}
C(\epsilon) \equiv \frac1{\pi} \Gamma^2\Bigl( \frac12 + \frac1{1 \!-\! \epsilon}
\Bigr) [ 2 (1 \!-\! \epsilon)]^{\frac{2}{1-\epsilon}} \; . \label{Cdef}
\end{equation}
\noindent Expression (\ref{T2def}) is exact for constant $\epsilon(n)$. When
the first slow roll parameter evolves there are very small nonlocal corrections
whose form is known \cite{Brooker:2017kjd} but whose net contribution is 
negligible for smooth potentials.

\subsection{Temporal Photon}

\begin{figure}[H]
\centering
\begin{subfigure}[b]{0.33\textwidth}
\centering
\includegraphics[width=\textwidth]{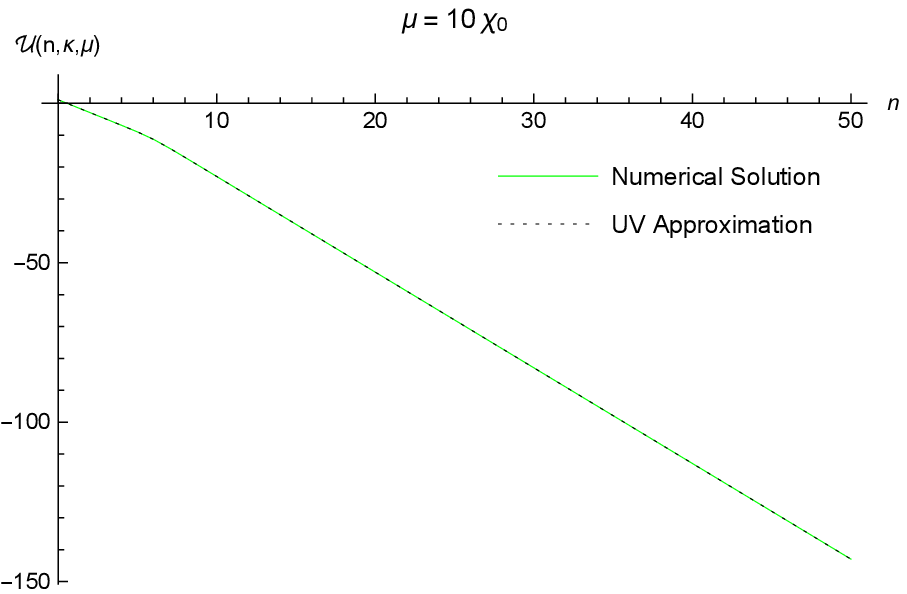}
\caption{$n_{\mu} < 0$}
\end{subfigure}
\begin{subfigure}[b]{0.33\textwidth}
\centering
\includegraphics[width=\textwidth]{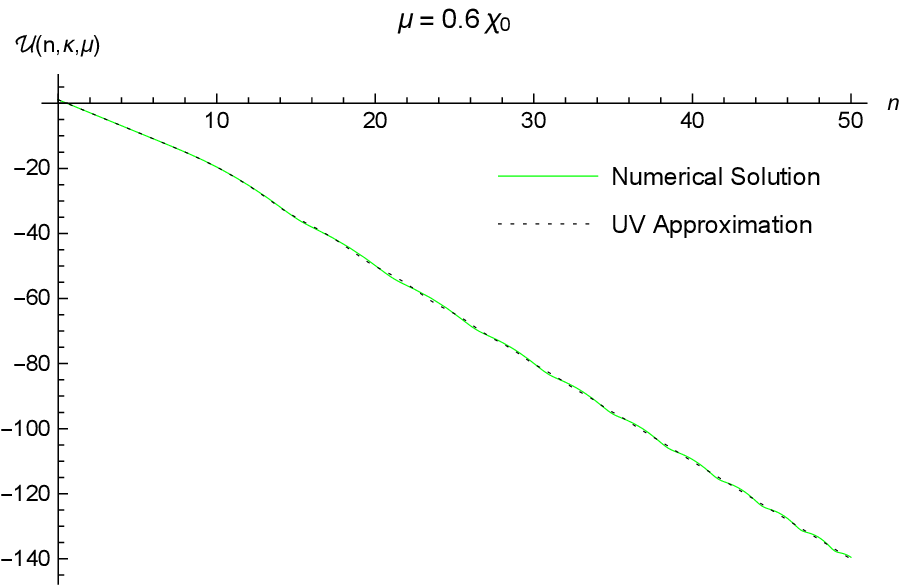}
\caption{$n_{\mu} < 0$}
\end{subfigure}
\begin{subfigure}[b]{0.33\textwidth}
\centering
\includegraphics[width=\textwidth]{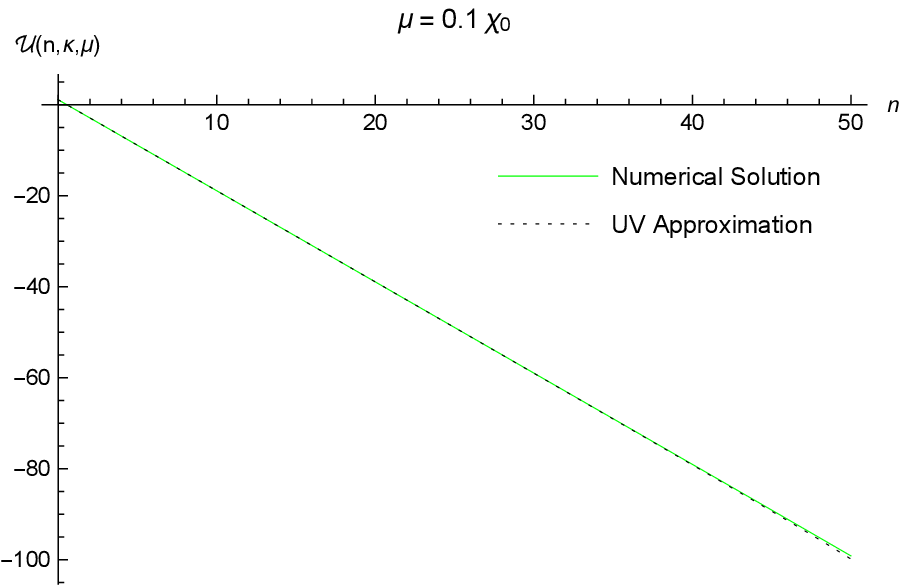}
\caption{$n_{\mu} > 50$}
\end{subfigure}%
\caption{\footnotesize Plots the temporal amplitude $\mathcal{U}(n,\kappa,\mu)$ 
and the ultraviolet approximation (\ref{U1def}) versus the e-folding $n$ for 
$\kappa = 3800 \chi_0$ (with $n_{\kappa} \simeq 8.3$) and three different values 
of $\mu$ with outside the range of inflation.}
\label{TempUV}
\end{figure}

The temporal photon amplitude is very similar to the massive scalar which 
was the subject of a previous study \cite{Kyriazis:2019xgj}. Like that system,
the functional form of the amplitude is controlled by two key events:
\begin{enumerate}
\item{First horizon crossing at $n_{\kappa}$ such that $\kappa e^{-n_{\kappa}}
= \chi(n_{\kappa})$; and}
\item{Mass domination at $n_{\mu}$ such that $\mu = \frac12 
\chi(n_{\mu})$.\footnote{The quadratic slow roll approximation 
(\ref{slowroll}) gives $n_{\mu} \simeq \frac12 \psi_0^2 [1 -
(2 \mu/\chi_0)^2]$.}}
\end{enumerate}
The ultraviolet is well approximated by the form that applies for constant
$\epsilon(n)$ and $\mu \propto \chi(n)$ \cite{Janssen:2009pb},
\begin{equation}
\mathcal{U}_{1}(n,\kappa,\mu) \equiv \ln\Biggl[ \frac{\frac{\pi}{2} 
z(n,\kappa)}{2 \kappa e^{(D-2) n}} \Bigl\vert H^{(1)}_{\nu_u(n,\mu)}
\Bigl( z(n,\kappa)\Bigr) \Bigr\vert^2 \Biggr] , \label{U1def}
\end{equation}
where the temporal index is,
\begin{equation}
\nu^2_u(n,\mu) \equiv \frac14 \Bigl( \frac{D \!-\! 3 \!+\! \epsilon(n)}{1 
\!-\! \epsilon(n)} \Bigr)^2 \!\!- \frac{\mu^2}{[1 \!-\! \epsilon(n)]^2 
\chi^2(n)} . \label{nuudef}
\end{equation}
Figure~\ref{TempUV} shows that the ultraviolet approximation is excellent
when matter domination comes either before or after inflation.

The ultraviolet regime is $\kappa e^{-n} \gg \{\chi(n),\mu\}$. To see how well
the ultraviolet approximation captures this regime we substitute the difference
into the exact evolution equation (\ref{scUeqn}) and expand in powers of $e^n 
\chi(n)/\kappa$ to find \cite{Kyriazis:2019xgj},
\begin{eqnarray}
\lefteqn{\mathcal{U}(n,\kappa,\mu) - \mathcal{U}_1(n,\kappa,\mu) = \Biggl\{ 
\Bigl( 5 \epsilon - 3 \epsilon^2\Bigr) \frac{\mu^2}{4 \chi^2} } \nonumber \\
& & \hspace{1.8cm} + \Bigl( \frac{D\!-\!2}{16}\Bigr) \Bigl[ (D - 9 + 7 \epsilon) 
\epsilon' - \epsilon''\Bigr] \Biggr\} \Bigl( \frac{\chi e^n}{\kappa}\Bigr)^4 \!+ 
O\Biggl( \Bigl(\frac{\chi e^n}{\kappa}\Bigr)^6 \Biggr) . \qquad \label{Uasymp}
\end{eqnarray}
This is suffices to give an exact result for the ultraviolet so we that can take 
the unregulated limit of $D = 4$ for the approximations which pertain for $n > 
n_{\kappa}$.

The various terms in equation (\ref{scUeqn}) behave differently before and after
first horizon crossing. Evolution before first horizon crossing is controlled by
the 4th and 7th terms,
\begin{equation}
\frac{2 \kappa^2 e^{-2n}}{\chi^2} - \frac{e^{-2(D-1)n - 2\mathcal{U}}}{2 \chi^2}
\simeq 0 \qquad \Longrightarrow \qquad \mathcal{U} \simeq -\ln(2\kappa) - 
(D\!-\!2) n \; . \label{scUUV}
\end{equation}
After first horizon crossing these terms rapidly redshift into insignificance.
We can take the unregulated limit ($D=4$), and equation (\ref{scUeqn}) becomes,
\begin{equation}
\mathcal{U}'' + \frac12 {\mathcal{U}'}^2 + (3 \!-\! \epsilon) \mathcal{U}'
+ 4 (1 \!-\! \epsilon) + \frac{2 \mu^2}{\chi^2} \simeq 0 \; . \label{scUlate}
\end{equation}
This is a nonlinear, first order equation for $\mathcal{U}'$. Following
\cite{Kyriazis:2019xgj} we make the ansatz,
\begin{equation}
\mathcal{U}' \simeq \alpha + \beta \tanh(\gamma) \; . \label{ansatz}
\end{equation}
Substituting (\ref{ansatz}) in (\ref{scUlate}) gives,
\begin{eqnarray}
\lefteqn{ \Bigl({\rm Eqn.\ \ref{scUlate}} \Bigr) = \alpha' + \frac12 \alpha^2
+ \frac12 \beta^2 + (3 \!-\! \epsilon) \alpha + 4 (1 \!-\! \epsilon) +
\frac{2 \mu^2}{\chi^2} } \nonumber \\
& & \hspace{2.7cm} + \Bigl[ (3 \!-\! \epsilon \!+\! \alpha) \beta + \beta'\Bigr]
\tanh(\gamma) + \beta \Bigl( \gamma' - \frac12 \beta\Bigr) {\rm sech}^2(\gamma) 
\; . \qquad \label{almost}
\end{eqnarray}Ansatz (\ref{ansatz}) does not quite solve (\ref{scUlate}), but 
the following choices reduce the residue to terms of order $\epsilon \times 
\tanh(\gamma)$,
\begin{equation}
\alpha = -3 \qquad , \qquad \frac14 \beta^2 = \frac14 + \frac{\epsilon}{2} - 
\frac{\mu^2}{\chi^2} \qquad , \qquad \gamma' = \frac12 \beta \; . 
\label{ansatzparams}
\end{equation}

\begin{figure}[H]
\centering
\begin{subfigure}[b]{0.33\textwidth}
\centering
\includegraphics[width=\textwidth]{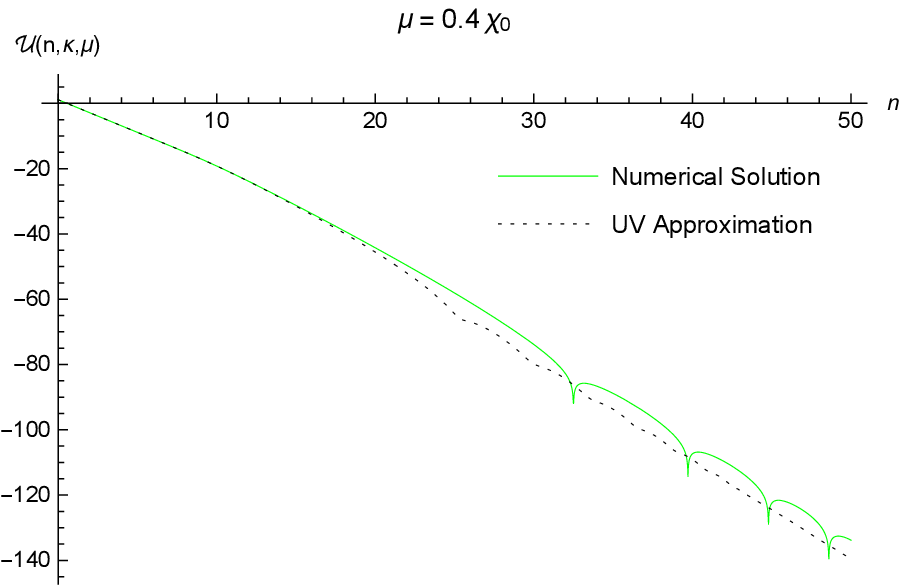}
\end{subfigure}
\begin{subfigure}[b]{0.33\textwidth}
\centering
\includegraphics[width=\textwidth]{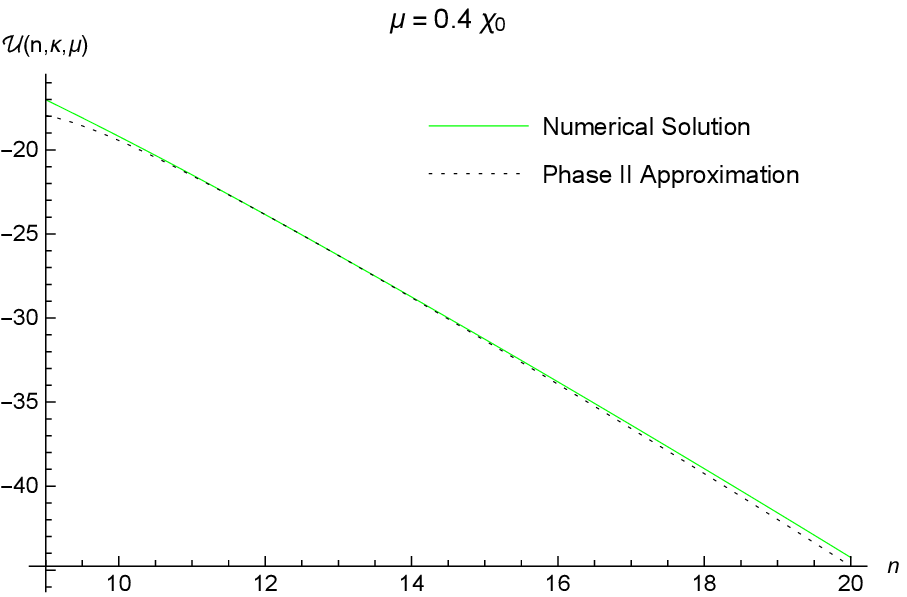}
\end{subfigure}
\begin{subfigure}[b]{0.33\textwidth}
\centering
\includegraphics[width=\textwidth]{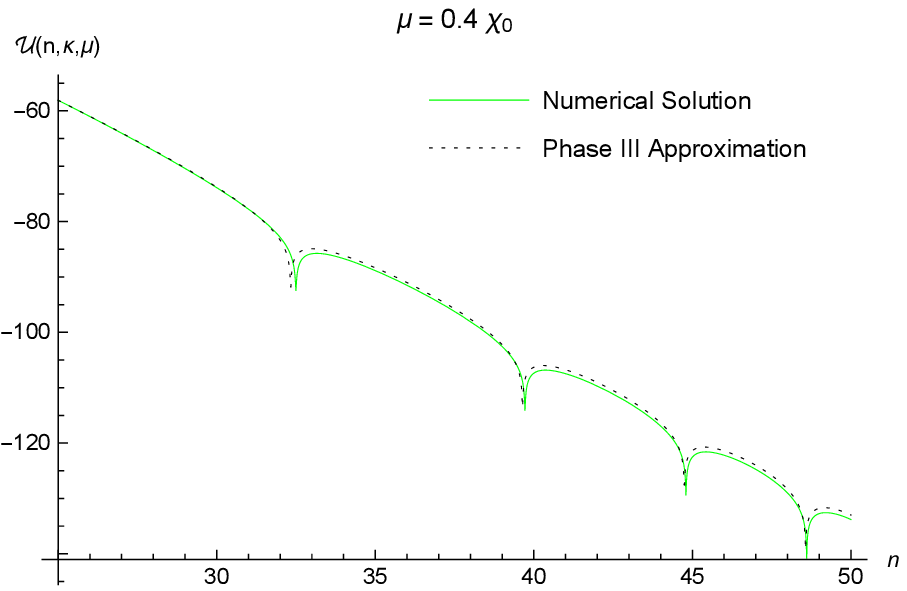}
\end{subfigure}
\caption{\footnotesize Plots the temporal amplitude 
$\mathcal{U}(n,3800 \chi_0, 0.4 \chi_0)$ and the three approximations: 
(\ref{U1def}), (\ref{U2def}) and (\ref{U3def}). For $\kappa = 3800 \chi_0$
horizon crossing occurs at $n_{\kappa} \simeq 8.3$; for $\mu = 0.4 \chi_0$ mass
domination occurs at $n_{\mu} \simeq 20.2$.}
\label{Temp3phasesA}
\end{figure}

\begin{figure}[H]
\centering
\begin{subfigure}[b]{0.33\textwidth}
\centering
\includegraphics[width=\textwidth]{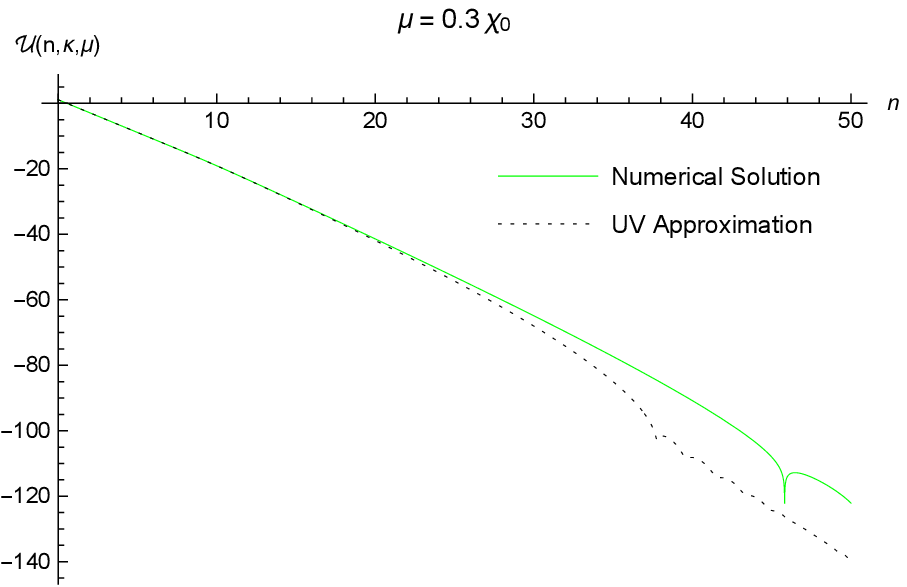}
\end{subfigure}
\begin{subfigure}[b]{0.33\textwidth}
\centering
\includegraphics[width=\textwidth]{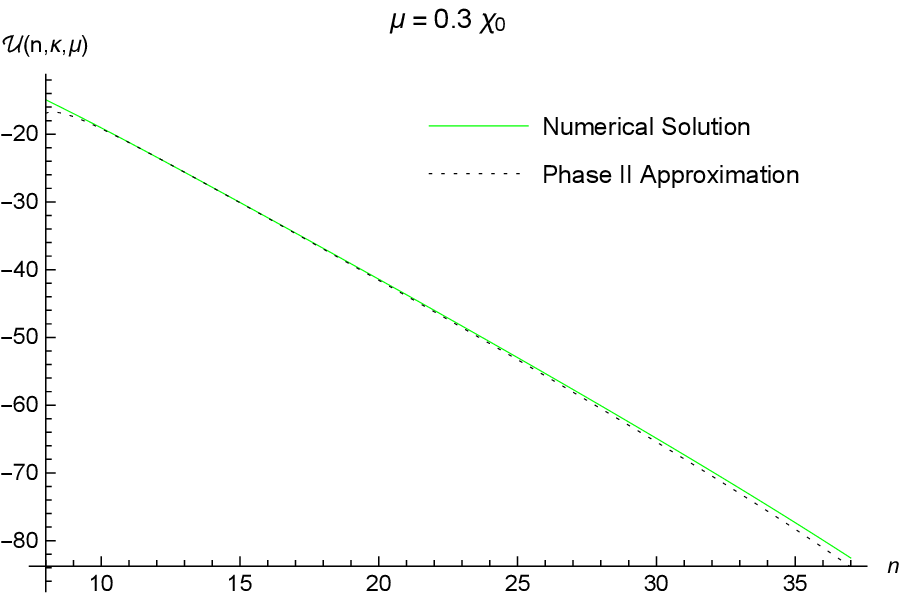}
\end{subfigure}
\begin{subfigure}[b]{0.33\textwidth}
\centering
\includegraphics[width=\textwidth]{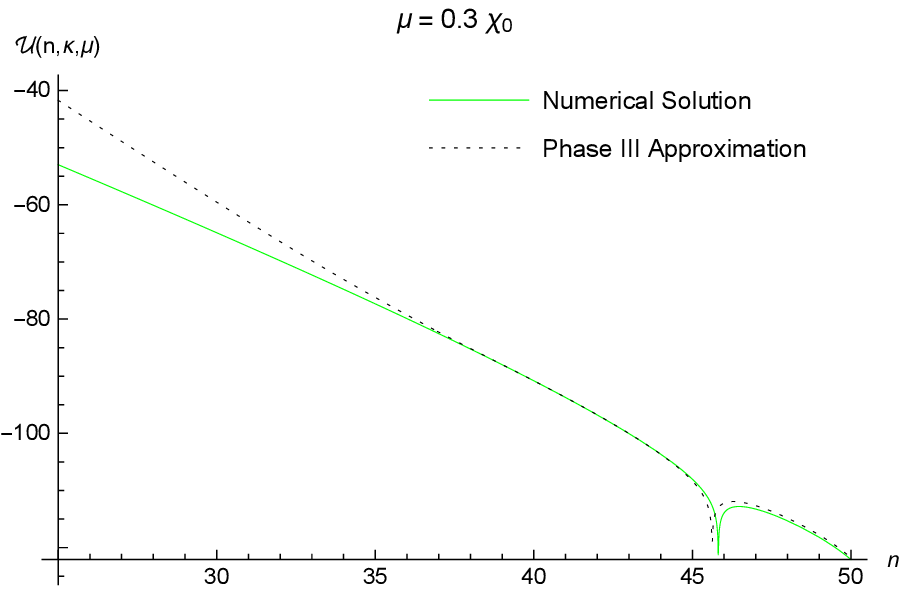}
\end{subfigure}
\caption{\footnotesize Plots the temporal amplitude 
$\mathcal{U}(n,3800 \chi_0, 0.3 \chi_0)$ and the three approximations: 
(\ref{U1def}), (\ref{U2def}) and (\ref{U3def}). For $\kappa = 3800 \chi_0$
horizon crossing occurs at $n_{\kappa} \simeq 8.3$; for $\mu = 0.3 \chi_0$ mass
domination occurs at $n_{\mu} \simeq 36.0$.}
\label{Temp3phasesB}
\end{figure}

Figures~\ref{Temp3phasesA} and \ref{Temp3phasesB} show how 
$\mathcal{U}(n,\kappa,\mu)$ behaves when mass domination comes after first horizon
crossing and before the end of inflation.
\noindent First comes a phase of slow decline followed by a period of
oscillations. From (\ref{ansatz}) with (\ref{ansatzparams}) we see that these
phases are controlled by a ``frequency'' defined as,
\begin{equation}
\omega^2_u(n,\mu) \equiv \frac14 + \frac{\epsilon(n)}{2} - 
\frac{\mu^2}{\chi^2(n)} \equiv -\Omega^2_u(n,\mu) \; . \label{tempomega}
\end{equation}
During the phase of slow decline $\omega^2_u(n,\mu) > 0$. Integrating 
(\ref{ansatz}) with (\ref{ansatzparams}) for this case gives,
\begin{eqnarray}
\lefteqn{\mathcal{U}_{2}(n,\kappa,\mu) = \mathcal{U}_{2} - 3 (n \!-\! n_2) + 
2 \ln\Biggl[\cosh\Bigl( \int_{n_2}^{n} \!\!\! dn' \omega_u(n',\mu)\Bigr) } 
\nonumber \\
& & \hspace{5cm} + \Bigl( \frac{3 \!+\! \mathcal{U}_{2}'}{2 \omega_u(n_2,\mu)} 
\Bigr) \sinh\Bigl( \int_{n_2}^{n} \!\!\! dn' \omega_u(n',\mu)\Bigr) \Biggr] ,
\qquad \label{U2def}
\end{eqnarray}
where $n_2 \equiv n_{\kappa} + 4$. The oscillatory phase is characterized by 
$\omega^2_u(n,\mu) < 0$. Integrating (\ref{ansatz}) with (\ref{ansatzparams})
for this case produces,
\begin{eqnarray}
\lefteqn{\mathcal{U}_{3}(n,\kappa,\mu) = \mathcal{U}_{3} - 3 (n \!-\! n_3) + 
2 \ln\Biggl[ \Biggl\vert\cos\Bigl( \int_{n_3}^{n} \!\!\! dn' \Omega_u(n',\mu)\Bigr) } 
\nonumber \\
& & \hspace{5cm} + \Bigl( \frac{3 \!+\! \mathcal{U}_{3}'}{2 \Omega_u(n_3,\mu)} 
\Bigr) \sin\Bigl( \int_{n_3}^{n} \!\!\! dn' \Omega_u(n',\mu)\Bigr) \Biggr\vert 
\Biggr] , \qquad \label{U3def}
\end{eqnarray}
where $n_3 \equiv n_{\mu} + 4$. Figures~\ref{Temp3phasesA} and 
\ref{Temp3phasesB} show that these approximations are excellent.

\begin{figure}[H]
\centering
\begin{subfigure}[b]{0.33\textwidth}
\centering
\includegraphics[width=\textwidth]{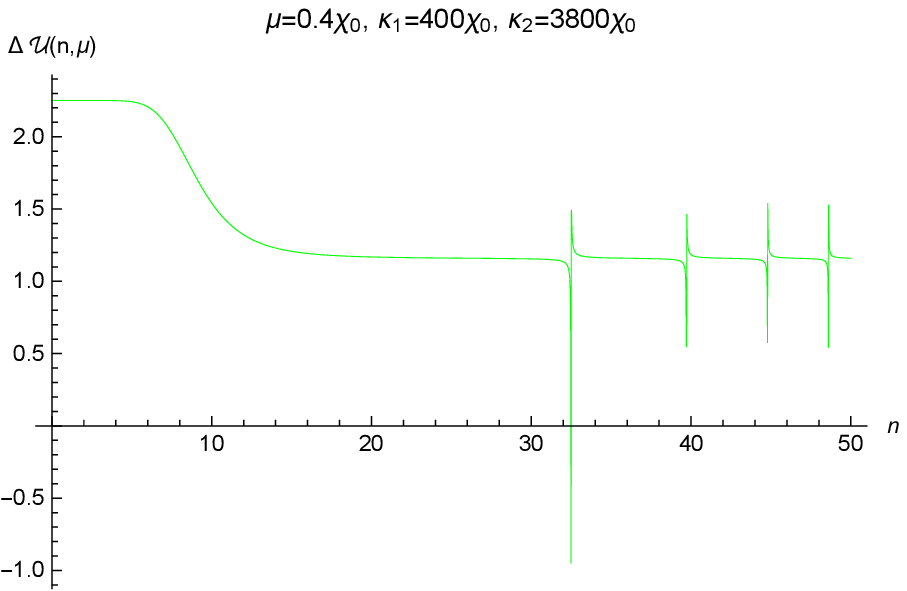}
\end{subfigure}
\begin{subfigure}[b]{0.33\textwidth}
\centering
\includegraphics[width=\textwidth]{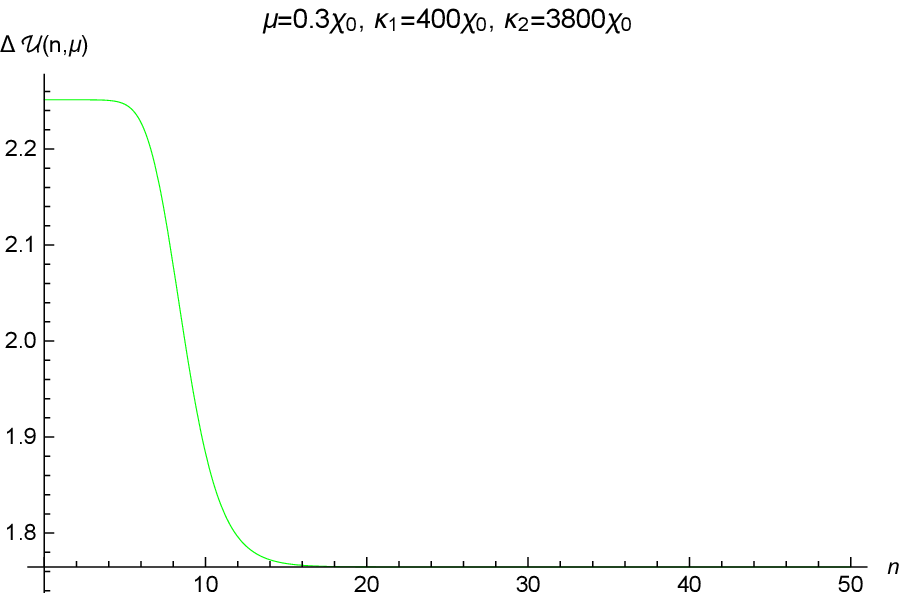}
\end{subfigure}
\begin{subfigure}[b]{0.33\textwidth}
\centering
\includegraphics[width=\textwidth]{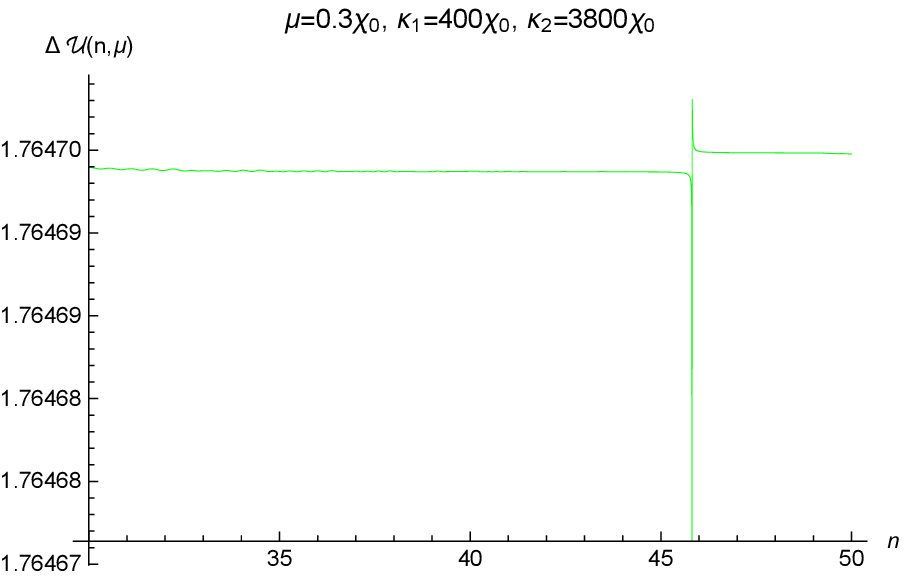}
\end{subfigure}
\caption{\footnotesize Plots the difference of the temporal amplitude 
$\Delta \mathcal{U} \equiv \mathcal{U}(n,\kappa_1,\mu) -
\mathcal{U}(n,\kappa_2,\mu)$ for $\kappa_1 = 400 \chi_0$ and $\kappa_2
= 3800 \chi_0$ with $\mu$ chosen so that all three approximations 
(\ref{U1def}), (\ref{U2def}) and (\ref{U3def}) are necessary.}
\label{Tempkdep}
\end{figure}

It is worth noting that the approximations (\ref{U2def}) and (\ref{U3def})
depend on $\kappa$ principally through the integration constants $\mathcal{U}_2
\equiv \mathcal{U}(n_2,\kappa,\mu)$ and $\mathcal{U}_3 \equiv 
\mathcal{U}(n_3,\kappa,\mu)$. Figure~\ref{Tempkdep} shows the difference
$\mathcal{U}(n,400 \chi_0,\mu) - \mathcal{U}(n,3800 \chi_0,\mu)$ for the same
two choices of $\mu$ in Figures~\ref{Temp3phasesA} and \ref{Temp3phasesB}. One
can see that the difference freezes into a constant after first horizon 
crossing to better than five significant figures!

\subsection{Spatially Transverse Photons}

\begin{figure}[H]
\centering
\begin{subfigure}[b]{0.33\textwidth}
\centering
\includegraphics[width=\textwidth]{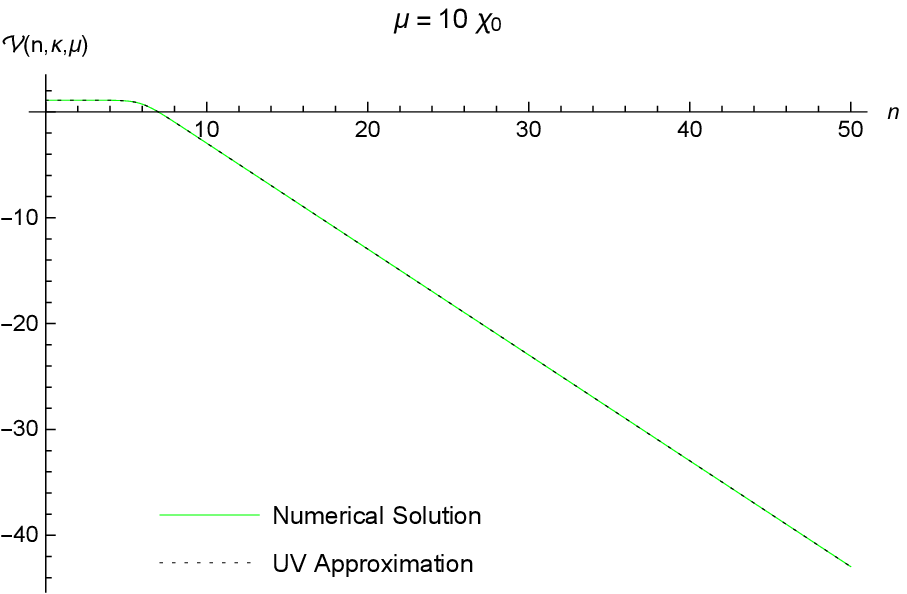}
\caption{$n_{\mu} < 0$}
\end{subfigure}
\begin{subfigure}[b]{0.33\textwidth}
\centering
\includegraphics[width=\textwidth]{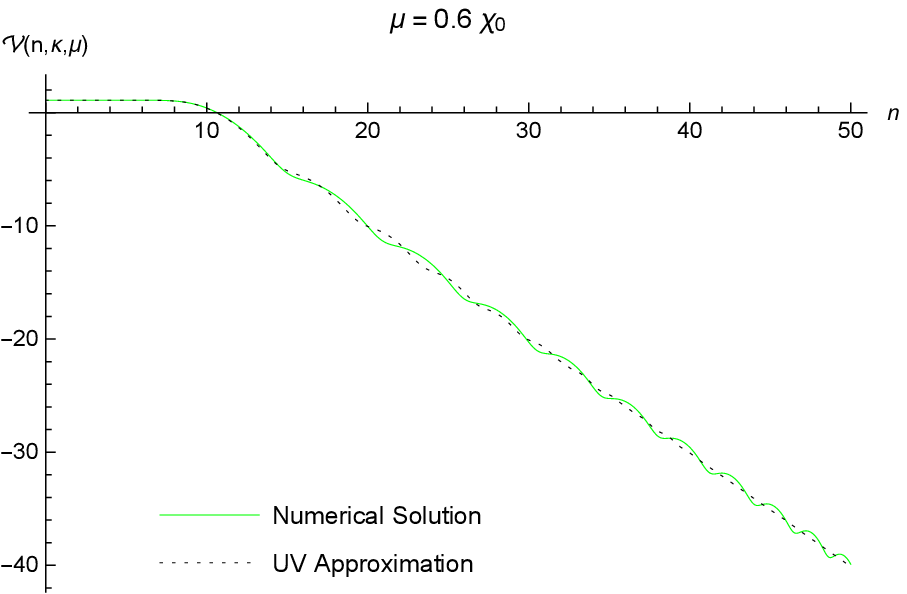}
\caption{$n_{\mu} < 0$}
\end{subfigure}
\begin{subfigure}[b]{0.33\textwidth}
\centering
\includegraphics[width=\textwidth]{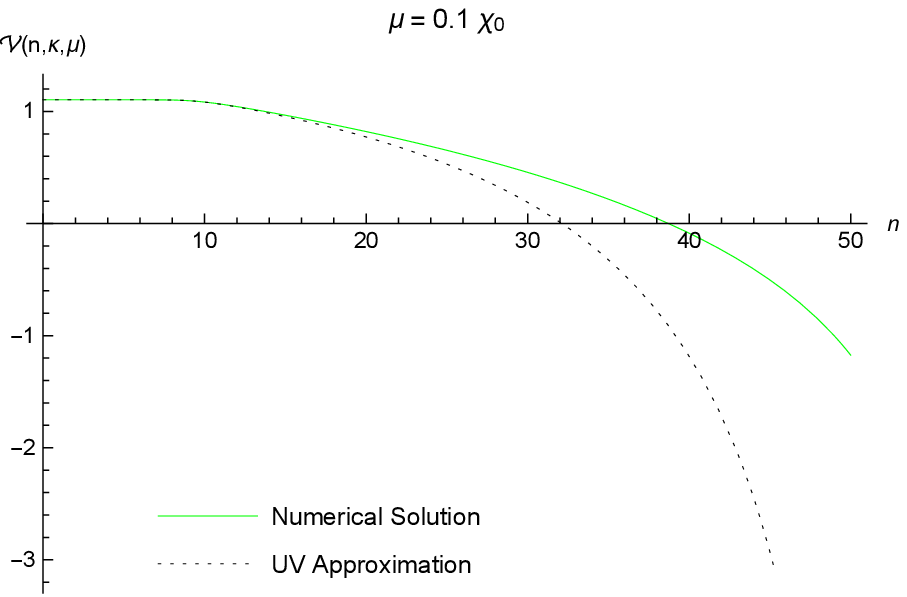}
\caption{$n_{\mu} > 50$}
\end{subfigure}%
\caption{\footnotesize Plots the transverse amplitude $\mathcal{V}(n,\kappa,\mu)$ 
and the ultraviolet approximation (\ref{V1def}) versus the e-folding $n$ for 
$\kappa = 3800 \chi_0$ (with $n_{\kappa} \simeq 8.3$) and three different values 
of $\mu$ with outside the range of inflation.}
\label{TransUV}
\end{figure}

The general considerations for the amplitude of spatially transverse photons 
are similar to those for temporal photons. Before first horizon crossing it 
is the 4th and last terms of equation (\ref{scVeqn}) which control the 
evolution,
\begin{equation}
\frac{2 \kappa^2 e^{-2n}}{\chi^2} - \frac{e^{-2(D-3)n - 2\mathcal{V}}}{
2 \chi^2} \simeq 0 \qquad \Longrightarrow \qquad \mathcal{V} \simeq 
-\ln(2\kappa) - (D \!-\! 4) n \; . \label{UVscVapprox}
\end{equation}
A more accurate approximation is,
\begin{equation}
\mathcal{V}_{1}(n,\kappa,\mu) \equiv \ln\Biggl[ \frac{\frac{\pi}{2} 
z(n,\kappa)}{2 \kappa e^{(D-4) n}} \Bigl\vert H^{(1)}_{\nu_v(n,\mu)}
\Bigl( z(n,\kappa)\Bigr) \Bigr\vert^2 \Biggr] , \label{V1def}
\end{equation}
where $z(n,\kappa)$ is the same as (\ref{znudef}) and the transverse index is,
\begin{equation}
\nu^2_v(n,\mu) \equiv \frac14 \Bigl( \frac{D \!-\! 3 \!-\! \epsilon(n)}{
1 \!-\! \epsilon(n)} \Bigr)^2 - \frac{\mu^2}{[1 \!-\! \epsilon(n)]^2 
\chi^2(n)} \; . \label{nuvdef}
\end{equation}
Note the slight (order $\epsilon$) difference between $\nu^2_u(n,\mu)$ and 
$\nu^2_v(n,\mu)$. Figure~\ref{TransUV} shows that (\ref{V1def}) is excellent
up to several e-foldings after first horizon crossing, and throughout inflation
for $n_{\mu} < 0$.
\noindent Expression (\ref{V1def}) also models the ultraviolet to high precision,
\begin{eqnarray}
\lefteqn{\mathcal{V}(n,\kappa,\mu) - \mathcal{V}_1(n,\kappa,\mu) = \Biggl\{ 
\Bigl( 5 \epsilon - 3 \epsilon^2\Bigr) \frac{\mu^2}{4 \chi^2} } \nonumber \\
& & \hspace{1.8cm} + \Bigl( \frac{D\!-\!4}{16}\Bigr) \Bigl[ (D + 3 - 7 \epsilon) 
\epsilon' + \epsilon''\Bigr] \Biggr\} \Bigl( \frac{\chi e^n}{\kappa}\Bigr)^4 \!+ 
O\Biggl( \Bigl(\frac{\chi e^n}{\kappa}\Bigr)^6 \Biggr) . \qquad \label{Vasymp}
\end{eqnarray}

\begin{figure}[H]
\centering
\begin{subfigure}[b]{0.33\textwidth}
\centering
\includegraphics[width=\textwidth]{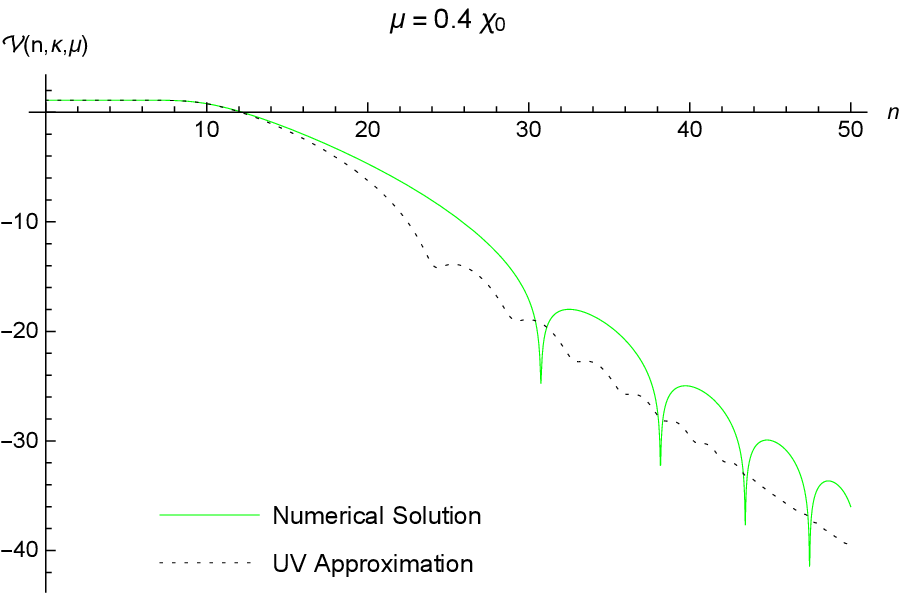}
\end{subfigure}
\begin{subfigure}[b]{0.33\textwidth}
\centering
\includegraphics[width=\textwidth]{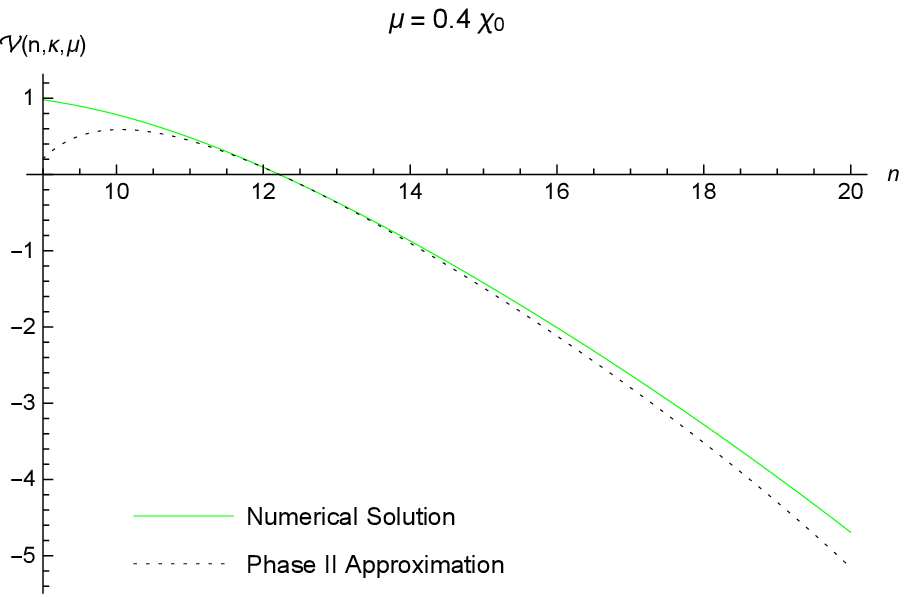}
\end{subfigure}
\begin{subfigure}[b]{0.33\textwidth}
\centering
\includegraphics[width=\textwidth]{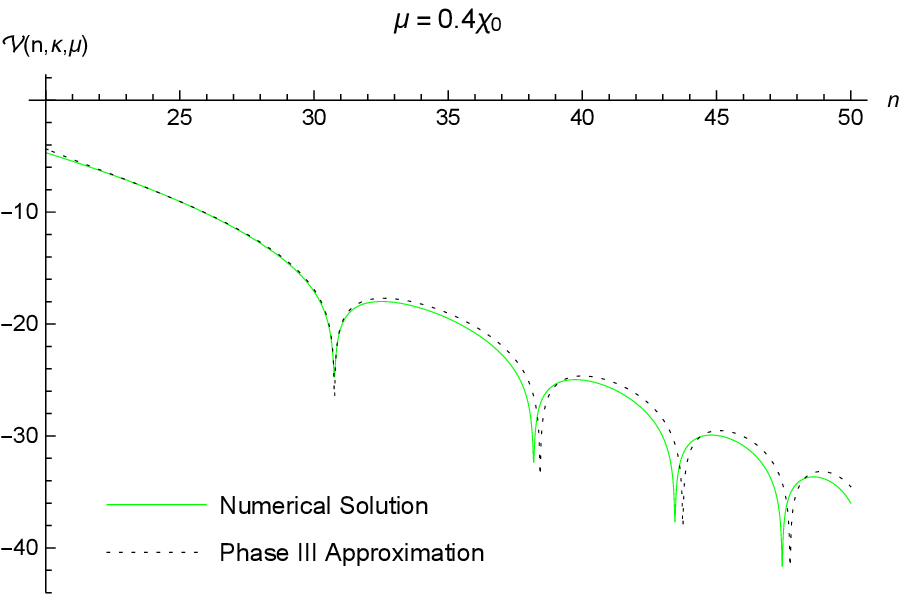}
\end{subfigure}
\caption{\footnotesize Plots the transverse amplitude 
$\mathcal{V}(n,3800 \chi_0, 0.4 \chi_0)$ and the three approximations: 
(\ref{V1def}), (\ref{V2def}) and (\ref{V3def}). For $\kappa = 3800 \chi_0$
horizon crossing occurs at $n_{\kappa} \simeq 8.3$; for $\mu = 0.4 \chi_0$ mass
domination occurs at $n_{\mu} \simeq 20.2$.}
\label{Trans3phasesA}
\end{figure}

Figure~\ref{Trans3phasesA} shows $\mathcal{V}(n,\kappa,\mu)$ for the case
where $n_{\mu}$ happens after first horizon crossing and before the end of
inflation. One sees the same phases of slow decline after first horizon 
crossing, followed by oscillations.
\noindent The second and third phases can be understood by noting that the two 
terms of expression (\ref{UVscVapprox}) redshift into insignificance after 
first horizon crossing. We can also set $D=4$ so that equation (\ref{scVeqn}) 
degenerates to,
\begin{equation}
\mathcal{V}'' + \frac12 {\mathcal{V}'}^2 + (1 \!-\! \epsilon) \mathcal{V}'
+ \frac{2 \mu^2}{\chi^2} \simeq 0 \; .
\end{equation}
The same ansatz (\ref{ansatz}) applies to this regime, with the parameter
choices,
\begin{equation}
\alpha = -1 \quad , \quad \frac14 \beta^2 = \frac14 - \frac{\mu^2}{\chi^2} 
\equiv \omega^2_v \equiv -\Omega^2_v \quad , \quad \gamma' = \frac12 \beta \; .
\label{Vparams}
\end{equation}
Just as there was an order $\epsilon$ difference between the temporal and 
transverse indices --- expressions (\ref{znudef}) and (\ref{nuvdef}), respectively 
--- so too there is an order $\epsilon$ difference between $\omega^2_u(n,\mu)$ and
$\omega^2_v(n,\kappa)$. 

\begin{figure}[H]
\centering
\begin{subfigure}[b]{0.33\textwidth}
\centering
\includegraphics[width=\textwidth]{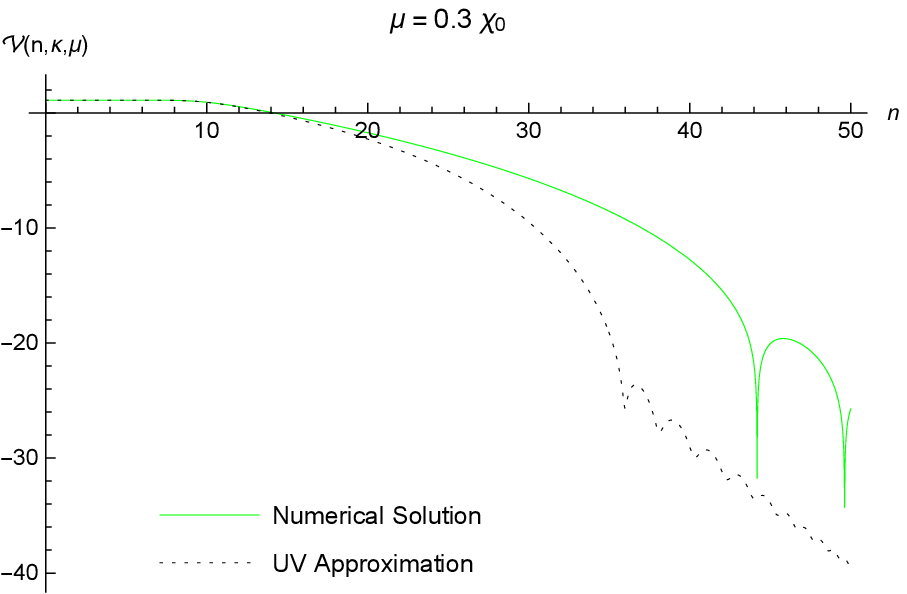}
\end{subfigure}
\begin{subfigure}[b]{0.33\textwidth}
\centering
\includegraphics[width=\textwidth]{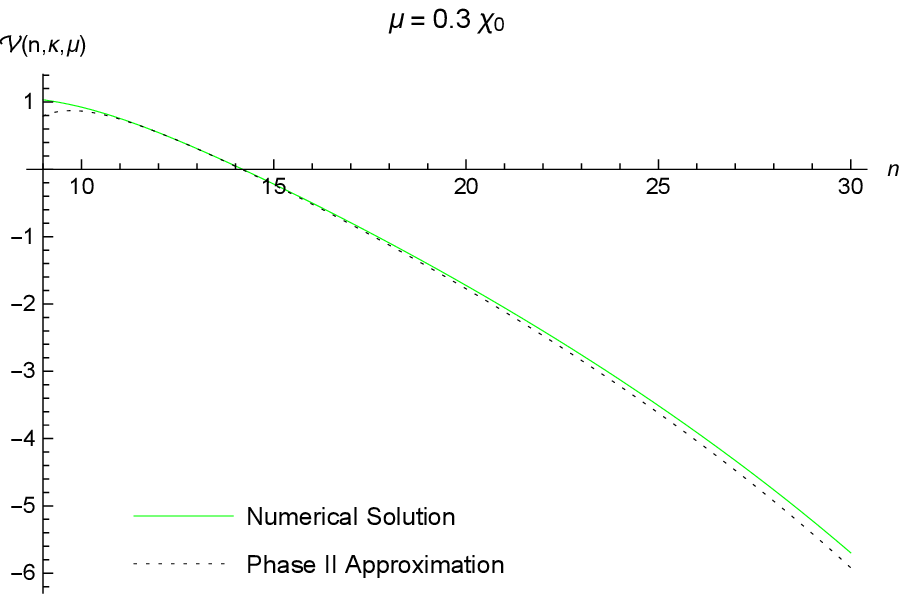}
\end{subfigure}
\begin{subfigure}[b]{0.33\textwidth}
\centering
\includegraphics[width=\textwidth]{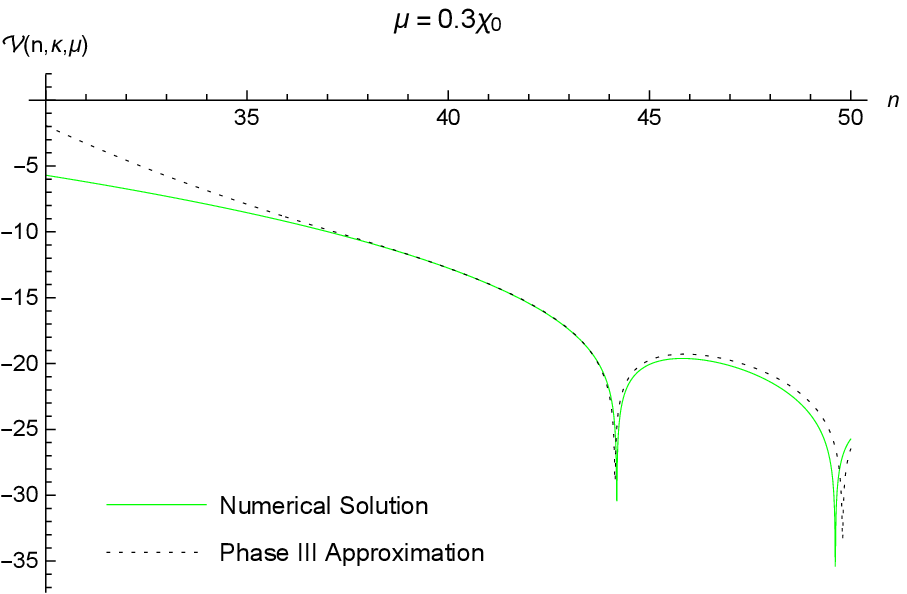}
\end{subfigure}
\caption{\footnotesize Plots the transverse amplitude 
$\mathcal{V}(n,3800 \chi_0, 0.3 \chi_0)$ and the three approximations: 
(\ref{V1def}), (\ref{V2def}) and (\ref{V3def}). For $\kappa = 3800 \chi_0$
horizon crossing occurs at $n_{\kappa} \simeq 8.3$; for $\mu = 0.3 \chi_0$ mass
domination occurs at $n_{\mu} \simeq 36.0$.}
\label{Trans3phasesB}
\end{figure}

Integrating (\ref{ansatz}) with (\ref{Vparams}) for $\omega^2_v(,\mu) > 0$ gives,
\begin{eqnarray}
\lefteqn{\mathcal{V}_{2}(n,\kappa,\mu) = \mathcal{V}_{2} - (n \!-\! n_2) + 
2 \ln\Biggl[\cosh\Bigl( \int_{n_2}^{n} \!\!\! dn' \omega_v(n',\mu)\Bigr) } 
\nonumber \\
& & \hspace{5cm} + \Bigl( \frac{1 \!+\! \mathcal{V}_{2}'}{2 \omega_v(n_2,\mu)} 
\Bigr) \sinh\Bigl( \int_{n_2}^{n} \!\!\! dn' \omega_v(n',\mu)\Bigr) \Biggr] ,
\qquad \label{V2def}
\end{eqnarray}
where $n_2 \equiv n_{\kappa} + 4$. Integrating (\ref{ansatz}) with (\ref{Vparams})
for $\omega^2_v(n,\mu) < 0$ results in,
\begin{eqnarray}
\lefteqn{\mathcal{V}_{3}(n,\kappa,\mu) = \mathcal{V}_{3} - (n \!-\! n_3) + 
2 \ln\Biggl[ \Biggl\vert\cos\Bigl( \int_{n_3}^{n} \!\!\! dn' \Omega_v(n',\mu)\Bigr) } 
\nonumber \\
& & \hspace{5cm} + \Bigl( \frac{1 \!+\! \mathcal{V}_{3}'}{2 \Omega_v(n_3,\mu)} 
\Bigr) \sin\Bigl( \int_{n_3}^{n} \!\!\! dn' \Omega_v(n',\mu)\Bigr) \Biggr\vert 
\Biggr] , \qquad \label{V3def}
\end{eqnarray}
where $n_3 \equiv n_{\mu} + 4$. Figures~\ref{Trans3phasesA} and \ref{Trans3phasesB}
demonstrate that the (\ref{V2def}) and (\ref{V3def}) approximations are excellent.

\begin{figure}[H]
\centering
\begin{subfigure}[b]{0.33\textwidth}
\centering
\includegraphics[width=\textwidth]{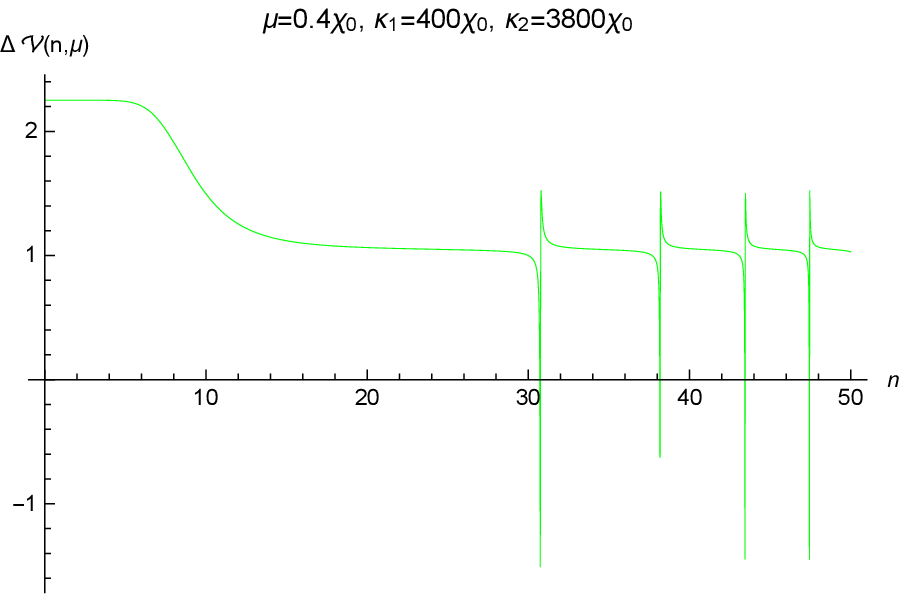}
\end{subfigure}
\begin{subfigure}[b]{0.33\textwidth}
\centering
\includegraphics[width=\textwidth]{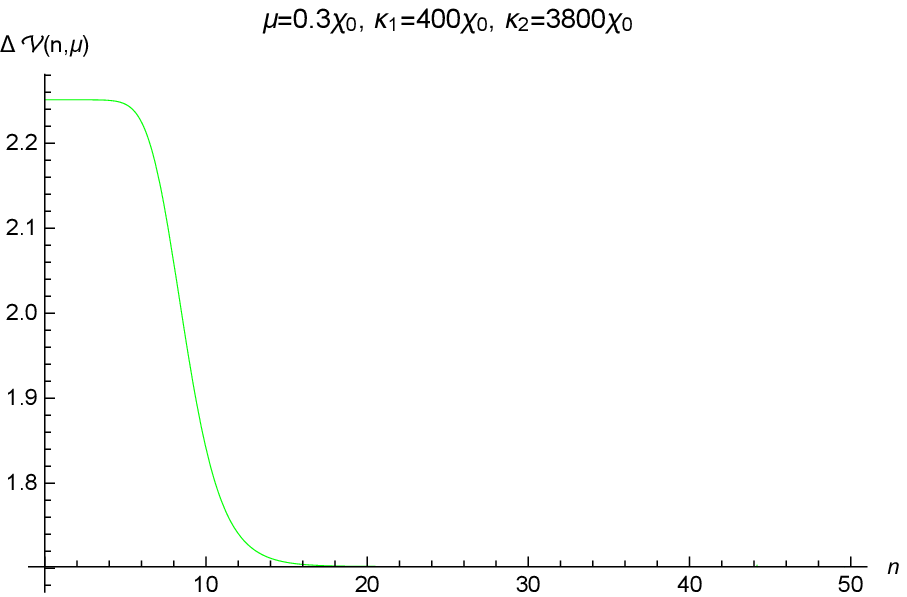}
\end{subfigure}
\begin{subfigure}[b]{0.33\textwidth}
\centering
\includegraphics[width=\textwidth]{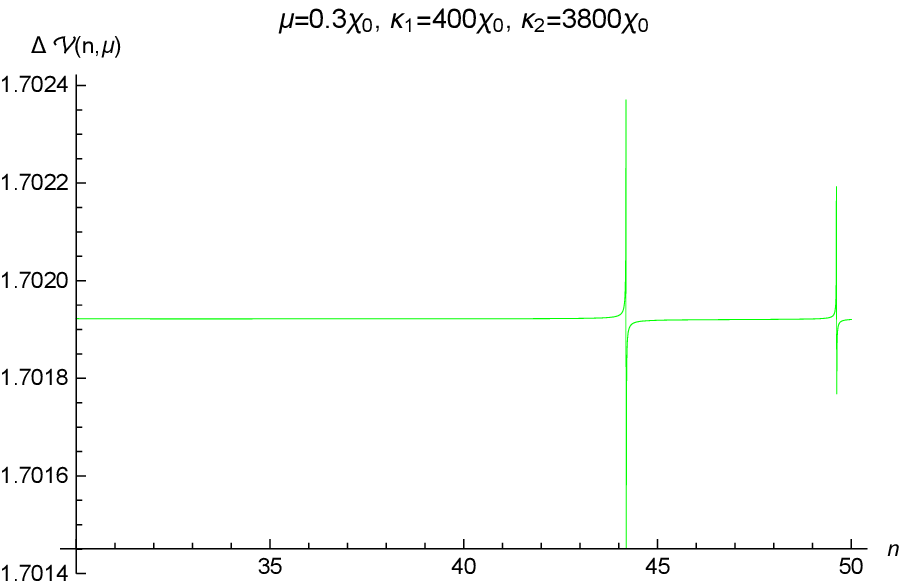}
\end{subfigure}
\caption{\footnotesize Plots the difference of the transverse amplitude 
$\Delta \mathcal{V} \equiv \mathcal{V}(n,\kappa_1,\mu) -
\mathcal{V}(n,\kappa_2,\mu)$ for $\kappa_1 = 400 \chi_0$ and $\kappa_2 =
3800 \chi_0$ with $\mu$ chosen so that all three approximations (\ref{V1def}), 
(\ref{V2def}) and (\ref{V3def}) are necessary.}
\label{Transkdep}
\end{figure}

Finally, we note that from Figure~\ref{Transkdep} that $\mathcal{V}'(n,\kappa,\mu)$
is nearly independent of $\kappa$ after first horizon crossing.
\noindent One consequence for the (\ref{V2def}) and (\ref{V3def}) 
approximations is that only the integration constants $\mathcal{V}_2$ and
$\mathcal{V}_3$ depend on $\kappa$.

\subsection{Plateau Potentials}

We chose the quadratic dimensionless potential $U(\psi\psi^*) = c^2 \psi \psi^*$ for 
detailed studies because it gives simple, analytic expressions (\ref{slowroll}) in
the slow roll approximation for the dimensionless Hubble parameter $\chi(n)$ and the 
first slow roll parameter $\epsilon(n)$. Setting $c \simeq 7.126 \times 10^{-6}$ makes
this model consistent with the observed values for the scalar amplitude and the 
scalar spectral index \cite{Aghanim:2018eyx}. On the other hand, the model's large
prediction of $r \simeq 0.14$ is badly discordant with limits on the tensor-to-scalar 
ratio \cite{Aghanim:2018eyx}. We shall therefore briefly consider how our analytic
approximations fare when used with the plateau potentials currently consistent with
observation. 

\begin{figure}[H]
\centering
\begin{subfigure}[b]{0.33\textwidth}
\centering
\includegraphics[width=\textwidth]{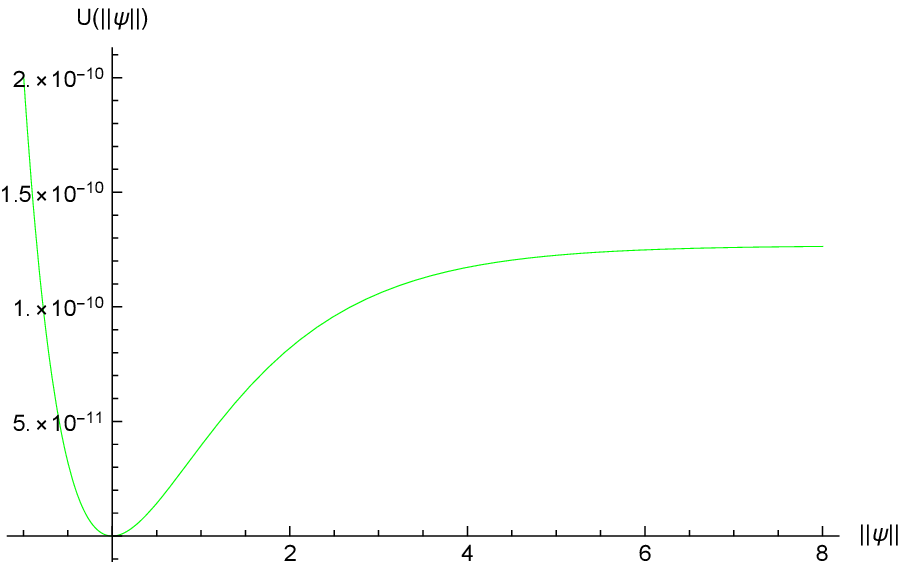}
\end{subfigure}
\begin{subfigure}[b]{0.33\textwidth}
\centering
\includegraphics[width=\textwidth]{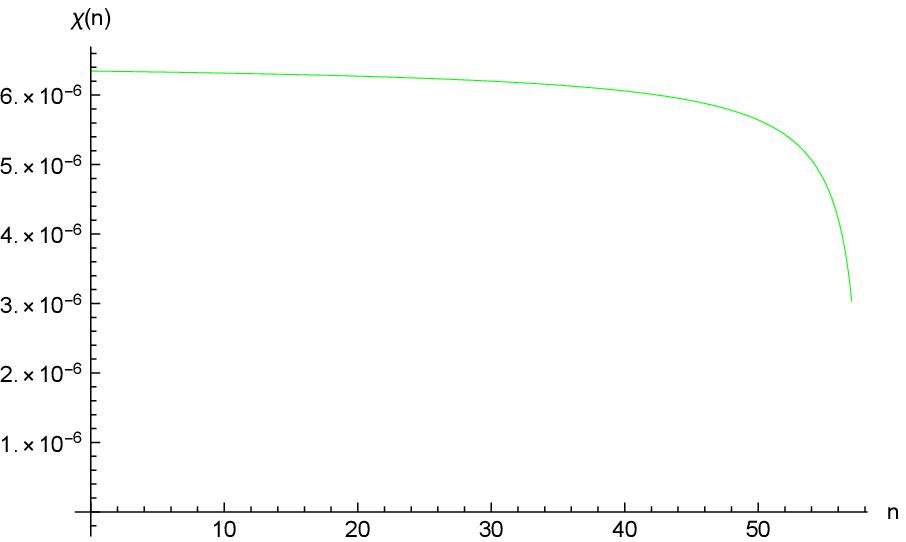}
\end{subfigure}
\begin{subfigure}[b]{0.33\textwidth}
\centering
\includegraphics[width=\textwidth]{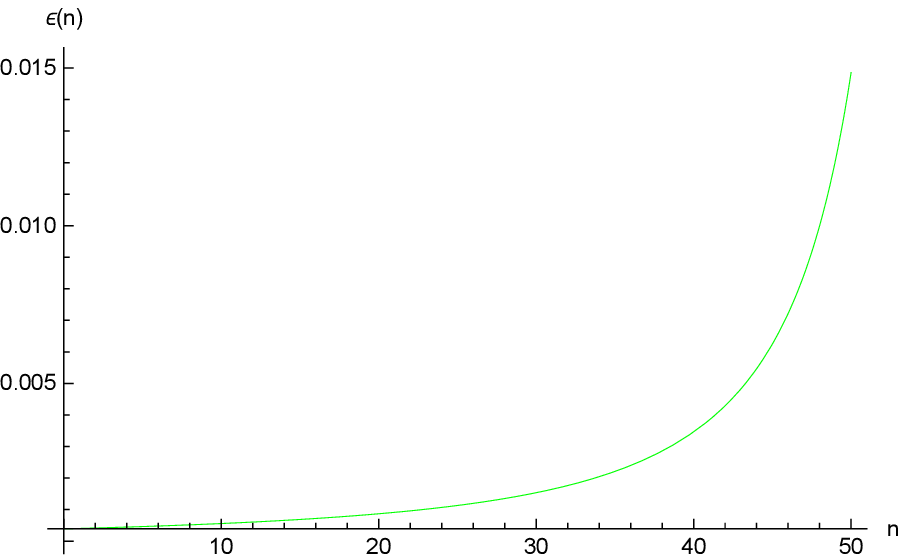}
\end{subfigure}
\caption{\footnotesize{Potential and geometry for the Einstein-frame representation 
of Starobinsky's original model of inflation \cite{Starobinsky:1980te}. The left shows
the dimensionless potential $U(\psi\psi^*)$ (\ref{StaroU}); the middle plot gives the
dimensionless Hubble parameter $\chi(n)$ and the right hand plot depicts the first
slow roll parameter $\epsilon(n)$. Inflation was assumed to start from $\psi_0 = 4.6$.}}
\label{StarobinskyA}
\end{figure}

The best known plateau potential is the Einstein-frame version of Staro\-binsky's
famous $R + R^2$ model \cite{Starobinsky:1980te}. Expressing the dimensionless 
potential for this model in our notation gives \cite{Brooker:2016oqa},
\begin{equation}
U(\psi\psi^*) = \frac34 M^2 \Bigl( 1 - e^{-\sqrt{\frac23} \, \vert\psi\vert}\Bigr)^2 
\qquad , \qquad M = 1.3 \times 10^{-5} \; . \label{StaroU}
\end{equation}
Somewhat over 50 e-foldings of inflation result if one starts from $\psi_0 = 4.6$, 
and the choice of $M = 1.3 \times 10^{-5}$ makes the model consistent with observation 
\cite{Aghanim:2018eyx}. Figure~\ref{StarobinskyA} shows why $r = 16 \epsilon$ is 
so small for this model: its dimensionless Hubble parameter $\chi(n)$ is nearly constant.  

\begin{figure}[H]
\centering
\begin{subfigure}[b]{0.5\textwidth}
\centering
\includegraphics[width=\textwidth]{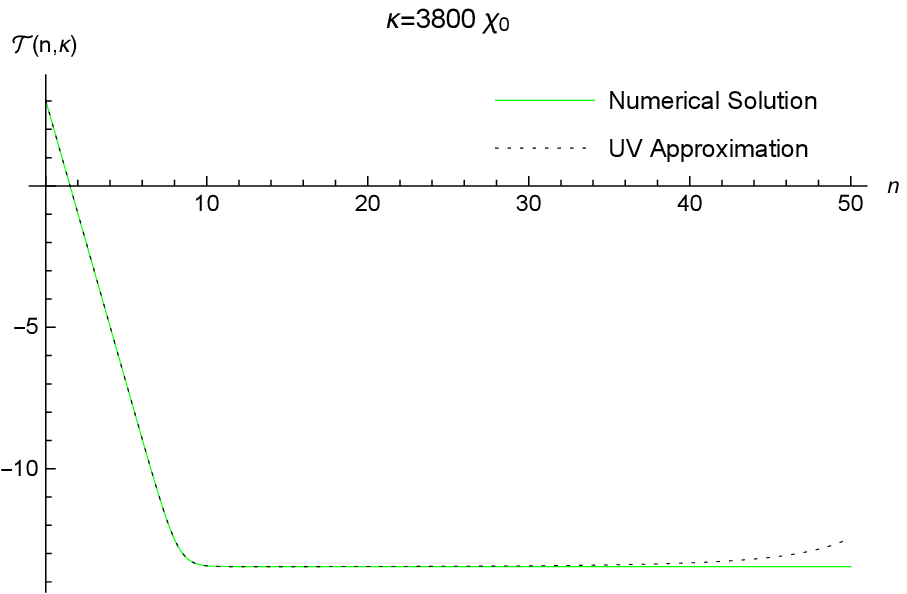}
\end{subfigure}
\begin{subfigure}[b]{0.5\textwidth}
\centering
\includegraphics[width=\textwidth]{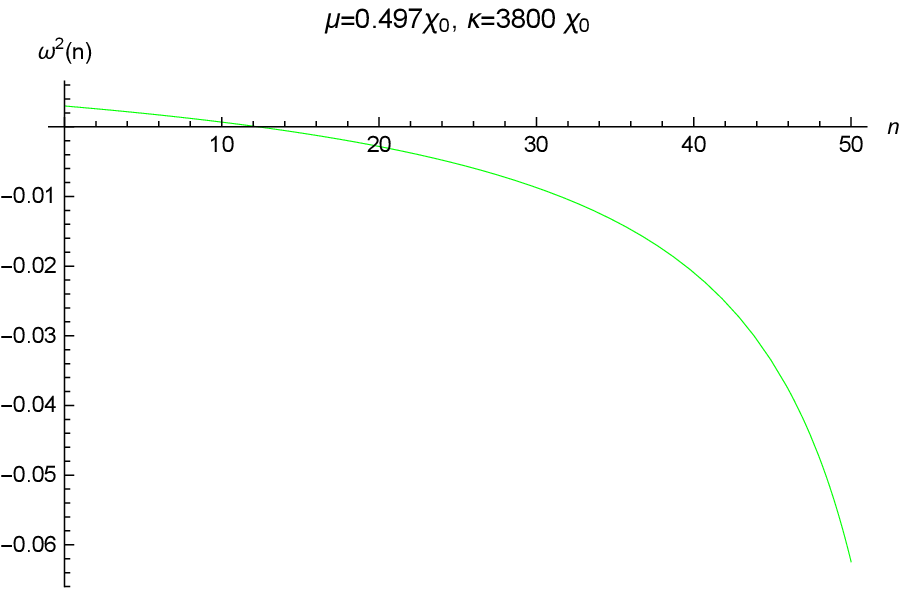}
\end{subfigure}
\caption{\footnotesize The left hand plot shows the amplitude 
$\mathcal{T}(n,\kappa)$ of the massless, minimally coupled scalar
for $\kappa = 3800 \chi_0$, which corresponds to $n_{\kappa} \simeq
8.3$. The right hand graph shows the frequency $\omega_u^2(n,\mu)
\simeq \omega^2_v(n,\mu)$ for $\mu = 0.497 \chi_0$ which passes 
through zero at $n_{\mu} \simeq 12$.} 
\label{StarobinskyB}
\end{figure}
\begin{figure}[H]
\centering
\begin{subfigure}[b]{0.5\textwidth}
\centering
\includegraphics[width=\textwidth]{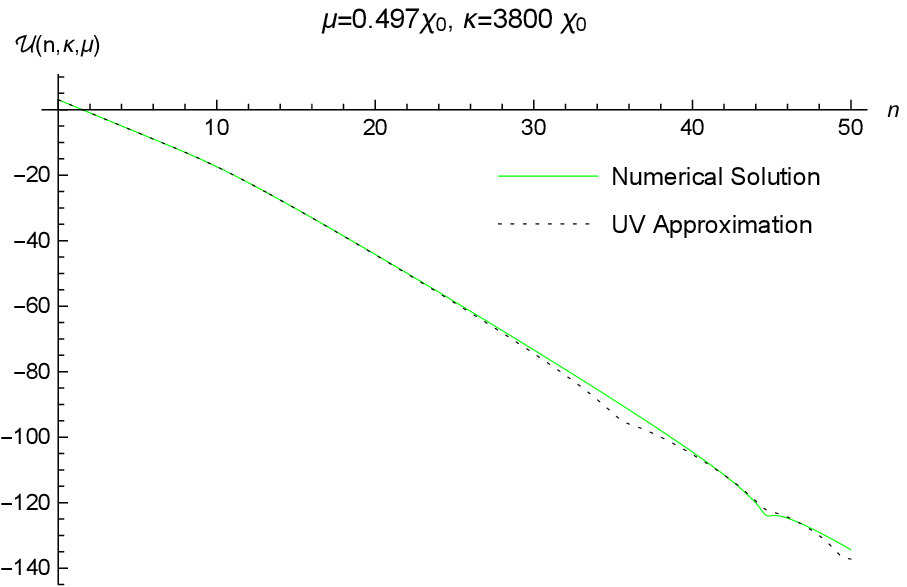}
\end{subfigure}
\begin{subfigure}[b]{0.5\textwidth}
\centering
\includegraphics[width=\textwidth]{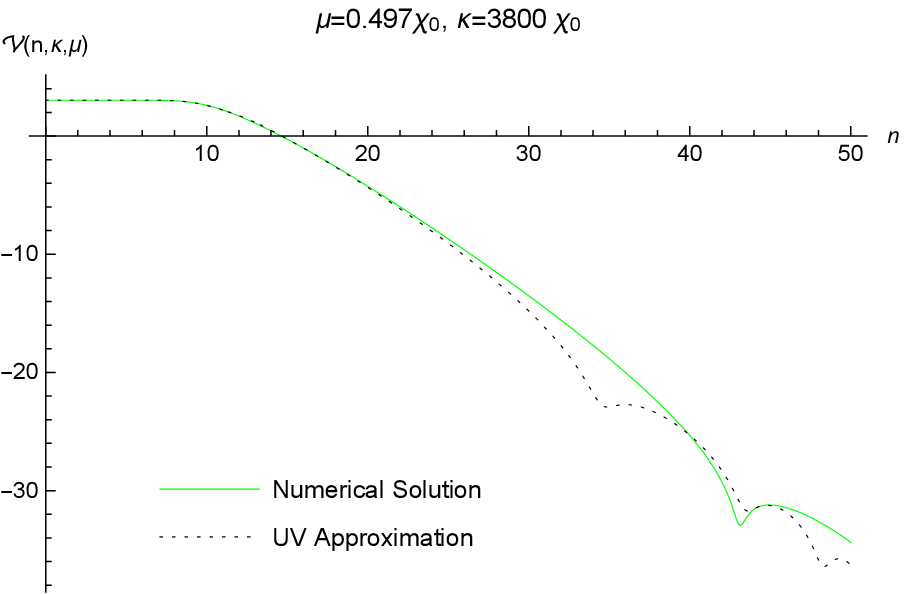}
\end{subfigure}
\caption{\footnotesize Plots of the temporal amplitude 
$\mathcal{U}(n,\kappa,\mu)$ (left) and the spatially transverse 
amplitude $\mathcal{V}(n,\kappa,\mu)$ (right) versus $n$ for the 
Starobinsky potential (\ref{StaroU}). For each amplitude $\kappa 
= 3800 \chi_0$ (which implies $n_{\kappa} \simeq 8.3$) and $\mu 
= 0.497 \chi_0$ (which implies $n_{\mu} \simeq 12$).}
\label{StarobinskyC}
\end{figure}

\noindent All our approximations pertain for this model, but the 
general effect of $\chi(n)$ being so nearly constant is to increase
the range over which the ultraviolet approximations pertain. The left 
hand plot of Figure~\ref{StarobinskyB} shows this for the MMC scalar 
amplitude $\mathcal{T}(n,\kappa)$. Because $\epsilon(n)$ is so small, 
the temporal and transverse frequencies are nearly equal $\omega^2_u(n,\mu) 
\simeq \omega^2_v(n,\mu)$ and nearly constant. The right hand plot of 
Figure~\ref{StarobinskyB} shows this for a carefully chosen value of 
$\mu = 0.497 \chi_0$ which causes mass domination to occur during 
inflation. For this case we can just see the second and third phases 
occur in Figure~\ref{StarobinskyC}.

\section{Effective Potential}

The purpose of this section is to evaluate the one photon loop contribution 
to the inflaton effective potential defined by equation (\ref{DeltaVdef}).
We begin by deriving some exact results for the trace of the coincident 
propagator, and we recall that $\mathcal{T}(n,\kappa)$ can be obtained from
$\mathcal{U}(n,\kappa,0)$. Then the ultraviolet approximations (\ref{U1def})
and (\ref{V1def}) are used to derive a divergent result whose renormalization
gives the part of the effective potential that depends locally on the geometry.
We give large field and small field expansions for this local part, and we
study its dependence on derivatives of $\epsilon(n)$. The section closes with
a discussion of the nonlocal part of the effective potential which derives 
from the late time approximations (\ref{U2def}), (\ref{U3def}), (\ref{V2def})
and (\ref{V3def}).

\subsection{Trace of the Coincident Photon Propagator}

At coincidence the mixed time-space components of the photon mode sum vanish,
and factors of $\widehat{k}_m \widehat{k}_n$ average to $\delta_{mn}/(D-1)$,
\begin{eqnarray}
\lefteqn{ i \Bigl[ \mbox{}_{\mu} \Delta_{\nu}\Bigr](x;x) = \int \!\! 
\frac{d^{D-1}k}{(2\pi)^{D-1}} \, \Biggl\{ \frac1{M^2} \left( 
\matrix{k^2 u u^* & 0 \cr 0 & \frac{\delta_{mn}}{D - 1} \mathcal{D} u
\mathcal{D} u^*}\right) } \nonumber \\
& & \hspace{1.8cm} - \frac1{M^2} \left( \matrix{\partial_0 t \partial_0 t^* & 0 \cr 
0 & \frac{\delta_{mn}}{D - 1} k^2 t t^*}\right) + \left( \matrix{0 & 0 \cr 0 &
(\frac{D-2}{D-1}) \delta_{mn} v v^*}\right) \Biggr\} . \qquad \label{coincprop}
\end{eqnarray}
Its trace is,
\begin{eqnarray}
\lefteqn{g^{\mu\nu} i \Bigl[ \mbox{}_{\mu} \Delta_{\nu}\Bigr](x;x) =} \nonumber \\
& & \hspace{1cm} \int \!\! \frac{d^{D-1}k}{(2\pi)^{D-1}} \, \Biggl\{ 
\frac{ \mathcal{D} u \mathcal{D} u^* \!-\! k^2 u u^* \!+\! \partial_0 t 
\partial_0 t^* \!-\! k^2 t t^*}{a^2 M^2} + \frac{(D \!-\!2) v v^*}{a^2} \Biggr\} 
. \qquad \label{coinctrace}
\end{eqnarray}
Relation (\ref{MMCtotemp}) allows us to replace the MMC scalar mode function 
$t(\eta,k)$ with the massless limit of the temporal mode function $u_0(\eta,k) 
\equiv u(\eta,k,0)$,
\begin{equation}
\partial_0 t \partial_0 t^* = k^2 u_0 u^*_0 \qquad , \qquad k^2 t t^* = 
\mathcal{D} u_0 \mathcal{D} u^*_0 \; . \label{ttouzero}
\end{equation}
Substituting (\ref{ttouzero}) in (\ref{coinctrace}) gives,
\begin{eqnarray}
\lefteqn{g^{\mu\nu} i \Bigl[ \mbox{}_{\mu} \Delta_{\nu}\Bigr](x;x) =} \nonumber \\
& & \hspace{0.8cm} \int \!\! \frac{d^{D-1}k}{(2\pi)^{D-1}} \, \Biggl\{ 
\frac{ \mathcal{D} u \mathcal{D} u^* \!-\! \mathcal{D} u_0 \mathcal{D} u_0^*
\!-\! k^2 (u u^* \!-\! u_0 u^*_0)}{a^2 M^2} + \frac{(D \!-\!2) v v^*}{a^2} \Biggr\} 
. \qquad \label{coinctrace2}
\end{eqnarray}
This second form (\ref{coinctrace2}) is very important because it demonstrates
the absence of any $1/M^2$ pole as an exact relation, before any approximations 
are made.

The mode equation for temporal photons implies,
\begin{eqnarray}
\lefteqn{\mathcal{D} u \mathcal{D} u^* = a^2 H^2 \Bigl[ u' {u'}^* + 
(D\!-\!2) (u u^*)' + (D \!-\! 2)^2 u u^*\Bigr] \; , } \\
& & \hspace{0.8cm} = (k^2 + a^2 M^2) u u^* + \frac{a^2 H^2}{2} \Bigl( 
\partial_n \!+\! D \!-\! 1 \!-\! \epsilon \Bigr) \Bigl(\partial_n \!+\! 2 D 
\!-\! 4\Bigr) (u u^*) \; . \qquad \label{tempsimp}
\end{eqnarray}
Using relations (\ref{tempsimp}) and (\ref{coinctrace2}) allows us to express 
the trace of the coincident photon propagator in terms of three coincident 
scalar propagators,
\begin{eqnarray}
\lefteqn{g^{\mu\nu} i \Bigl[ \mbox{}_{\mu} \Delta_{\nu}\Bigr](x;x) =
i\Delta_u(x;x) + \frac{(D \!-\! 2)}{a^2} i\Delta_v(x;x) } \nonumber \\
& & \hspace{1cm} + \frac{H^2}{2 M^2} \Bigl( \partial_n \!+\! D \!-\! 1 \!-\! 
\epsilon \Bigr) \Bigl(\partial_n \!+\! 2 D \!-\! 4\Bigr) \Bigl[i\Delta_u(x;x) 
- i\Delta_{u_0}(x;x) \Bigr] \; . \qquad \label{coinctrace3}
\end{eqnarray}
The disappearance of any factors of $k^2$ from the Fourier mode sums in
(\ref{coinctrace3}), coupled with the ultraviolet expansions (\ref{Uasymp}) and
(\ref{Vasymp}), means that the phase 1 approximations $\mathcal{U}_1(n,\kappa,\mu)$
and $\mathcal{V}_1(n,\kappa,\mu)$ exactly reproduce the ultraviolet divergence
structures.

Two of the scalar propagators in expression (\ref{coinctrace3}) are,
\begin{eqnarray}
\lefteqn{i\Delta_u(x;x') \equiv \int \!\! \frac{d^{D-1}k}{(2\pi)^{D-1}} \Biggl\{ 
\theta(\Delta \eta) u(\eta,k,M) u^*(\eta',k,M) e^{i \vec{k} \cdot \Delta \vec{x}} }
\nonumber \\
& & \hspace{4.5cm} + \theta(-\Delta \eta) u^*(\eta,k,M) u(\eta',k,M) e^{-i\vec{k} 
\cdot \Delta \vec{x}} \Biggr\} , \qquad \label{tempprop} \\
\lefteqn{i\Delta_v(x;x') \equiv \int \!\! \frac{d^{D-1}k}{(2\pi)^{D-1}} \Biggl\{ 
\theta(\Delta \eta) v(\eta,k,M) v^*(\eta',k,M) e^{i \vec{k} \cdot \Delta \vec{x}} }
\nonumber \\
& & \hspace{4.5cm} + \theta(-\Delta \eta) v^*(\eta,k,M) v(\eta',k,M) e^{-i\vec{k} 
\cdot \Delta \vec{x}} \Biggr\} . \qquad \label{transprop}
\end{eqnarray}
The third scalar propagator $i\Delta_{u_0}(x;x')$ is just the $M \rightarrow 0$
limit of $i\Delta_{u}(x;x')$. The coincidence limits of each propagator can be
expressed in terms of the corresponding amplitude,
\begin{equation}
\frac{i\Delta_u(x;x)}{\sqrt{8\pi G} } = \int \!\! \frac{d^{D-1}k}{(2\pi)^{D-1}} \,
e^{\mathcal{U}(n,\kappa,\mu)} \quad , \quad \frac{i\Delta_v(x;x)}{\sqrt{8\pi G} } 
= \int \!\! \frac{d^{D-1}k}{(2\pi)^{D-1}} \, e^{\mathcal{V}(n,\kappa,\mu)} \; .
\label{coincmsums}
\end{equation}

Expression(\ref{coinctrace3}) is exact but not immediately useful because we 
lack explicit expressions for the coincident propagators (\ref{coincmsums}).
It is at this stage that we must resort to the analytic approximations developed 
in section 3. Recall that the phase 1 approximation is valid until roughly 4 
e-foldings after horizon crossing. If one instead thinks of this as a condition 
on the dimensionless wave number $\kappa \equiv \sqrt{8\pi G} \, k$ at fixed $n$, 
it means that $\kappa > \kappa_{n-4}$, where we define $\kappa_n$ as the 
dimensionless wave number which experiences horizon crossing at e-folding $n$. 
Taking as an example the temporal photon contribution we can write,
\begin{eqnarray}
\lefteqn{e^{\mathcal{U}(n,\kappa,\mu)} \simeq \theta\Bigl( \kappa \!-\! 
\kappa_{n-4}\Bigr) e^{\mathcal{U}_{1}(n,\kappa,\mu)} + \theta\Bigl( \kappa_{n-4}
- \kappa \Bigr) e^{\mathcal{U}_{2,3}(n,\kappa,\mu)} \; , } \\
& & \hspace{2cm} = e^{\mathcal{U}_{1}(n,\kappa,\mu)} + \theta\Bigl( \kappa_{n-4}
- \kappa \Bigr) \Biggl[e^{\mathcal{U}_{2,3}(n,\kappa,\mu)} -
e^{\mathcal{U}_{1}(n,\kappa,\mu)} \Biggr] . \qquad \label{ampapprox}
\end{eqnarray}
Substituting the approximation (\ref{ampapprox}) into expression 
(\ref{coincmsums}) allows us to write,
\begin{equation}
i\Delta_u(x;x) \simeq L_u(n) + N_u(n) \; , 
\label{locnonloc}
\end{equation}
where we define the local ($L$) and nonlocal ($N$) contributions as,
\begin{eqnarray}
L_u(n) &\!\!\! \equiv \!\!\!& \sqrt{8\pi G} \! \int \!\! 
\frac{d^{D-1}k}{(2\pi)^{D-1}} \, e^{\mathcal{U}_{1}(n,\kappa,\mu)} \; , 
\label{localprop} \\
N_u(n) &\!\!\! \equiv \!\!\!& \sqrt{8\pi G} \! \int \!\! 
\frac{d^{3}k}{(2\pi)^{3}} \, \theta\Bigl( \kappa_{n-4} - \kappa \Bigr) 
\Biggl[e^{\mathcal{U}_{2,3}(n,\kappa,\mu)} - e^{\mathcal{U}_{1}(n,\kappa,\mu)} 
\Biggr] . \qquad \label{nonlocalprop}
\end{eqnarray}
Note that we have taken the unregulated limit ($D = 4$) in expression 
(\ref{nonlocalprop}) because it is ultraviolet finite. The same considerations
apply as well for the coincident spatially transverse photon propagator 
$i\Delta_v(x;x')$, and for the massless limit of the temporal photon propagator 
$i\Delta_{u_0}(x;x)$.

\subsection{The Local Contribution}

The local contribution for each of the coincident propagators (\ref{coincmsums})
comes from using the phase 1 approximation (\ref{localprop}). For the temporal 
modes the amplitude is approximated by expression (\ref{U1def}), whereupon we 
change variables to $z$ using $k =
(1-\epsilon) H a z$, and then employ integral $6.574\; \# 2$ of 
\cite{Gradshteyn:1965},
\begin{equation}
L_u(n) = \frac{[(1 \!-\! \epsilon) H]^{D-2}}{(4\pi)^{\frac{D}2}}
\times \frac{\Gamma(\frac{D-1}{2} \!+\! \nu_u) \Gamma(\frac{D-1}{2} \!-\! 
\nu_u)}{\Gamma(\frac12 \!+\! \nu_u) \Gamma(\frac12 \!-\! \nu_u)} \times
\Gamma\Bigl(1 \!-\! \frac{D}2\Bigr) \; . \label{Lu}
\end{equation}
Recall that the index $\nu_u(n,\mu)$ is defined in expression (\ref{nuudef}).
Of course the massless limit is,
\begin{equation}
L_{u_0}(n) = \frac{[(1 \!-\! \epsilon) H]^{D-2}}{(4\pi)^{\frac{D}2}}
\times \frac{\Gamma(\frac{D-1}{2} \!+\! \nu_{u_0}) \Gamma(\frac{D-1}{2} \!-\! 
\nu_{u_0})}{\Gamma(\frac12 \!+\! \nu_{u_0}) \Gamma(\frac12 \!-\! \nu_{u_0})} 
\times \Gamma\Bigl(1 \!-\! \frac{D}2\Bigr) \; , \label{Lu0}
\end{equation}
where the index is,
\begin{equation}
\nu_{u_0}(n) \equiv \nu_u(n,0) = \frac12 \Bigl( \frac{D \!-\! 3 \!+\! \epsilon(n)}{
1 \!-\! \epsilon(n)} \Bigr) \; . \label{nuu0def}
\end{equation}
The phase 1 approximation (\ref{V1def}) for the transverse amplitude contains 
two extra scale factors which serve to exactly cancel the inverse scale 
factors that are evident in the transverse contribution to the trace of the 
coincident photon propagator (\ref{coinctrace3}). Hence we have,
\begin{equation}
\frac{L_v(n)}{a^2} = \frac{[(1 \!-\! \epsilon) H]^{D-2}}{
(4\pi)^{\frac{D}2}}\times \frac{\Gamma(\frac{D-1}{2} \!+\! \nu_v) 
\Gamma(\frac{D-1}{2} \!-\! \nu_v)}{\Gamma(\frac12 \!+\! \nu_v) 
\Gamma(\frac12 \!-\! \nu_v)} \times \Gamma\Bigl(1 \!-\! \frac{D}2\Bigr) 
\; , \label{Lv}
\end{equation}
where the transverse index $\nu_v(n,\mu)$ is given in (\ref{nuvdef}).

Each of the local contributions (\ref{Lu}), (\ref{Lu0}) and (\ref{Lv}) is 
proportional to the same divergent Gamma function,
\begin{equation}
\Gamma\Bigl(1 \!-\! \frac{D}2\Bigr) = \frac{2}{D \!-\! 4} + 
O\Bigl( (D\!-\!4)^0 \Bigr) \; .
\end{equation}
Each also contains a similar ratio of Gamma functions,
\begin{eqnarray}
\lefteqn{ \frac{ \Gamma(\frac{D-1}2 \!+\! \nu) \Gamma(\frac{D-1}2 \!-\! \nu)}{
\Gamma(\frac12 \!+\! \nu) \Gamma(\frac12 \!-\! \nu)} = \Bigl[ 
\Bigl(\frac{D\!-\!3}{2}\Bigr)^2 \!-\! \nu^2\Bigr] \!\times\! 
\frac{ \Gamma(\frac{D-3}2 \!+\! \nu) \Gamma(\frac{D-3}2 \!-\! \nu)}{
\Gamma(\frac12 \!+\! \nu) \Gamma(\frac12 \!-\! \nu)} \; , } \\
& & \hspace{-0.7cm} = \! \Bigl[ \Bigl(\frac{D\!-\!3}{2}\Bigr)^2 \!\!\!-\! \nu^2\Bigr]
\Biggl\{ \! 1 + \Bigl[ \psi\Bigl( \frac12 \!+\! \nu\!\Bigr) \!+\! \psi\Bigl(
\frac12 \!-\! \nu\!\Bigr) \Bigr] \Bigl( \frac{D \!-\! 4}{2}\Bigr) \!+\! O\Bigl( 
(D\!-\!4)^2\Bigr) \! \Biggr\} . \qquad 
\end{eqnarray}
These considerations allow us to break up each of the three terms in 
(\ref{coinctrace3}) into a potentially divergent part plus a manifestly finite part. 
For $i\Delta_u(x;x) \rightarrow L_u(n)$ this decomposition is,
\begin{eqnarray}
\lefteqn{ L_u = \frac{[(1 \!-\! \epsilon) H]^{D-4}}{(4 \pi)^{\frac{D}2}} 
\Biggl[ M^2 - \frac{(D \!-\! 2) H^2}{2} \Bigl( (D \!-\! 3) \epsilon -
\frac12 (D\!-\! 4) \epsilon^2\Bigr) \Biggr] \Gamma\Bigl(1 \!-\! \frac{D}2\Bigr) }
\nonumber \\
& & \hspace{1.5cm} + \frac1{16 \pi^2} \Bigl[M^2 - \epsilon H^2\Bigr] \Bigl[
\psi\Bigl( \frac12 \!+\! \nu_u\Bigr) + \psi\Bigl( \frac12 \!-\! \nu_u\Bigr) \Bigr]
+ O( D\!-\! 4) \; . \qquad \label{1stterm}
\end{eqnarray}
For $(D-2) i\Delta_v(x;x) \rightarrow (D-2) L_v(n)$ we have,
\begin{eqnarray}
\lefteqn{ (D \!-\!2) L_v = \frac{[(1 \!-\! \epsilon) H]^{D-4}}{(4 \pi)^{\frac{D}2}} 
\Biggl[ (D \!-\! 2) M^2 - \frac{(D \!-\! 2) (D \! -\! 4) H^2}{2} \Bigl( (D \!-\! 3) 
\epsilon } \nonumber \\
& & \hspace{-0.5cm} - \frac{(D\!-\! 2) \epsilon^2}{2}\Bigr) \Biggr] \Gamma\Bigl(1 
\!-\! \frac{D}2\Bigr) + \frac{2 M^2}{16 \pi^2} \Bigl[ \psi\Bigl( \frac12 \!+\! 
\nu_v\Bigr) + \psi\Bigl( \frac12 \!-\! \nu_v\Bigr) \Bigr] + O( D\!-\! 4) \; . 
\qquad \label{2ndterm}
\end{eqnarray}
And the final term in (\ref{coinctrace3}) --- the one with derivatives --- becomes,
\begin{eqnarray}
\lefteqn{ \frac{H^2}{2 M^2} \Bigl( \partial_n \!+\! D \!-\! 1 \!-\! \epsilon\Bigr)
\Bigl(\partial_n \!+\! 2 D \!-\! 4\Bigr) \Bigl[ L_u \!-\! L_{u_0}\Bigr] = \frac{H^2}{2} 
\Bigl( \partial_n \!+\! D \!-\! 1 \!-\! \epsilon\Bigr) } \nonumber \\
& & \hspace{0cm} \times \Bigl(\partial_n \!+\! 2 D \!-\! 4\Bigr) \frac{[(1 \!-\! 
\epsilon) H]^{D-4}}{(4 \pi)^{\frac{D}2}} \Gamma\Bigl(1 \!-\! \frac{D}2\Bigr) + 
\frac{H^2}{32 \pi^2} \Bigl(\partial_n \!+\! 3 \!-\! \epsilon\Bigr) \Bigl( \partial_n 
\!+\! 4\Bigr) \qquad \nonumber \\
& & \hspace{1cm} \times \Biggl\{ \psi\Bigl( \frac12 \!+\! \nu_u\Bigr) + \psi\Bigl(
\frac12 \!-\! \nu_u\Bigr) - \frac{\epsilon H^2}{M^2} \Bigl[ \psi\Bigl(\frac12 \!+\! 
\nu_u\Bigr) - \psi\Bigl(\frac12 \!+\! \nu_{u_0}\Bigr) \nonumber \\
& & \hspace{4.5cm} + \psi\Bigl(\frac12 \!-\! \nu_u\Bigr) - \psi\Bigl(\frac12 \!-\! \nu_{u_0}
\Bigr) \Bigr] \Biggr\} + O(D \!-\! 4) \; . \qquad \label{3rdterm}
\end{eqnarray}
Note that the difference $\psi(\frac12 \pm \nu_u) - \psi(\frac12 \pm \nu_{u_0})$ 
is of order $M^2$ so expression (\ref{3rdterm}) has no $1/M^2$ pole. Note also that 
the $1/\epsilon$ pole in $\psi(\frac12 - \nu_{u_0}) = \psi(\frac{-\epsilon}{1 - \epsilon})$
is canceled by an explicit multiplicative factor of $\epsilon$.

The potentially divergent terms (the ones proportional to $\Gamma(1 - \frac{D}2)$) in 
expressions (\ref{1stterm}), (\ref{2ndterm} and (\ref{3rdterm}) sum to give,
\begin{eqnarray}
\lefteqn{ (\ref{1stterm})_{\rm div} + (\ref{2ndterm})_{\rm div} + 
(\ref{3rdterm})_{\rm div} = \frac{[(1 \!-\! \epsilon) H]^{D-4}}{(4\pi)^{\frac{D}2}}
\Bigl[ (D\!-\!1) M^2 + \frac12 R\Bigr] \Gamma\Bigl(1 \!-\! \frac{D}2\Bigr) }
\nonumber \\
& & \hspace{0.7cm} + \frac{H^2}{16 \pi^2} \Bigl[3 - 12 \epsilon + 4 \epsilon^2 - 2 
\epsilon' -\frac{(6 \epsilon' \!+\! \epsilon'')}{1 \!-\! \epsilon} - \Bigl( 
\frac{\epsilon'}{1 \!-\! \epsilon}\Bigr)^2 \Bigr] + O(D \!-\! 4) \; , \qquad
\end{eqnarray}
where we recall that the $D$-dimensional Ricci scalar is $R = (D-1) (D - 2 \epsilon) 
H^2$. Comparison with expression (\ref{DeltaVdef}) for $\Delta V'(\varphi \varphi^*)$
reveals that we can absorb the divergences with the following counterterms,
\begin{equation}
\delta \xi = -\frac{\Gamma(1 \!-\! \frac{D}2) s^{D-4}}{(4\pi)^{\frac{D}2}} \times 
\frac12 q^2 \qquad , \qquad \delta \lambda = -\frac{\Gamma(1 \!-\! \frac{D}2) 
s^{D-4}}{(4\pi)^{\frac{D}2}} \times 4 (D \!-\! 1) q^4 \; , \label{cterms}
\end{equation}
where $s$ is the renormalization scale. Up to finite renormalizations, these choices
agree with previous results \cite{Miao:2015oba,Prokopec:2007ak,Miao:2019bnq}, in the 
same gauge and using the same regularization, on de Sitter background.

Substituting expressions (\ref{1stterm}), (\ref{2ndterm}), (\ref{3rdterm}) and
(\ref{cterms}) into the definition (\ref{DeltaVdef}) of $\Delta V'(\varphi \varphi^*)$
and taking the unregulated limit gives the local contribution,
\begin{eqnarray}
\lefteqn{ \Delta V'_{\rm L}(\varphi \varphi^*) = \frac{q^2 H^2}{16 \pi^2} \Biggl\{ 
\frac{(6 M^2 \!+\! R)}{2 H^2} \ln\Bigl[ \frac{(1 \!-\! \epsilon)^2 H^2}{s^2}\Bigr] 
\!+\! 3 \!-\! 12 \epsilon \!+\! 4 \epsilon^2 \!-\! 2 \epsilon' \!-\! \frac{(6 \epsilon' 
\!+\! \epsilon'')}{1 \!-\! \epsilon} } \nonumber \\
& & \hspace{-0.5cm} - \Bigl( \frac{\epsilon'}{1 \!-\! \epsilon}\Bigr)^2 
+ \frac{M^2}{H^2} \Biggl[ \psi\Bigl( \frac12 \!+\! \nu_u\Bigr) +
\psi\Bigl( \frac12 \!-\! \nu_u\Bigr) + 2 \psi\Bigl( \frac12 \!+\! \nu_v\Bigr) + 2
\psi\Bigl( \frac12 \!-\! \nu_v\Bigr) \Biggr] \nonumber \\
& & \hspace{-0.5cm} + \frac12 \Bigl[ (\partial_n \!+\! 3 \!-\! \epsilon) (\partial_n 
\!+\! 4) - 2 \epsilon \Bigr] \Bigl[ \psi\Bigl( \frac12 \!+\! \nu_u\Bigr) \!+\! 
\psi\Bigl( \frac12 \!-\! \nu_u\Bigr) \Bigr] - (\partial_n \!+\! 3 \!-\! \epsilon)
(\partial_n \!+\! 4) \nonumber \\
& & \hspace{1.8cm} \times \frac{\epsilon H^2}{2 M^2} \Bigl[ \psi\Bigl( \frac12 
\!+\! \nu_u\Bigr) \!-\! \psi\Bigl( \frac1{1 \!-\! \epsilon}\Bigr) + \psi\Bigl( 
\frac12 \!-\! \nu_u\Bigr) \!-\! \psi\Bigl( \frac{-\epsilon}{1 \!-\! \epsilon}\Bigr) 
\Bigr] \Biggr\} . \qquad \label{VLprime}
\end{eqnarray}
It is worth noting that there are no singularities at $\epsilon = 1$, or when
either $1/(1 - \epsilon)$ or $-\epsilon/(1 - \epsilon)$ become non-positive integers
\cite{Janssen:2008px}. The effective potential is obtained by integrating 
(\ref{VLprime}) with respect to $\varphi \varphi^*$. The result is best expressed 
using the variable $z \equiv q^2 \varphi \varphi^*/H^2$,
\begin{eqnarray}
\lefteqn{ \Delta V_{\rm L} = \frac{H^4}{16 \pi^2} \Biggl\{\! \Bigl[
3 z^2 \!+\! \frac{R z}{2 H^2} \Bigr]\! \ln\Bigl[ \frac{ (1 \!-\! \epsilon)^2 H^2}{s^2}
\Bigr] \!+\! \Bigl[3 \!-\! 12 \epsilon \!+\! 4 \epsilon^2 \!-\! 2 \epsilon' \!-\! 
\frac{(6 \epsilon' \!+\! \epsilon'')}{1 \!-\! \epsilon} \Bigr] z } \nonumber \\
& & \hspace{-0.5cm} - \frac{{\epsilon'}^2 z}{(1 \!-\! \epsilon)^2} + 2 \! 
\int_{0}^{z} \!\!\! dx \, x \Biggl[ \psi\Bigl( \frac12 \!+\! \alpha\! \Bigr) 
\!+\! \psi\Bigl( \frac12 \!-\! \alpha \!\Bigr) \!+\! 2 \psi\Bigl( \frac12 \!+\! 
\beta \!\Bigr) \!+\! 2 \psi\Bigl( \frac12 \!-\! \beta \! \Bigr) \Biggr] 
\nonumber \\
& & \hspace{-0.5cm} + \frac12 \Bigl[ (\partial_n \!+\! 3 \!-\! 3 \epsilon) 
(\partial_n \!+\! 4 \!-\! 2 \epsilon) - 2 \epsilon\Bigr] \! \int_{0}^{z} \!\!\! dx 
\Bigl[ \psi\Bigl( \frac12 \!+\! \alpha(x)\Bigr) + \psi\Bigl( \frac12 \!-\! \alpha(x) 
\Bigr) \Bigr] \nonumber \\
& & \hspace{-0.5cm} - (\partial_n \!+\! 3 \!-\! 3 \epsilon) (\partial_n \!+\! 4 
\!-\! 2 \epsilon) \! \int_{0}^{z} \!\! \frac{dx \epsilon}{4 x} \Bigl[ \psi\Bigl( 
\frac12 \!+\! \alpha(x) \Bigr) - \psi\Bigl( \frac1{1 \!-\! \epsilon}\Bigr) 
\nonumber \\
& & \hspace{6.5cm} + \psi\Bigl( \frac12 \!-\! \alpha(x)\Bigr) - \psi\Bigl( 
\frac{-\epsilon}{1 \!-\! \epsilon}\Bigr) \Bigr] \Biggr\} , \qquad \label{VL}
\end{eqnarray}
where the $x$-dependent indices are,
\begin{equation}
\alpha(x) \equiv \sqrt{\frac14 + \frac{\epsilon \!-\! 2 x}{(1 \!-\! \epsilon)^2}}
\qquad , \qquad \beta(x) \equiv \sqrt{ \frac14 - \frac{2 x}{(1 \!-\! \epsilon)^2}} 
\; . \label{alphabeta}
\end{equation}
Note that the term inside the square brackets on the last line of (\ref{VL})
vanishes for $x = 0$, so the integrand is well defined at $x = 0$.

\subsection{Large Field \& Small Field Expansions}

\begin{figure}[H]
\centering
\begin{subfigure}[b]{0.5\textwidth}
\centering
\includegraphics[width=\textwidth]{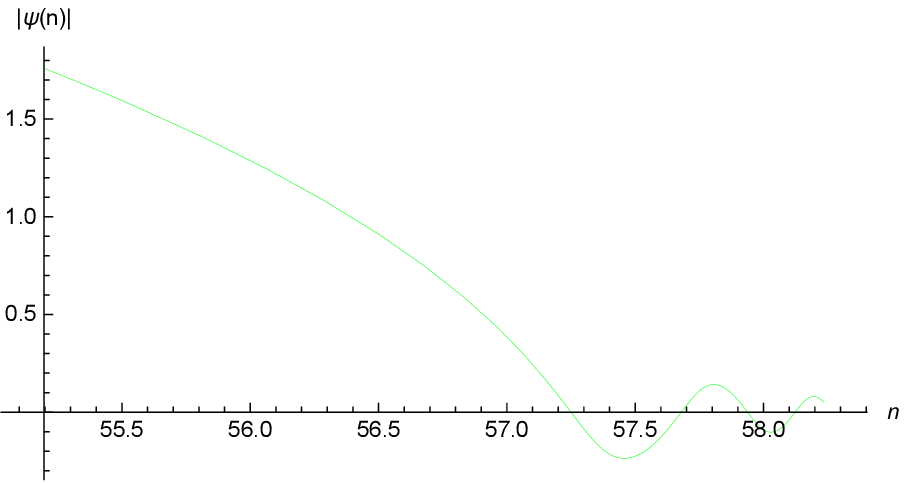}
\end{subfigure}
\begin{subfigure}[b]{0.5\textwidth}
\centering
\includegraphics[width=\textwidth]{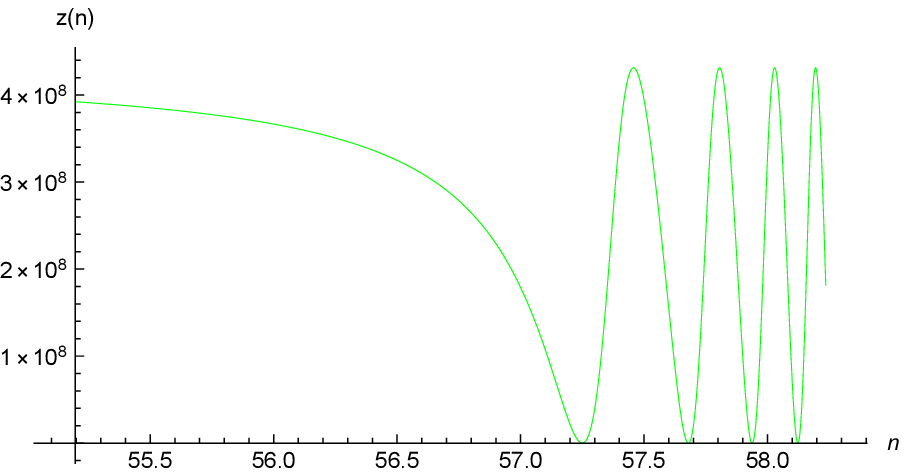}
\end{subfigure}
\caption{\footnotesize Plots of the dimensionless inflaton field $\psi(n)$ 
and the ratio $z \equiv q^2 \psi^2/\chi^2$ after the end of inflation for
the quadratic potential. Here we chose $q^2 = \frac1{137}$.}
\label{psiandz}
\end{figure}

Expression (\ref{VL}) depends principally on the quantity $z = q^2 \varphi 
\varphi^*/H^2$. During inflation $z$ is typically quite large, whereas it 
touches $0$ after the end of inflation. Figure~\ref{psiandz} shows this 
for the quadratic potential, and the results are similar for the 
Starobinsky potential (\ref{StaroU}). It is therefore desirable to expand
the potential $\Delta V_L(\varphi \varphi^*)$ for large $z$ and for small $z$.

\begin{figure}[H]
\centering
\begin{subfigure}[b]{0.33\textwidth}
\centering
\includegraphics[width=\textwidth]{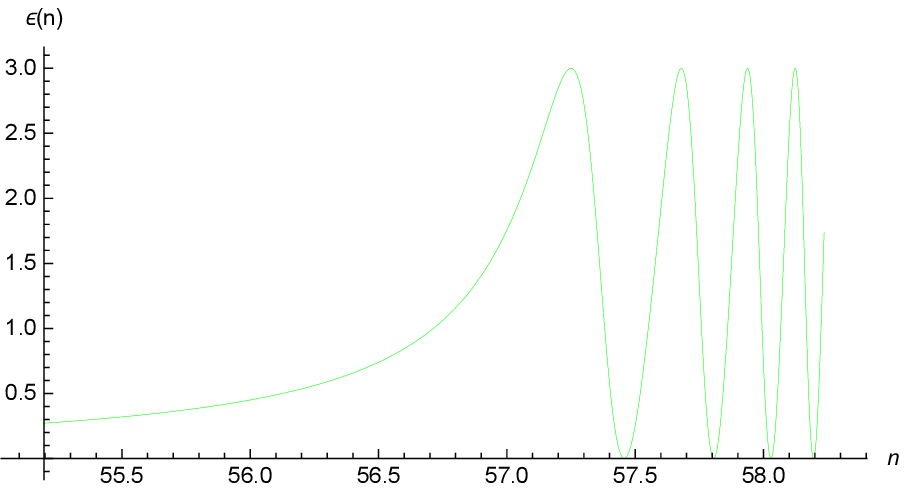}
\end{subfigure}
\begin{subfigure}[b]{0.33\textwidth}
\centering
\includegraphics[width=\textwidth]{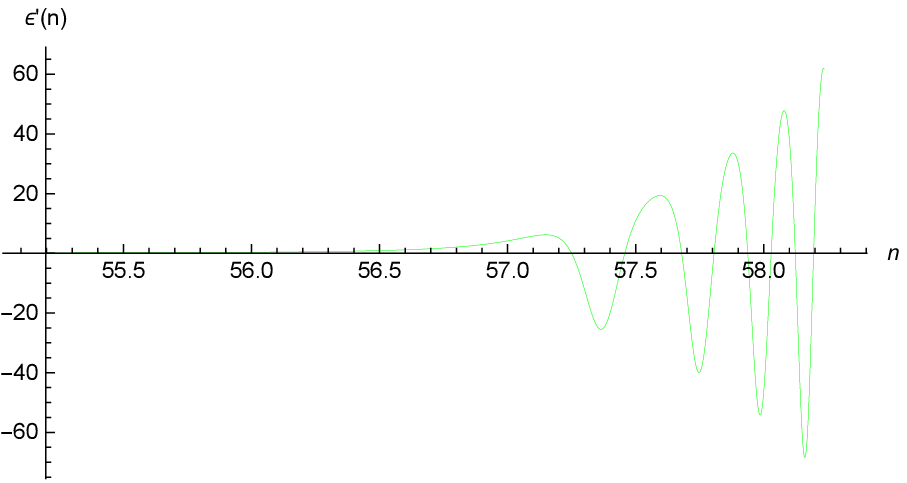}
\end{subfigure}
\begin{subfigure}[b]{0.33\textwidth}
\centering
\includegraphics[width=\textwidth]{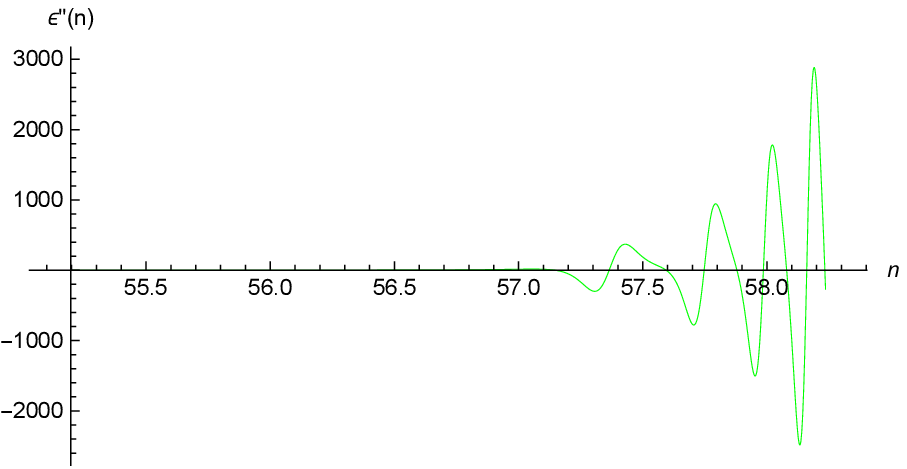}
\end{subfigure}
\caption{\footnotesize Plots of the first slow roll and its derivatives after
the end of inflation for the quadratic potential.}
\label{epsilon}
\end{figure}

The large field regime follows from the large argument expansion of the digamma
function,
\begin{equation}
\psi(x) = \ln(x) - \frac1{2 x} - \frac1{12 x^2} + \frac1{120 x^4} - \frac1{256 x^6}
+ O\Bigl( \frac1{x^8}\Bigr) \; . \label{psiexp}
\end{equation}
Substituting (\ref{psiexp}) in (\ref{VL}), and performing the various integrals
gives,
\begin{eqnarray}
\lefteqn{\Delta V_L = \frac{H^4}{16 \pi^2} \Biggl\{3 z^2 \ln\Bigl( \frac{2 q^2 
\varphi \varphi^*}{s^2}\Bigr) \!-\! \frac32 z^2 + \frac{R z}{2 H^2} \ln\Bigl( 
\frac{2 q^2 \varphi \varphi^*}{s^2}\Bigr) \!-\! (4 \!+\! 8 \epsilon \!-\! 3 
\epsilon^2) z } \nonumber \\
& & \hspace{0.5cm} -\epsilon' z - \Bigl[\frac34 \epsilon (1 \!-\! \epsilon) (2 \!-\! 
\epsilon) + \frac78 (1 \!-\! \epsilon) \epsilon' + \frac18 \epsilon''\Bigr]
\ln^2(2 z) + O\Bigl( \ln(z)\Bigr) \Biggr\} . \qquad \label{largezexp}
\end{eqnarray}
The leading contribution of (\ref{largezexp}) agrees with the famous flat space
result of Coleman and Weinberg \cite{Coleman:1973jx},
\begin{equation}
\Delta V \longrightarrow \frac{3 (q^2 \varphi \varphi^*)^2}{16 \pi^2} 
\ln\Bigl( \frac{2 q^2 \varphi \varphi^*}{s^2}\Bigr) \; .
\end{equation}
The first three terms of (\ref{largezexp}) could be subtracted using allowed 
counterterms of the form $F(\varphi \varphi^*,R)$ \cite{Woodard:2006nt}. A
prominent feature of the remaining terms is the presence of derivatives of
the first slow roll parameter. These derivatives are typically very small
during inflation but Figure~\ref{epsilon} shows that they can be quite large
after the end of inflation.

The small field expansion derives from expanding the digamma functions in
expression (\ref{VL}) in powers of $x$,
\begin{eqnarray}
\psi\Bigl( \frac12 \!+\! \alpha(x)\Bigr) & = & \psi\Bigl( \frac1{1 \!-\! \epsilon}
\Bigr) - \psi'\Bigl( \frac1{1 \!-\! \epsilon} \Bigr) \frac{2x}{1 \!-\! \epsilon^2}
+ O(x^2) \; , \\
\psi\Bigl( \frac12 \!-\! \alpha(x)\Bigr) & = & \psi\Bigl( \frac{-\epsilon}{1 \!-\! 
\epsilon}\Bigr) + \psi'\Bigl( \frac{-\epsilon}{1 \!-\! \epsilon} \Bigr) 
\frac{2x}{1 \!-\! \epsilon^2} + O(x^2) \; , \\
\psi\Bigl( \frac12 \!+\! \beta(x)\Bigr) & = & -\gamma - \frac{\pi^2}{6} \frac{2x}{
(1 \!-\! \epsilon)^2} + O(x^2) \; , \\
\psi\Bigl( \frac12 \!-\! \beta(x)\Bigr) & = & -\frac{(1 \!-\! \epsilon)^2}{2 x}
+ 1 -\gamma + \Bigl[1 \!+\! \frac{\pi^2}{6}\Bigr] \frac{2 x}{(1 \!-\! \epsilon)^2}
+ O(x^2) \; . \qquad
\end{eqnarray}
The result is,
\begin{eqnarray}
\lefteqn{ \Delta V_L = \frac{H^4}{16 \pi^2} \Biggl\{ \Biggl[ \frac{R}{2 H^2}
\ln\Bigl[ \frac{(1 \!-\! \epsilon)^2 H^2}{s^2} \Bigr] \!+\! 1 \!-\! 8 \epsilon 
\!+\! 2 \epsilon^2 \!-\! 2 \epsilon' - \frac{(6 \epsilon' \!+\! \epsilon'')}{1
\!-\! \epsilon} \!-\! \frac{{\epsilon'}^2}{(1 \!-\! \epsilon)^2} \Biggr] z }
\nonumber \\
& & \hspace{2cm} + \frac12 \Bigl[ (\partial_n \!+\! 3 \!-\! 3 \epsilon)
(\partial_n \!+\! 4 \!-\! 2 \epsilon) \!-\! 2 \epsilon\Bigr] \Bigl[
\psi\Bigl( \frac1{1 \!-\! \epsilon}\Bigr) \!+\! \psi\Bigl( \frac{-\epsilon}{1
\!-\! \epsilon}\Bigr) \Bigr] z \nonumber \\
& & \hspace{0cm} + \frac12 (\partial_n \!+\! 3 \!-\! 3 \epsilon) (\partial_n 
\!+\! 4 \!-\! 2 \epsilon) \Bigl[ \psi'\Bigl( \frac1{1 \!-\! \epsilon}\Bigr) 
\!-\! \psi'\Bigl( \frac{-\epsilon}{1 \!-\! \epsilon}\Bigr) \Bigr] 
\frac{\epsilon z}{1 \!-\! \epsilon^2} + O(z^2) \Biggr\} . \qquad 
\label{smallzexp}
\end{eqnarray}
Note that the $1/\epsilon$ pole from $\psi(\frac{-\epsilon}{1 - \epsilon})$ on
the penultimate line of (\ref{smallzexp}) cancels against the double pole from
$\psi'( \frac{-\epsilon}{1 - \epsilon})$ on the last line.

\subsection{The Nonlocal Contribution}

The nonlocal contribution to the effective potential is obtained by 
substituting the nonlocal contribution (\ref{nonlocalprop}) to each coincident
propagator in (\ref{coinctrace3}), and then into expression (\ref{DeltaVdef}),
\begin{eqnarray}
\lefteqn{ \Delta V'_{\rm N}(\varphi \varphi^*) = q^2 N_u(n) + 2 q^2 e^{-2n} 
N_v(n) } \nonumber \\
& & \hspace{4cm} + \frac{q^2 H^2}{2 M^2} \Bigl(\partial_n \!+\! 3 \!-\! 
\epsilon \Bigr) \Bigl(\partial_n \!+\! 4\Bigr) \Bigl[ N_u(n) \!-\! N_{u_0}(n)
\Bigr] . \qquad \label{nonlocal1}
\end{eqnarray}
The nonlocal contributions to the various propagators are,
\begin{eqnarray}
N_u(n) &\!\!\! = \!\!\!& \int_{0}^{\kappa_{n-4}} \!\! \frac{d\kappa \, 
\kappa^2}{32 \pi^3 G} \Biggl[e^{\mathcal{U}_{2,3}(n,\kappa,\mu)} - 
e^{\mathcal{U}_{1}(n,\kappa,\mu)} \Biggr] , \qquad \label{newNu} \\
N_{u_0}(n) &\!\!\! = \!\!\!& \int_{0}^{\kappa_{n-4}} \!\! \frac{d\kappa \, 
\kappa^2}{32 \pi^3 G} \Biggl[e^{\mathcal{U}_{2}(n,\kappa,0)} - 
e^{\mathcal{U}_{1}(n,\kappa,0)} \Biggr] , \qquad \label{newNu0} \\
N_v(n) &\!\!\! = \!\!\!& \int_{0}^{\kappa_{n-4}} \!\! \frac{d\kappa \, 
\kappa^2}{32 \pi^3 G} \Biggl[e^{\mathcal{V}_{2,3}(n,\kappa,\mu)} - 
e^{\mathcal{V}_{1}(n,\kappa,\mu)} \Biggr] . \qquad \label{newNv}
\end{eqnarray}
The nonlocal nature of these contributions derives from the integration over
$\kappa$, which can be converted to an integration over $n_{\kappa}$,
\begin{equation}
\kappa \equiv e^{n_{\kappa}} \chi(n_{\kappa}) \qquad \Longrightarrow \qquad
\frac{d\kappa}{\kappa} = \Bigl[ 1 \!-\! \epsilon(n_{\kappa})\Bigr] dn_{\kappa}
\; . \label{kappaton}
\end{equation}
After this is done, any factors of $\kappa$ depend on the earlier geometry.

A number of approximations result in huge simplification. First, note from 
Figures~\ref{Temp3phasesA} and \ref{Temp3phasesB} that the ultraviolet 
approximation (\ref{U1def}) for $\mathcal{U}(n,\kappa,\mu)$ is typically more
negative than the late time approximations (\ref{U2def}) and (\ref{U3def}).
Figures~\ref{Trans3phasesA} and \ref{Trans3phasesB} show that the same rule 
applies to $\mathcal{V}(n,\kappa,\mu)$. Hence we can write,
\begin{equation}
N_u(n) \simeq \int_{0}^{\kappa_{n-4}} \!\! \frac{d\kappa \, \kappa^2}{16 \pi^3 G} 
\, e^{\mathcal{U}_{2,3}(n,\kappa,\mu)} \qquad , \qquad 
N_v(n) \simeq \int_{0}^{\kappa_{n-4}} \!\! \frac{d\kappa \, \kappa^2}{16 \pi^3 G} 
\, e^{\mathcal{V}_{2,3}(n,\kappa,\mu)} \; . \label{simp1}
\end{equation} 
Second, because the temporal and transverse frequencies are nearly equal, we 
can write,
\begin{equation}
\omega^2_u(n,\mu) \simeq \omega^2_v(n,\mu) \qquad \Longrightarrow \qquad
\mathcal{U}(n,\kappa,\mu) \simeq \mathcal{V}(n,\kappa,\mu) - 2 n \; . \label{simp2}
\end{equation}
When the mass vanishes there is so little difference between the ultraviolet
approximation (\ref{U1def}) and its late time extension (\ref{U2def}) that we
can ignore this contribution, $N_{u_0}(n) \simeq 0$. Next, Figures~\ref{Tempkdep} 
and \ref{Transkdep} imply that the late time approximations for 
$\mathcal{U}(n,\kappa,\mu)$ and $\mathcal{V}(n,\kappa,\mu)$ inherit their $\kappa$
dependence from the ultraviolet approximation at $n \simeq n_{\kappa} + 4$, which 
is itself independent of $\mu$,
\begin{equation}
n > n_{\kappa} + 4 \qquad \Longrightarrow \qquad \mathcal{U}_{2,3}(n,\kappa,\mu)
\simeq \mathcal{U}_1(n_{\kappa} \!+\! 4,\kappa,0) + f_{2,3}(n,\mu) \; ,
\label{simp4}
\end{equation}
where $f_{2,3}(n,\mu)$ can be read off from expressions (\ref{U2def}) and (\ref{U3def})
by omitting the $\kappa$-dependent integration constants. Finally, we can use the 
slow roll form (\ref{T2def}) for the amplitude reached after first horizon crossing 
and before the mass dominates,
\begin{equation}
e^{\mathcal{U}_1(n_{\kappa} + 4,\kappa,0)} \simeq \frac{\chi^2(n_{\kappa})}{2 \kappa^3}
\times C\Bigl( \epsilon(n_{\kappa}) \Bigr) 
\; .
\end{equation}
Putting it all together gives,
\begin{eqnarray}
\lefteqn{\Delta V_N(\varphi \varphi^*) \simeq 3 q^2 \!\! \int_{0}^{n-4} 
\!\!\!\!\!\!\! d n_{\kappa} \frac{ [1 \!-\! \epsilon(n_{\kappa})] \chi^2(n_{\kappa})
C(n_{\kappa}))}{32 \pi^3 G} \times e^{f_{2,3}(n,\mu)} } \nonumber \\
& & \hspace{-0.5cm} + \frac{q^2 \chi^2(n)}{2 \mu^2} ( \partial_n \!+\! 3 \!-\!
\epsilon )( \partial_n \!+\! 4) \!\! \int_{0}^{n-4} \!\!\!\!\!\! 
d n_{\kappa} \frac{ [1 \!-\! \epsilon(n_{\kappa})] \chi^2(n_{\kappa})
C(\epsilon(n_{\kappa}))}{32 \pi^3 G} \!\times\! e^{f_{2,3}(n,\mu)} \!. 
\qquad \label{nonlocal2}
\end{eqnarray}

\section{Conclusions}

In section 2 we derived an exact, dimensionally regulated, Fourier mode 
sum (\ref{propagator}) for the Lorentz gauge propagator of a massive photon on 
an arbitrary cosmological background (\ref{geometry}). Our result is expressed 
in terms of mode functions $t(\eta,k)$, $u(\eta,k,M)$ and $v(\eta,k,M)$ whose 
defining relations are (\ref{MMCeqn}), (\ref{tempeqn}) and (\ref{spaceeqn}), which
respectively represent massless minimally coupled scalars, massive temporal
photons, and massive spatially transverse photons. The photon propagator can 
also be expressed as a sum (\ref{scalarsum}) of bi-vector differential 
operators acting on the scalar propagators $i\Delta_t(x;x')$, $i\Delta_u(x;x')$ 
and $i\Delta_v(x;x')$ associated with the three mode functions. Because Lorentz 
gauge is an exact gauge there should be no linearization instability, even on 
de Sitter, such as occurs for Feynman gauge \cite{Kahya:2005kj,Kahya:2006ui}.

In section 3 we converted to a dimensionless form with time represented 
by the number of e-foldings $n$ since the beginning of inflation, and the wave 
number, mass and Hubble parameter all expressed in reduced Planck units, $\kappa 
\equiv \sqrt{8\pi G} \,k$, $\mu \equiv \sqrt{8\pi G} \, M$ and $\chi(n) \equiv 
\sqrt{8 \pi G} \, H(\eta)$. Analytic approximations were derived for the amplitudes
$\mathcal{T}(n,\kappa)$, $\mathcal{U}(n,\kappa,\mu)$ and $\mathcal{V}(n,\kappa,\mu)$
associated with each of the mode functions. Which approximation to use is controlled 
by first horizon crossing at $\kappa = e^{n_{\kappa}} \chi(n_{\kappa})$ and mass
domination at $\mu = \frac12 \chi(n_{\mu})$. Until shortly after first horizon
crossing we employ the ultraviolet approximations (\ref{T1def}), (\ref{U1def}) and
(\ref{V1def}). After first horizon crossing and before mass domination the 
appropriate approximations are (\ref{T2def}), (\ref{U2def}) and (\ref{V2def}).
And after mass domination (which $\mathcal{T}(n,\kappa)$ never experiences) the
amplitudes are well approximated by (\ref{U3def}) and (\ref{V3def}). The validity
of these approximations was checked against explicit numerical solutions for 
inflation driven by the simple quadratic model, and by the phenomenologically 
favored plateau model (\ref{StaroU}).

In section 4 we applied our approximations to compute the effective potential 
induced by photons coupled to a charged inflaton. Our result consists of a 
part (\ref{VL}) which depends locally on the geometry (\ref{geometry}) and a
numerically smaller part (\ref{nonlocal2}) which depends on the past history.
The local part was expanded both for the case of large field strength 
(\ref{largezexp}), and for small field strength (\ref{smallzexp}). The 
existence of the second, nonlocal contribution, was conjectured on the 
basis of indirect arguments \cite{Miao:2015oba} that have now been explicitly 
confirmed. Another conjecture that has been confirmed is the rough validity of
extrapolating de Sitter results \cite{Allen:1983dg,Prokopec:2007ak} from the
constant Hubble parameter of de Sitter background to the time dependent one
of a general cosmological background (\ref{geometry}). However, we now have
good approximations for the dependence on the first slow roll parameter 
$\epsilon(n)$.

We have throughout considered the inflaton field in the vector mass $M^2 \equiv
2 q^2 \varphi \varphi^*$ to be constant because this is how the ``effective 
potential'' is defined. It would be easy to relax this assumption with only 
minor changes in the result. In particular, equations (\ref{DeltaVdef}) and
(\ref{coinctrace3}) would still pertain. Our approximations for the propagators
would remain, so that renormalization would be unaffected. The only thing that
would change is that the $\partial_n$ derivatives in expression (\ref{coinctrace3})
would now act on $\psi$ as well as $\epsilon$ and $\chi$. This would produce 
some factors of $\epsilon$ and its first derivative through the relation,
\begin{equation}
\psi' {\psi'}^* = \frac12 (D\!-\!2) \epsilon \; .
\end{equation}

Our most important result is probably the fact that electromagnetic
corrections to the effective potential depend upon first and second derivatives
of the first slow roll parameter. One consequence is that the effective
potential from electromagnetism responds more strongly to changes in the 
geometry than for scalars \cite{Kyriazis:2019xgj} or spin one half 
fermions \cite{Sivasankaran:2020dzp}. This can be very important
during reheating (see Figure~\ref{epsilon}); it might also be significant if 
features occur during inflation. Another consequence is that there cannot be 
perfect cancellation between the positive effective potentials induced by 
bosons and the negative potentials induced by fermions \cite{Miao:2020zeh}.
Note that the derivatives of $\epsilon$ come exclusively from the 
constrained part of the photon propagator --- the $t(\eta,k)$ and $u(\eta,k)$
modes --- which is responsible for long range electromagnetic interactions. 
Dynamical photons --- the $v(\eta,k)$ modes --- produce no derivatives
at all. These statements can be seen from expression (\ref{coinctrace3}), 
which is exact, independent of any approximation.

We close with a speculation based on the correlation between the spin of 
the field and the number of derivatives it induces in the effective potential: 
scalars produce no derivatives \cite{Kyriazis:2019xgj}, spin one half fermions 
induce one derivative \cite{Sivasankaran:2020dzp}, and this paper has shown
that spin one vectors give two derivatives. It would be interesting to see if 
the progression continues for gravitinos (which ought to induce three derivatives)
and gravitons (which would induce four derivatives). Of course gravitons do not 
acquire a mass through coupling to a scalar inflaton, but they do respond to it, 
and the mode equations have been derived in a simple gauge \cite{Iliopoulos:1998wq,
Abramo:2001dc}. Until now it was not possible to do much with this system because
it can only be solved exactly for the case of constant $\epsilon(n)$, however,
we now have a reliable approximation scheme that can be used for arbitrary
$\epsilon(n)$. Further, we have a worthy object of study in the graviton 1-point
function, which defines how quantum 0-point fluctuations back-react to change
the classical geometry. At one loop order it consists of the same sort of 
coincident propagator we have studied in this paper. On de Sitter background the
result is just a constant times the de Sitter metric \cite{Tsamis:2005je}, which
must be absorbed into a renormalization of the cosmological constant if ``$H$''
is to represent the true Hubble parameter. Now suppose that the graviton 
propagator for general first slow roll parameter consists of a local part with 
up to 4th derivatives of $\epsilon(n)$ plus a nonlocal part. That sort of result
could {\it not} be absorbed into any counterterm. So perhaps there is one loop
back-reaction after all \cite{Geshnizjani:2002wp}, and de Sitter represents a 
case of unstable equilibrium?

\newpage

\centerline{\bf Acknowledgements}

This work was partially supported by Taiwan MOST grants 
108-2112-M-006-004 and 107-2119-M-006-014; by NSF grants PHY-1806218 
and PHY-1912484; and by the Institute for Fundamental Theory at the 
University of Florida.

\end{document}